\documentclass[aps, pra,11pt,showpacs]{revtex4-1}
\usepackage{amsmath}    
\usepackage{graphicx}   
\usepackage{verbatim}   
\usepackage{color}      
\usepackage{subfigure}  
\usepackage{hyperref}   
\usepackage{dcolumn}
\begin{document}

\title{Analysis of evanescently-coupled pairs of spin-polarised vertical-cavity surface-emitting lasers}

\author{M.P. Vaughan$^{1}$}\email[Corresponding author: ]{mpvaug@essex.ac.uk}
\author{H. Susanto$^{2}$}
\author{I.D. Henning$^{1}$}
\author{M.J. Adams$^{1}$}

\affiliation{$^{1}$School of Computer Science and Electronic Engineering, University of Essex, Wivenhoe Park, Colchester CO4 3SQ, United Kingdom}

\affiliation{$^{2}$Department of Mathematical Sciences, University of Essex, Wivenhoe Park, Colchester CO4 3SQ, United Kingdom}

\date{\today}

\begin{abstract}
 A general model for the dynamics of arrays of coupled, spin-polarised lasers is derived, which is shown to reduce to both the spin flip model in a single cavity and the coupled mode model for a pair of guides in the appropriate limit. The general model is able to deal with waveguides of any geometry with any number of supported normal modes. A unique feature of the model is that it allows for independent polarisation of the pumping in each laser. The particular geometry is shown to be introduced via `overlap factors', which are a generalisation of the optical confinement factor. These factors play an important role in determining the laser dynamics. The model is specialised to the case of a general double-guided structure, which is then analysed and simulated numerically. For this case it is found that increasing the ellipticity of the pumping tends to enhance the regions where stable solutions are predicted in the plane of pumping strength versus guide separation.    
\end{abstract}

\pacs{42.82.Et, 02.30.Oz, 05.45.-a, 42.55.Px}

% 42.82.Et Waveguides, couplers, and arrays
% 02.30.Oz   Bifurcation theory (see also 47.20.Ky in fluid dynamics)
%  05.45.−a Nonlinear dynamics and chaos 
% 42.55.Px Semiconductor lasers; laser diodes
% 67.57.Lm Spin dynamics
% 85.75.−d Magnetoelectronics; spintronics: devices exploiting spin polarized transport or integrated magnetic fields
% 42.60.−v Laser optical systems: design and operation
% 42.60.Mi Dynamical laser instabilities; noisy laser behavior

% 42.82.Et    Waveguides, couplers, and arrays
% 02.30.Oz   Bifurcation theory
% 05.45.−a   Nonlinear dynamics and chaos
% 42.55.Px  Semiconductor lasers; laser diodes

\maketitle

%\tableofcontents

\section{Introduction}
The spin flip model (SFM)~\cite{san1995light} is now well-established as a quantitative description of the effects of electron spin and light polarisation in vertical cavity surface-emitting lasers (VCSELs) with quantum well (QW) active regions. The basic SFM consists of four coupled rate equations (two for spin-polarised carriers and two for polarised field components) and includes rates of carrier recombination, photon field decay and electron spin relaxation (spin relaxation of holes is usually assumed to be instantaneous). The nonlinear dispersion that couples the carrier concentrations to the phases of the optical fields is described by the linewidth enhancement factor, and the field interactions due to nonlinear anisotropies are included via rates of birefringence and dichroism. For conventional VCSELs, the SFM has been applied to explain experimental results of polarisation switching (PS)~\cite{martin1997polarizationProperties,travagnin1996role,travagnin1997erratum}. An “extended SFM”~\cite{balle1999mechanisms} that accounts for thermal effects and includes a realistic spectral dependence of the gain and the index of refraction of the QWs has been used~\cite{sondermann2004experimental} to explain experimental results on elliptically polarised dynamical states that occur in the polarisation dynamics of VCSELs in the vicinity of one type of PS. For a more complete discussion of polarisation dynamics in VCSELs the reader is referred to~\cite{Panajotov2012Polarization}.

A further development of the extended SFM~\cite{mulet2002spatio} includes a description of the spatial variation of the electromagnetic modes and the carrier densities. The variation in the longitudinal direction is dealt with by integration over the length of the VCSEL cavity whilst the radial and azimuthal variation is described by accurate solutions of the wave equation. The model assumes a given functional dependence of the guiding mechanisms (built-in refractive index and thermal lensing) as well as the spatial dependence of the current density. The transverse mode behaviour of gain-guided, bottom and top-emitter VCSELs were studied and it was shown that the stronger the thermal lens, the stronger the tendency toward multimode operation, which indicates that high lateral uniformity of the temperature is required in order to maintain single mode operation in gain-guided VCSELs. Also, close-to-threshold numerical simulations showed that, depending on the current profile, thermal lensing strength and relative detuning, different transverse modes could be selected.

Another version of the extended SFM~\cite{masollera2008modeling} includes a rate equation for the temperature of the active region, which takes into account decay to a fixed substrate temperature, Joule heating and heating due to non-radiative recombination. The temperature dependence of the PS point is characterised in terms of various model parameters, such as the room-temperature gain-cavity offset, the substrate temperature, and the size of the active region.

The SFM has also been widely applied to describe the behaviour of spin-VCSELs whose output polarisation can be controlled by injection of spin-polarised electrons using either electrical or optical pumping (for a review with more details, see~\cite{gerhardt2012spin}). In the latter case the polarisation of the optical pump is included~\cite{gahl1999polarization} to reveal its effect on the output polarisation~\cite{gerhardt2006enhancement,adams2009parametric,adams2018algebraic}. The SFM has also been used~\cite{li2010birefringence,lindemann2016frequency,torre2017high} to explain experimental results on high-speed polarisation oscillations that result from competition between the spin-flip processes, dichroism and birefringence.

It is clear from this brief summary of the SFM and its applications that the structures studied have been limited to single lasers, either conventional electrically driven VCSELs or spin-VCSELs which may be pumped electrically or optically. In the present contribution we seek to extend the range of application to include structures where two or more evanescently-coupled lasers are arranged in parallel to form arrays with the possibility of different lasers having differing pumping polarisation. To the best of our knowledge this configuration has not been analysed previously, although there is one report~\cite{hendriks1996phase} of an experiment where optical pumping with orthogonally polarised beams was used to study the interaction between two VCSELs as a function of their separation. There is of course a vast literature on laser arrays because of their important practical applications as high-power sources (including, most recently, for 3D sensing in smartphones~\cite{ebeling2018vertical}) and very sophisticated models of VCSEL arrays have been developed~\cite{czyszanowski2013spatial}. Arrays of coupled lasers are also of fundamental interest in view of the range of nonlinear dynamics that they can exhibit (see, for example,~\cite{blackbeard2014synchronisation} and references cited therein). Although there is a need sometimes to stabilise the polarisation of such arrays of VCSELs, the possibility of manipulating the output polarisation of an array by means of independent pumping polarisations has not yet been considered. This, together with the issue of how the array dynamics is affected by this pumping arrangement, is the motivation for the present study.

In Section~\ref{sec:general}, we derive a model for guided mode lasers of general geometry with any number of guides and any number of normal modes. An important aspect of this model is the introduction of the \emph{overlap factors}, discussed in detail in Section~\ref{sec:overlap}. These are calculated by integrating products of the spatial mode solutions of the Helmholtz equation over the active regions. As such, they represent a generalisation of the optical confinement factor. It is through these factors that the particular geometry of the waveguide is introduced and their effect on the laser dynamics can be quite significant, as indicated in Ref~\cite{vaughan2019stability} in comparison to the coupled mode model~\cite{adams2017effects}. 

\begin{figure}[!ht]
\centering
\subfigure[\ Slab waveguide.]{
\includegraphics[width=0.4\textwidth]{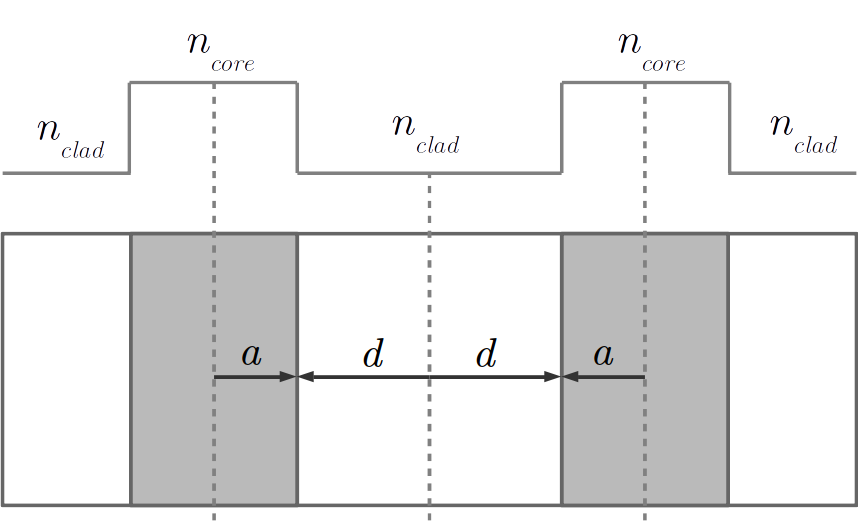}}
\subfigure[\ Circular waveguide.]{
\includegraphics[width=0.4\textwidth]{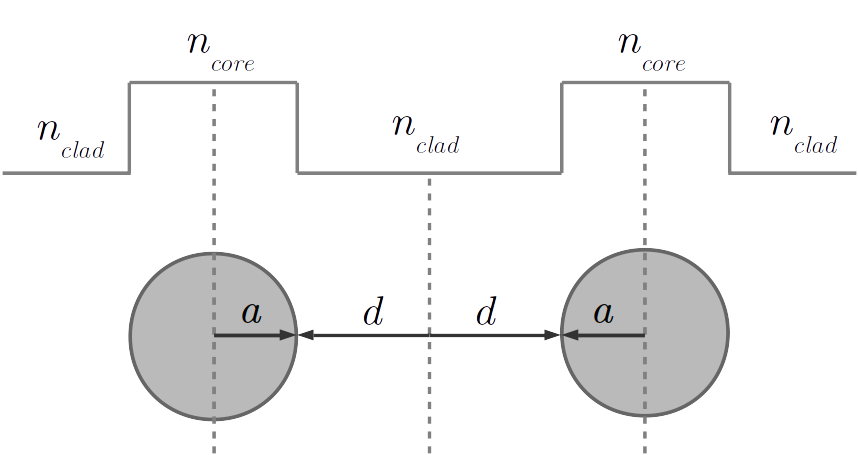}}
\caption{\label{fig:waveguides} The refractive index profiles of the double guided structures used in this work. Here, the distance between the guides is given as $2d$, whilst $a$ is used both for the half-width of a slab guide and the radius of a circular guide. Elsewhere in this work, we use $n_{1} = n_{core}$ and $n_{2} = n_{clad}$ for brevity.}
\end{figure}

\begin{figure}[!ht]
\centering
\includegraphics[width=0.45\textwidth]{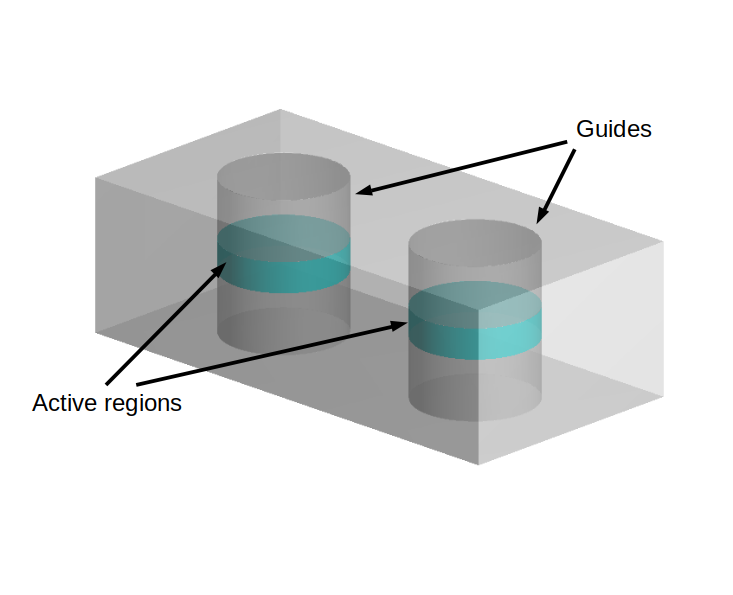}
\caption{\label{fig:VCSELS} A 3D schematic of two coupled circular waveguides encapsulating the essence of the application to a pair of VCSEL cavities. Shown are the cylindrical waveguide regions incorporating the active areas. Pumping is assumed to be confined to these regions. Note that we have omitted the Bragg stack mirrors and substrate from this figure.}
\end{figure}

Familiarity with the overlap factors should give the necessary physical intuition into their properties and limiting behaviour that we frequently exploit in the derivation of the double-guided model in Section~\ref{sec:double}. Here, we specialise to the case of just two guides and consider only the lowest two normal modes. This model is still quite general in regards to the waveguide geometry that may be simulated, although it is particularly appropriate for the case of symmetric waveguides. In this paper, we look at two particular cases: equal slab guides and equal circular guides, both with real, stepped refractive index profiles as shown schematically in Fig.~\ref{fig:waveguides}. The application to coupled VCSELs with circular guides is indicated in the schematic of Fig.~\ref{fig:VCSELS}, omitting the Bragg mirrors, substrate and other structural details. Note that in Figs~\ref{fig:waveguides} and ~\ref{fig:VCSELS}, a resonant cavity is assumed with propagation in the z-direction, i.e. normal to the plane of optical confinement. No further account is taken of the z-direction in what follows and the values of parameters appearing in the analysis are assumed to be averaged over the cavity length. For widely separated guides, we show in Section~\ref{sec:limit} that the model reduces to both the SFM~\cite{san1995light,martin1997polarization, gahl1999polarization} and coupled mode model~\cite{adams2017effects} in the appropriate limits.

Having established the mathematical model, we investigate the effect of varying the optical pump polarisation in each guide via numerical simulation in Section~\ref{sec:results}. A novel feature of this model is that it allows us to examine the spatial variation of the circularly polarised components of the optical intensity and the optical ellipticity throughout the waveguide structure. Examples of this are given in Section~\ref{sec:spatial}. In Section~\ref{sec:stability} we give some introductory examples of stability boundaries in the plane of total pump power and normalised guide separation. This illustrates how we can use this model to investigate the effect of independently varying the pump polarisation in each guide. More generally, we may also vary the overall pump power or adjust the relative sizes of each guide, thereby introducing an effective frequency detuning. Such investigations are deferred for future study.   

\section{The general model}\label{sec:general}
\subsection{The optical rate equations}
In any waveguiding structure defined by a spatially-dependent relative permittivity $\epsilon(\mathbf{r})$, we will have optical mode solutions $\Phi_{k}(\mathbf{r})$ satisfying the Helmholtz equation

\begin{equation}
 \left[\nabla^{2} + \frac{\Omega^{2}\epsilon(\mathbf{r})}{c^{2}}\right]\Phi_{k}(\mathbf{r}) = 0, \label{eq:Helmholtz}
\end{equation}

\noindent where $\Omega$ is a reference frequency taken to be the average of the modal frequencies and $k$ is the transverse  mode index. The $\Phi_{k}(\mathbf{r})$ are known as the normal modes or, sometimes, supermodes of the waveguide. For modes of a given order, the orthogonal polarisations have almost exactly the same spatial profile (having checked this numerically for cases of interest) and we shall assume this to be precisely true. Thus, each mode $\Phi_{k}(\mathbf{r})$ may be associated with two polarisations. Later, we shall explicitly formulate this in terms of left and right-circularly polarised light.

After cancelling a phase factor $e^{i\beta z}$, where $\beta$ is the propagation constant along the cavity length (see Appendix~\ref{app:optical}), the total optical field may then be written as a superposition of the normal modes as

\begin{equation}
 \mathbf{E}(x,y, t) = \sum_{k}\mathbf{E}_{k}(t)\Phi_{k}(x,y)e^{-i\nu_{k}t}, \label{eq:E}
\end{equation}

\noindent where $\mathbf{E}_{k}(t)$ is a time dependent Jones vector incorporating the polarisation and $\nu_{k}$ is the modal frequency, determined via solution of the Helmholtz equation for the mode. Hereafter, we shall use $\mathbf{r} = x\mathbf{e}_{x} = y\mathbf{e}_{y}$ for brevity wherever we need to denote spatial coordinates, but it should be remembered that $\mathbf{r}$ is confined to the $x-y$ plane.

Starting from the general form of Maxwell's wave equation and applying the slowly varying envelope approximation (SVEA), as described in Appendix~\ref{app:optical}, we may obtain a set of optical rate equations for the complex amplitudes of the normal modes 

\begin{align}
 \frac{\partial E_{k,\pm}}{\partial t} &= \left[i\left(\nu_{k}  - \Omega\right) - \frac{1}{2\tau_{p}}\right]E_{k,\pm} - \left[\gamma_{a} + i\gamma_{p}\right]E_{k,\mp} \nonumber \\
&+ \sum_{k'}\frac{c}{2n_{g}}\left(1 + i\alpha\right)E_{k',\pm}e^{i\Delta\nu_{kk'}t}\sum_{i}\overline{g}_{\pm}^{(i)}\Gamma_{kk'}^{(i)}. \label{eq:wave_complex}
\end{align}

\noindent Here, the $\pm$ subscripts denote the right ($+$) and left ($-$) circularly polarised components, $\tau_{p}$ is photon lifetime, $c$ is the speed of light, $n_{g}$ is the group refractive index and $\alpha$ is the linewidth enhancement factor, defined in terms of the change in the real and imaginary components of the electric susceptibility, $\Delta\chi_{\pm}'$ and $\Delta\chi_{\pm}''$ respectively, by $\alpha = -\Delta\chi_{\pm}'/\Delta\chi_{\pm}''$. Note that we have adopted this sign convention for consistency with the SFM model~\cite{san1995light,martin1997polarization, gahl1999polarization} and is opposite to that used in Ref~\cite{adams2017effects}. Hence the values of $\alpha$ used in this work take the opposite sign to that in the latter reference.

Each polarisation component is coupled to the other via the birefringence rate $\gamma_{p}$ and dichroism rate $\gamma_{a}$. The time-dependent exponential factor involves the difference between modal frequencies $\Delta\nu_{kk'} = \nu_{k} - \nu_{k'}$.

The summation over $i$ in the last term of \eqref{eq:wave_complex} is over the optically confining guides. Here, we have defined optical overlap factors $\Gamma_{kk'}^{(i)}$ for the $i$th guide via 

\begin{equation*}
 \overline{g}_{\pm}^{(i)}\Gamma_{kk'}^{(i)} \equiv \int_{(i)} g_{\pm}(\mathbf{r})\Phi_{k}(\mathbf{r})\Phi_{k'}(\mathbf{r})~d^{2}\mathbf{r},
\end{equation*}

\noindent where $\overline{g}_{\pm}^{(i)}$ is the average gain for each polarisation in guide $(i)$ and the integral is over the $i$th guide. In practice, we take the gain to be spatially constant over a guide and zero outside it. Hence, in this paper, the overlap factors are defined simply by

\begin{equation}
 \Gamma_{kk'}^{(i)} \equiv \int_{(i)} \Phi_{k}(\mathbf{r})\Phi_{k'}(\mathbf{r})~d^{2}\mathbf{r} \label{eq:Gamma_kk}
\end{equation}

\noindent and we normalise the spatial profiles so that 

\begin{equation*}
 \int \left|\Phi_{k}(\mathbf{r})\right|^{2}~d^{2}\mathbf{r} = 1, 
\end{equation*}

\noindent where the integral is over all space.

\subsection{The carrier rate equations}
The rate equations for spin-polarised populations of carriers may be  derived from the optical Bloch equations. The general result for the spatially dependent carrier concentrations $N_{\pm}(\mathbf{r},t)$ and circularly polarised optical fields $E_{\pm}(\mathbf{r},t)$, including spin relaxation may be found to be given by

\begin{equation}
 \frac{\partial N_{\pm}}{\partial t} =  -\frac{N_{\pm}}{\tau_{N}} + \Lambda_{\pm} - \gamma_{J}\left(N_{\pm} - N_{\mp}\right) - \frac{c}{n_{g}}g_{\pm}\left(N_{\pm}\right)\left|E_{\pm}\right|^{2}, \label{eq:dNpm}
\end{equation}

\noindent where $\tau_{N}$ is the carrier lifetime, $\Lambda_{\pm}$ is the pumping rate and $\gamma_{J}$ is the spin relaxation rate. The $\pm$ subscripts on $N$ refer to spin up ($+$) and spin-down ($-$) carriers, which couple directly with right ($+$) and left ($-$) circularly polarised photons respectively. Here, we assume that all carrier pumping, whether that be optical or electrical, is confined to the active region, and hence the effects of lateral diffusion are neglected at this time.

Note that we take $\left|E_{\pm}(\mathbf{r},t)\right|^{2} = S(\mathbf{r},t)$ to be the photon density and hence $E_{\pm}$ does not have dimensions of electric field in \eqref{eq:dNpm}. This is unproblematic, since the optical rate equations \eqref{eq:wave_complex} may be multiplied by any arbitrary factor to match the dimensions of $E_{\pm}$ in \eqref{eq:dNpm} without changing the dynamics.

The earlier assumption that the gain is spatially constant over a given guide and zero between guides requires a similar assumption for the carrier concentrations. We shall assume a linear gain model of the form $g(N) = g_{0}(N- N_{0})$, where $N_{0}$ is the transparency concentration, so if $g(N)$ is a step function, $N \le N_{0}$ outside the active regions. With the pumping confined to the active regions and no spatial diffusion, we may take any optical loss in the cladding regions to have been absorbed into the cavity loss rate $\kappa$. We may therefore take the carrier concentration in this region to be exactly $N_{0}$.

Taking the spatial dependence to be in the $x-y$ plane only, we may then put

\begin{equation}
 N_{\pm}(t,x,y) = \sum_{i}N_{\pm}^{(i)}(t)\xi^{(i)}(x,y), \label{eq:Nsum}
\end{equation}

\noindent where $N^{(i)}(t)$ is the time dependent carrier concentration in the $i$th guide and $\xi^{(i)}(x,y)$ is a step function. Since we may subtract $N_{0}$ from either side (which we do on normalisation), we may take this as effectively giving zero outside the active regions. Applying this assumption, we find that the rate equations for the spin-polarised concentrations in the $(i)$th guide are given by
\begin{align}
 \frac{\partial N_{\pm}^{(i)}}{\partial t} &=  -\frac{N_{\pm}^{(i)}}{\tau_{N}} + \Lambda_{\pm}^{(i)} - \gamma_{J}\left(N_{\pm}^{(i)} - N_{\mp}^{(i)}\right)  \nonumber \\
 &-  \frac{c}{n_{g}}\sum_{k,k'}E_{k,\pm}^{*}E_{k',\pm}\overline{g}_{\pm}^{(i)}\Gamma_{kk'}^{(i)}e^{i\Delta\nu_{kk'}t}, \label{eq:dNpmi}
\end{align}
where the $(i)$ superscripts on a quantity label the values of that quantity in each guide. The details of the derivation are given in Appendix~\ref{app:carrier}.

In this study, we assume that $|\Delta\nu_{k,k'}|\ll\nu_{k},\nu_{k'}$. This assumption is physically relevant provided that the coupled waveguides are well separated (relative to a characteristic length). In that case, the modal frequency of the symmetric and anti-symmetric modes will become very similar. However, the assumption may not be so accurate for the frequency difference between different orders of transverse modes. Our main interest lies only in the symmetric and anti-symmetric versions of the lowest order mode of a two-guide structure, in which case the assumption is well-justified. Under the assumption, the non-autonomous equations \eqref{eq:wave_complex} and \eqref{eq:dNpmi} will be simplified to \begin{align}
\frac{\partial \tilde{E}_{k,\pm}}{\partial t} &=   \left[i\left(\nu_{k} - \Omega\right) - \frac{1}{2\tau_{p}}\right]\tilde{E}_{k,\pm} - \left[\gamma_{a} + i\gamma_{p}\right]\tilde{E}_{k,\mp} \nonumber \\
&+   \sum_{k'}\frac{c}{2n_{g}}\left(1 + i\alpha\right)\tilde{E}_{k',\pm}\sum_{i}\overline{g}_{\pm}^{(i)}\Gamma_{kk'}^{(i)}. \label{eq:optical_rate}
\end{align}
and
\begin{align}
 \frac{\partial N_{\pm}^{(i)}}{\partial t} &=  -\frac{N_{\pm}^{(i)}}{\tau_{N}} + \Lambda_{\pm}^{(i)} - \gamma_{J}\left(N_{\pm}^{(i)} - N_{\mp}^{(i)}\right)  \nonumber \\
 &-  \frac{c}{n_{g}}\sum_{k,k'}\tilde{E}_{k,\pm}^{*}\tilde{E}_{k',\pm}\overline{g}_{\pm}^{(i)}\Gamma_{kk'}^{(i)}. \label{eq:carrier_rate}
\end{align}
Equations \eqref{eq:optical_rate} and \eqref{eq:carrier_rate} then represent the general model for any number of normal modes and any number of confining guides.

In the following, instead of the model \eqref{eq:wave_complex} and \eqref{eq:dNpmi}, we will analyse Eqs.\ \eqref{eq:dNpmi} and \eqref{eq:carrier_rate}. Nevertheless, the analysis of the latter equations will still be valid in recognising unstable solutions of the former ones with a critical eigenvalue $\lambda$ that is much larger than $|\Delta\nu_{k,k'}|$. That is because before the factor $\mathrm{exp}(i\Delta\nu_{k,k'} t)$, that is slowly varying, starts to have any effect in the system, the unstable solution will already show its instability.

Stable solutions of Eqs.\ \eqref{eq:dNpmi} and \eqref{eq:carrier_rate} will also correspond to stable solutions of the model \eqref{eq:wave_complex} and \eqref{eq:dNpmi} if the time frame is of order $\mathcal{O}(1/|\Delta\nu_{k,k'}|)$. In this way, we also conjecture that if all eigenvalues of a solution are far away from the imaginary axis, then the presence of the slowly varying phase $\mathrm{exp}(i\Delta\nu_{k,k'} t)$ should not change the eigenvalues much.

\section{The overlap factors}\label{sec:overlap}

\subsection{Equal guides}\label{sec:equal}

\begin{figure}[!ht]
\centering
\subfigure[\ Symmetric and anti-symmetric modes for $d/a = 0.5$.]{
\includegraphics[width=0.45\textwidth]{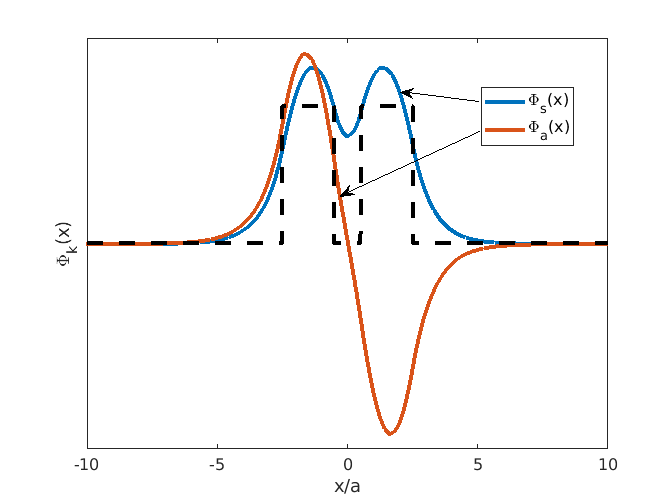}}
\subfigure[\ Products of spatial modes for $d/a = 0.5$.]{
\includegraphics[width=0.45\textwidth]{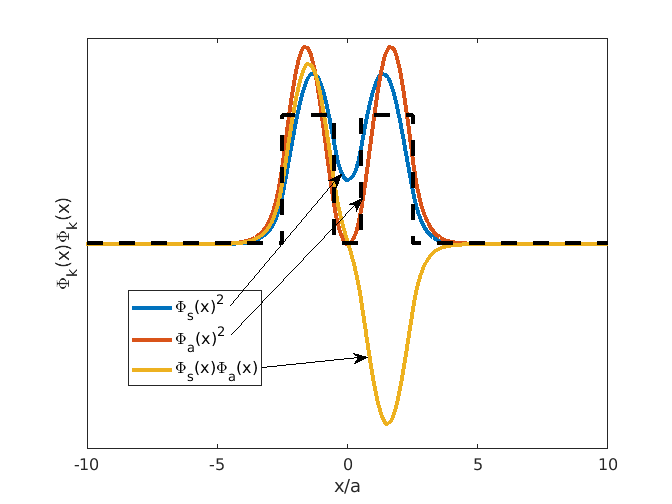}}
\caption{\label{fig:Phi_05} Spatial modes and their products for an equal width double slab guide with $v = \pi/2$, $a = 4~\mu\mathrm{m}$ and $d/a = 0.5$. For such closely spaced guides, the spatial profile of the products $\Phi_{s}^{2}(x)$ and $\Phi_{a}^{2}(x)$ are noticeably different.}
\end{figure}

The overlap factors defined by \eqref{eq:Gamma_kk} are calculated from the spatial modal solutions of the Helmholtz equation $\Phi(\mathbf{r})$. Details of the solutions used in this work are given in Appendix~\ref{app:Helmholtz}. For explanatory purposes, it suffices to consider the 1D solutions of a pair of slab guides. Firstly, we shall just consider equal guides with the same refractive index $n_{1}$ in the core regions and $n_{2}$ elsewhere, although our treatment of the rate equations in Section~\eqref{sec:double} is general enough to deal with twin guides of any geometry. The parameters used in our example calculations are listed in Table~\ref{tab:waveguide}. 

\begin{table}
\caption{\label{tab:waveguide}Waveguide parameters used in the solution of the Helmholtz equation.}
\begin{ruledtabular}
\begin{tabular}{llll}
Parameter & Value & Unit & Description\\
\hline
$n_{1}$  & 3.400971  &  & Core refractive index\\
$n_{2}$  & 3.4  &  & Cladding refractive index\\
$a$ & 4 & $\mu\mathrm{m}$ & Half guide width / radius
\end{tabular}
\end{ruledtabular}
\end{table}

The guides may be characterised by a normalised decay constants $u$ (in the core regions) and $w$ (in the cladding regions)

\begin{equation}
 u = a\sqrt{\left(\frac{n_{1}\nu_{k}}{c}\right)^{2} - \beta^{2}} \label{eq:u}
\end{equation}

\noindent and

\begin{equation}
 w = a\sqrt{\beta^{2} - \left(\frac{n_{2}\nu_{k}}{c}\right)^{2}}, \label{eq:w}
\end{equation}

\noindent where $a$ is either the half-width of a slab waveguide or the radius of a circular guide, as illustrated in Fig.~\ref{fig:waveguides}. These decay constants can also be combined into a conventional `normalised frequency' $v$, defined as

\begin{equation}
 v = \sqrt{u^{2} + w^{2}}. \label{eq:v}
\end{equation}

\noindent In practice, we shall take \eqref{eq:u} to \eqref{eq:v} to refer to the value for a single isolated guide.

\begin{figure}[!ht]
\centering
\subfigure[\ Symmetric and anti-symmetric modes for $d/a = 5$.]{
\includegraphics[width=0.45\textwidth]{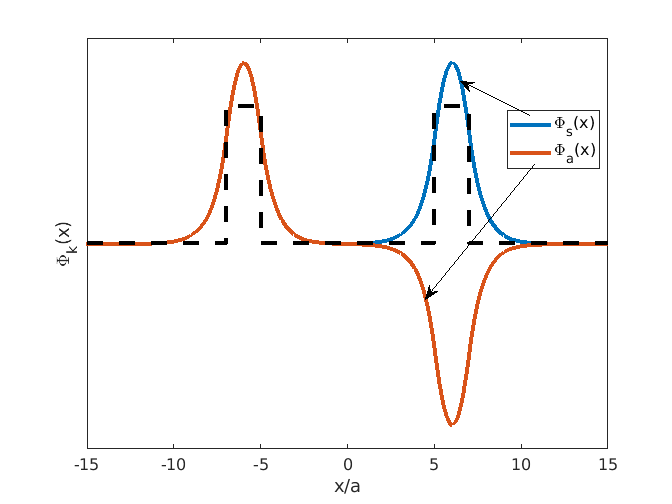}}
\subfigure[\ Products of spatial modes for $d/a = 5$.]{
\includegraphics[width=0.45\textwidth]{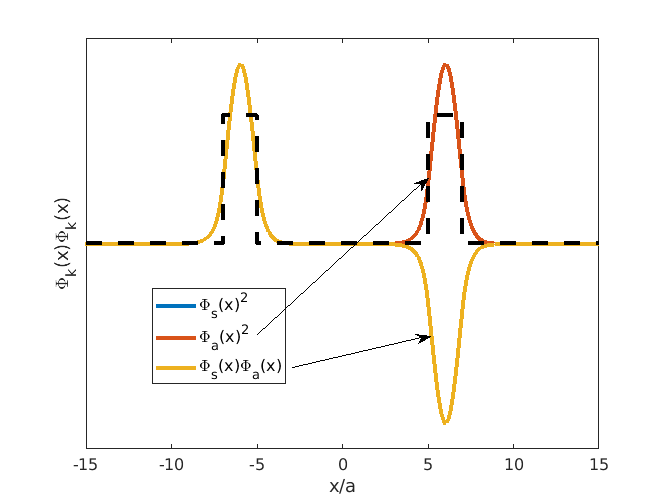}}
\caption{\label{fig:Phi_5da} Spatial modes and their products for an equal width double slab guide with $v = \pi/2$, $a = 4~\mu\mathrm{m}$ and $d/a = 5$. In this case the spatial profiles approach those of isolated, single guides. The spatial profile of the products $\Phi_{s}^{2}(x)$ and $\Phi_{a}^{2}(x)$ now approximately coincide. The modulus of $\Phi_{s}(x)\Phi_{a}(x)$ also becomes the same as these products but the sign of the product is different in each guide.}
\end{figure}

In a solitary guide, for values of $v < \pi/2$, only one guided mode is supported. In a double guide, only the lowest two modes are supported. We therefore refer to such structures as being `weakly-guiding'. Figures~\ref{fig:Phi_05}~$(a)$ and \ref{fig:Phi_5da}~$(a)$ show examples of the two lowest modes for light of wavelength $\lambda = 1.3~\mu\mathrm{m}$ in slab waveguides with $a = 4~\mu\mathrm{m}$ for $d/a = 0.5$ and $d/a = 5$ respectively ($2d$ is the edge-to-edge distance between the guides). The dashed lines indicate the refractive index profile of the structure. 

Due to the symmetry of the guides, the lowest mode has even parity and the second lowest, odd parity. We refer to these as the `symmetric' and `anti-symmetric' modes respectively and label them by suffixes $s$ and $a$ respectively. Note that the anti-symmetric mode $\Phi_{a}(x)$ always goes through zero in between the guides whereas the $\Phi_{s}(x)$ does not. This qualitative behaviour persists even when we break the symmetry of the guides, so that we may still use $s$ and $a$ as labels, although they would then be distinguished by topology rather than geometric symmetry.   

The overlap factors $\Gamma_{k'k}^{(i)}$ are found by integrating the products of the spatial modes over each guide, as in \eqref{eq:Gamma_kk}. These products are shown in Figs.~\ref{fig:Phi_05}~$(b)$ and \ref{fig:Phi_5da}~$(b)$ for the same structures. We can see in Fig.~\ref{fig:Phi_05}~$(b)$ that for the closely spaced guides, the products $\Phi_{s}^{2}(x)$, $\Phi_{a}^{2}(x)$ and the modulus $|\Phi_{s}(x)\Phi_{a}(x)|$ are noticeably different. Hence, we note that, in general,

\begin{equation*}
 \Gamma_{ss}^{(i)} \ne \Gamma_{aa}^{(i)} \ne \left|\Gamma_{sa}^{(i)}\right|.
\end{equation*}

\noindent However, in the case of equal guides, by symmetry, we do have

\begin{equation*}
 \Gamma_{ss}^{(1)} = \Gamma_{ss}^{(2)}~\mathrm{and}~\Gamma_{aa}^{(1)} = \Gamma_{aa}^{(2)}~~~(\mathrm{equal~guides}).
\end{equation*}

\noindent Also by symmetry the integral of $|\Phi_{s}(x)\Phi_{a}(x)|$ will be the same in each guide, although the sign will be opposite, so

\begin{equation*}
 \Gamma_{sa}^{(1)} = -\Gamma_{sa}^{(2)}.
\end{equation*}

As the separation between the guides gets larger, the spatial profiles become like those of isolated guides, as we see in Figs.~\ref{fig:Phi_5da}~$(a)$ and \ref{fig:Phi_5da}~$(b)$. The difference between these profiles and those of an isolated guide are (i) the normalisation - the integral of the squared modulus of the spatial modes will be half that of the optical confinement factor - and (ii) the sign of the anti-symmetric mode is flipped in one of the guides. In fact, as the separation tends to infinity, we may obtain the wave functions of the isolated guides by adding and subtracting the modes via

\begin{equation*}
 \Phi_{1}(x) = \lim_{d\to\infty} \frac{1}{\sqrt{2}}\left(\Phi_{s}(x) + \Phi_{a}(x)\right)~~~(\mathrm{equal~guides}) 
\end{equation*}

\noindent and 

\begin{equation*}
 \Phi_{2}(x) = \lim_{d\to\infty} \frac{1}{\sqrt{2}}\left(\Phi_{s}(x) - \Phi_{a}(x)\right)~~~(\mathrm{equal~guides}). 
\end{equation*}

\noindent This is the basis for the definition of the `composite modes' in terms of the normal modes defined in \eqref{eq:Eplus} and \eqref{eq:Eminus} defined in the next section. We then have

\begin{equation*}
 \int_{(1)} \left|\Phi_{1}(x)\right|^{2}dx = \int_{(2)} \left|\Phi_{2}(x)\right|^{2}dx = \Gamma_{S}~~~(\mathrm{equal~guides}),
\end{equation*}

\noindent where $\Gamma_{S}$ is the optical confinement factor of a single guide. In this limit, we also have

\begin{equation*}
 \lim_{d\to\infty}\Gamma_{ss}^{(i)} = \lim_{d\to\infty}\Gamma_{aa}^{(i)} = \lim_{d\to\infty}\left|\Gamma_{sa}^{(i)}\right| = \frac{\Gamma_{S}}{2}~~~(\mathrm{equal~guides}). 
\end{equation*}

\begin{figure}[!ht]
\centering
\includegraphics[width=0.45\textwidth]{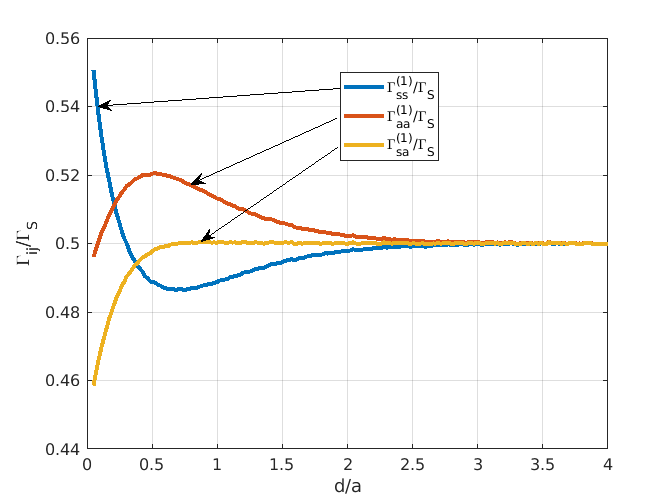}
\caption{\label{fig:overlaps_equal} Plot of the overlap factors for a weakly-guiding ($v = \pi/2$) symmetric slab structure as a function of spatial separation between the guides. Here, $2d$ is the edge-to-edge separation between the guides and $2a$ is the guide width ($8~\mu\mathrm{m}$ as in Figs.~\ref{fig:Phi_05} and ~\ref{fig:Phi_5da}). Only the overlap factors for guide $(1)$ are shown. The factors for guide $(2)$ are the same except that $\Gamma_{sa}^{(2)}$ is negative.}
\end{figure}

Figure~\ref{fig:overlaps_equal} shows the variation of the overlap factors for equal width slab guides with $v = \pi/2$ as a function of guide separation. The guide widths and all other parameters are the same as used for the modes shown in Figs.~\ref{fig:Phi_05} and ~\ref{fig:Phi_5da}. The overlap factors are divided by $\Gamma_{S}$ and we clearly see the tendency of all values to $\Gamma_{S}/2$ at large separation. Only the factors for guide $(1)$ are shown since the values for guide $(2)$ are the same except for the change of sign on $\Gamma_{sa}^{(2)}$. 

For such a weakly guiding structure, there is significant variation of the factors for $d/a <2$. For more strongly guiding structures, $v > \pi/2$, this variation from $\Gamma_{sa}^{(2)}$ is greatly reduced. 

\subsection{Unequal guides}

\begin{figure}[!ht]
\centering
\subfigure[\ Spatial modes for unequal guides with $d/a = 1$.]{
\includegraphics[width=0.45\textwidth]{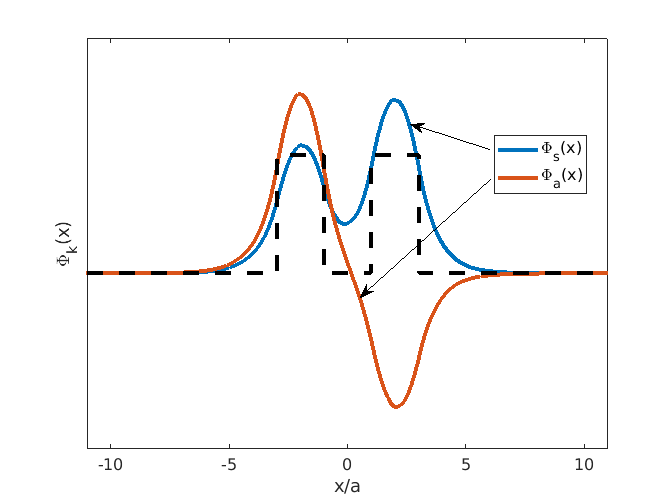}}
\subfigure[\ Spatial modes for unequal guides with $d/a = 2$.]{
\includegraphics[width=0.45\textwidth]{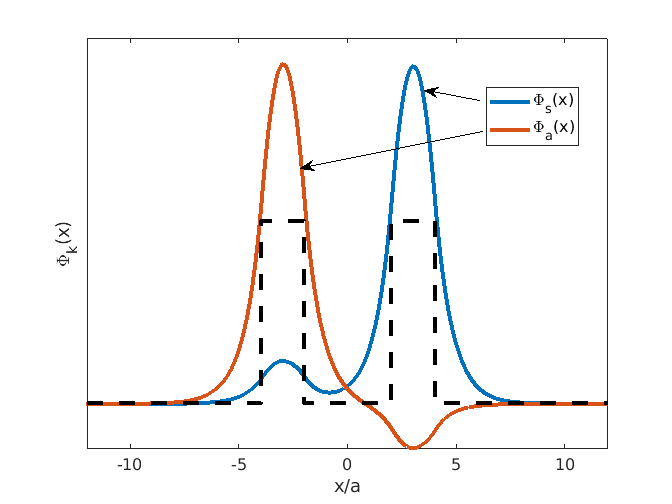}}
\caption{\label{fig:Phi_unequal} Spatial modes for slab guides with widths $w_{1} = 7.9~\mu\mathrm{m}$ and $w_{2} = 8.1~\mu\mathrm{m}$ (the wider guide is on the right). In this case, we define $a = (w_{1} + w_{2})/4$. The normalised frequency for an averaged guide is $v = \pi/2$ as for the equal width guides in Figs.~\ref{fig:Phi_5da} and \ref{fig:Phi_05}. Note that as the guide separation increases, the $s$ mode becomes more greatly confined to the wider guide, whilst the $a$ mode is confined to the narrower.}
\end{figure}

For double-guided structures with unequal guiding regions, we lose the symmetric relations previously found. Figure~\ref{fig:Phi_unequal} illustrates two examples with guide widths $w_{1} = 7.9~\mu\mathrm{m}$ and $w_{2} = 8.1~\mu\mathrm{m}$, where the wider guide is on the right. In these cases, we have put $a = (w_{1} + w_{2})/4$, whilst the same refractive index difference has been used as for the equal width guides. It can be clearly seen that the component of each mode is different in each guide and that it is now the case that  

\begin{equation*}
 \Gamma_{ss}^{(1)} \ne \Gamma_{ss}^{(2)}~\mathrm{and}~\Gamma_{aa}^{(1)} \ne \Gamma_{aa}^{(2)}~~~(\mathrm{unequal~guides}).
\end{equation*}

\noindent We also note that each mode becomes more localised to a particular guide with increasing separation, with the $s$-labelled mode tending to the wider guide. This may be understood from basic waveguiding theory, since the $s$ mode has the lower frequency and the frequency of the lowest mode of a single guide decreases with width. As the separation between the guides tends to infinity, each normal mode approaches the mode of a single isolated guide. Hence, labelling the narrow and wide guides $(1)$ and $(2)$ respectively,

\begin{equation*}
 \lim_{d\to\infty}\Gamma_{ss}^{(1)} = 0,~~~\lim_{d\to\infty}\Gamma_{ss}^{(2)} = \Gamma_{2}, 
\end{equation*}

\begin{equation*}
 \lim_{d\to\infty}\Gamma_{aa}^{(1)} = \Gamma_{1},~~~\lim_{d\to\infty}\Gamma_{aa}^{(2)} = 0
\end{equation*}

\noindent and

\begin{equation*}
 \lim_{d\to\infty}\Gamma_{sa}^{(1)} =\lim_{d\to\infty}\Gamma_{sa}^{(2)} = 0 ~~~(\mathrm{unequal~guides}), 
\end{equation*}

\noindent where $\Gamma_{1}$ and $\Gamma_{2}$ are the optical confinement factors of single guides of width $w_{1}$ and $w_{2}$.

\begin{figure}[!ht]
\centering
\includegraphics[width=0.45\textwidth]{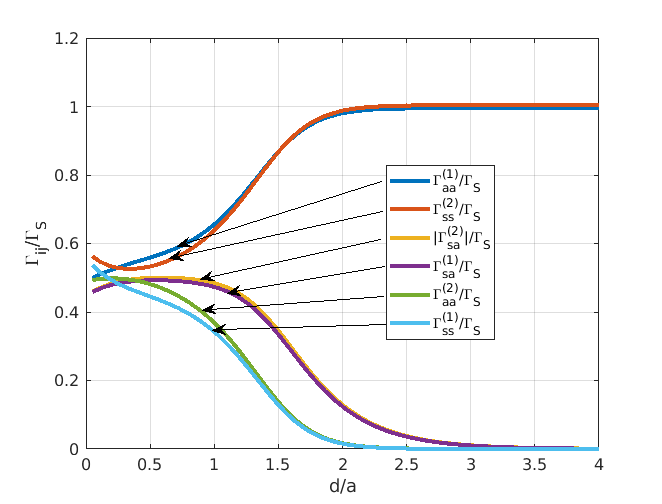}
\caption{\label{fig:overlaps_unequal} Plot of the overlap factors for a weakly-guiding ($v = \pi/2$) non-symmetric slab structure as a function of spatial separation between the guides. Here, $2d$ is the edge-to-edge separation between the guides and $2a = (w_{1} + w_{2})/2$ is the average guide width, using $w_{1} = 7.9~\mu\mathrm{m}$ and $w_{2} = 8.1~\mu\mathrm{m}$ as in Fig.~\ref{fig:Phi_unequal}. In contrast to the case of symmetric guides, in this case the $s$ mode tends to occupy guide $(2)$ (the wider guide) and the $a$ mode occupies guide $(1)$. As the separation increase, the corresponding overlap factors tend to the optical confinement factor, whilst all other factors tend to zero. Here, the modulus of $\Gamma_{sa}^{(2)}$ is shown as this factor is negative.}
\end{figure}

These behaviours are illustrated in Fig.~\ref{fig:overlaps_unequal}, which shows the variation of the overlap factors for a non-symmetric slab guide as a function of spatial separation. The calculations here use the same guide widths as for the modes shown in Fig.~\ref{fig:Phi_unequal}. Note that since $\Gamma_{S}$ is the optical confinement factor of the averaged isolated guide, the limiting values of $\Gamma_{ss}^{(2)}/\Gamma_{S}$ and $\Gamma_{aa}^{(1)}/\Gamma_{S}$ are not exactly unity. 

\subsection{Circular guides}
For circular guides, we used a commercial eigensolver to find solutions of the Helmholtz equation for symmetric structures at various guide separations. These solutions are for the two polarisation components of the lowest-order symmetric and antisymmetric mode. To interpolate between these results, we found that the overlap factors could be fitted very well by the following empirical formulae:

\begin{equation}
 \Gamma_{ss}^{(i)}(d) = \frac{\Gamma_{S}}{2}\left(1 - e^{-a_{1}d}\right) + \Gamma_{1}e^{-b_{1}d}, \label{eq:Gssfit}
\end{equation}

\begin{equation}
 \Gamma_{aa}^{(i)}(d) = \frac{\Gamma_{S}}{2}\left(1 + e^{-a_{1}d}\right) - \Gamma_{2}e^{-b_{1}d} \label{eq:Gaafit}
\end{equation}

\noindent and

\begin{equation}
 \left|\Gamma_{sa}^{(i)}(d)\right| = \frac{\Gamma_{S}}{2}\left(1 + e^{-a_{2}d}\right) - \Gamma_{3}e^{-b_{2}d}. \label{eq:Gsafit}
\end{equation}

\noindent The parameters used are listed in Table~\ref{tab:circular}. Here, $\Gamma_{S}$ is again the optical confinement factor associated with the lowest mode of a single isolated guide. For circular guides, this is HE11 (LP01) mode.

In Section~\ref{sec:redCMM}, we discuss the reduction of the normal mode model to the coupled mode model. It is found there that the coupling coefficient $\mu$ is given by the difference in normal mode frequencies $\mu = (\nu_{s} - \nu_{a})/2$. Using the calculated values of these frequencies, the coupling coefficient for equal circular guides is found to be well-approximated by the Ogawa~\cite{ogawa1977simplified} expression

\begin{equation*}
 \mu \propto \frac{1}{d^{1/2}}\exp\left(\frac{2wd}{a}\right), 
\end{equation*}

\noindent where $w$ is given by \eqref{eq:w}. The constant of proportionality may be found by fitting this to the calculated value of $(\nu_{s} - \nu_{a})/2$ at $d/a = 1$.

% \begin{table}
% \caption{\label{tab:circular}Parameters used in fitting functions for circular waveguides. Here, the cladding refractive index is $n_{2} = 3.4$ and the core refractive index is $n_{1} = n_{2} + \Delta n$.}
% \begin{ruledtabular}
% \begin{tabular}{llll}
% parameter & $\Delta n = 0.002$ & $\Delta n = 0.000971$  & unit \\
%   \hline
%   $a_{1}$ & 0.420 & 0.200 & $\mu\mathrm{m}^{-1}$ \\
%   $a_{2}$ & 0.800 & 0.399  & $\mu\mathrm{m}^{-1}$ \\
%   $b_{1}$ & 0.453 & 0.247 & $\mu\mathrm{m}^{-1}$ \\
%   $b_{2}$ & 0.8145 & 0.441 & $\mu\mathrm{m}^{-1}$ \\
%   $\Gamma_{S}$ & 0.8002 & 0.5766  &  \\
%   $\Gamma_{1}$ & 0.400 & 0.346 &  \\
%   $\Gamma_{2}$ & 0.390 & 0.300  &  \\
%   $\Gamma_{3}$ & 0.432 & 0.3156 &  \\
% \end{tabular}
% \end{ruledtabular}
% \end{table}

\begin{table}
\caption{\label{tab:circular}Parameters used in fitting functions for circular waveguides. Here, the cladding refractive index is $n_{2} = 3.4$ and the core refractive index is $n_{1} = n_{2} + \Delta n$.}
\begin{ruledtabular}
\begin{tabular}{lll}
parameter  & $\Delta n = 0.000971$  & unit \\
  \hline
  $a_{1}$  & 0.200 & $\mu\mathrm{m}^{-1}$ \\
  $a_{2}$  & 0.399  & $\mu\mathrm{m}^{-1}$ \\
  $b_{1}$  & 0.247 & $\mu\mathrm{m}^{-1}$ \\
  $b_{2}$  & 0.441 & $\mu\mathrm{m}^{-1}$ \\
  $\Gamma_{S}$  & 0.5766  &  \\
  $\Gamma_{1}$  & 0.346 &  \\
  $\Gamma_{2}$  & 0.300  &  \\
  $\Gamma_{3}$  & 0.3156 &  \\
\end{tabular}
\end{ruledtabular}
\end{table}

\section{Double-guided structure}\label{sec:double}

\subsection{Real form of the rate equations}
\subsubsection{Optical rate equations}
In this paper, we confine our attention to structures involving only two weakly-confining guides supporting only two guided modes. For equal width guides, these modes will be symmetric (even parity) and anti-symmetric (odd parity), which will be denoted by `$s$' and `$a$' respectively. In the more general case of unequal guides, it will be convenient to retain this notation, where the modes may be distinguished by the fact that the amplitude of the `$a$' mode goes through zero between the guides, whilst the `$s$' mode does not. The treatment we shall follow in this section will be valid for the general case of unequal guides, although it is of particular use for the symmetric case, which we focus on in this paper.

For a double-guided structure, we may put reference frequency to $\Omega = (\nu_{s} + \nu_{a})/2$. The optical rate equations of \eqref{eq:optical_rate} may now be written

\begin{align}
\frac{\partial \tilde{E}_{k,\pm}}{\partial t} &= \left[i\frac{\nu_{k} - \nu_{k'}}{2} - \frac{1}{2\tau_{p}}\right]\tilde{E}_{k,\pm} \nonumber \\
&+ \frac{c}{2n_{g}}\left(1 + i\alpha\right)\tilde{\Sigma}_{k,\pm} - \left[\gamma_{a} + i\gamma_{p}\right]\tilde{E}_{k,\mp}, \label{eq:dEdtnew}
\end{align}

\noindent where $k =s, a$ for the symmetric and anti-symmetric modes respectively and

\begin{equation}
 \tilde{\Sigma}_{k,\pm} = \sum_{k'}\tilde{E}_{k',\pm}\sum_{i}\overline{g}_{\pm}^{(i)}\Gamma_{kk'}^{(i)}, \label{eq:tildeS}
\end{equation}

\noindent where $k' =a, s$. These are then the equations for the evolution of the normal modes. However, solutions in terms of the normal modes do not lend themselves well to physical intuition. When thinking of optical guides in close proximity, it is more natural to think of the optical intensity in each guide. To this end, it is convenient to introduce new optical field variables, defined by

\begin{equation}
E_{1,\pm} = \frac{1}{\sqrt{2}}\left(\tilde{E}_{s,\pm} + \tilde{E}_{a,\pm}\right) \label{eq:Eplus}
\end{equation}

\noindent and

\begin{equation}
E_{2,\pm} = \frac{1}{\sqrt{2}}\left(\tilde{E}_{s,\pm} - \tilde{E}_{a,\pm}\right). \label{eq:Eminus}
\end{equation}

\noindent The motivation for this is that the squared modulus of these `composite' modes becomes the optical intensity in each guide at infinite separation, which greatly aids visualisation of the laser dynamics.

Defining the $\phi_{21\pm\pm}$ as the phase difference between $E_{2\pm}$ and $E_{1\pm}$ and $\phi_{kk+-}$ as the phase difference between $E_{k+}$ and $E_{k-}$, we find in Appendix~\ref{app:real} that the optical rate equations may be written in real form as

\begin{widetext}

\begin{align}
\frac{\partial |E_{1,\pm}|}{\partial t}  &=  \left[-\frac{1}{2\tau_{p}} + \frac{c\Gamma_{S}}{2n_{g}}G_{12\pm}\right]|E_{1,\pm}| + \left[\frac{c\Gamma_{S}}{2n_{g}}\Delta G_{\pm}\left(\cos(\phi_{21\pm\pm}) - \alpha\sin(\phi_{21\pm\pm})\right)-\mu \sin(\phi_{21\pm\pm})\right]|E_{2,\pm}| \nonumber \\
&- \left[\gamma_{a}\cos(\phi_{11+-}) + \gamma_{p}\sin(\phi_{11+-})\right]|E_{1,\mp}| \label{eq:dabsEdt1}
\end{align}

\begin{align}
\frac{\partial |E_{2,\pm}|}{\partial t}  &=  \left[-\frac{1}{2\tau_{p}} + \frac{c\Gamma_{S}}{2n_{g}}G_{21\pm}\right]|E_{2,\pm}| + \left[\frac{c\Gamma_{S}}{2n_{g}}\Delta G_{\pm}\left(\cos(\phi_{21\pm\pm}) + \alpha\sin(\phi_{21\pm\pm})\right) + \mu \sin(\phi_{21\pm\pm})\right]|E_{1,\pm}| \nonumber \\
&- \left[\gamma_{a}\cos(\phi_{22+-}) + \gamma_{p}\sin(\phi_{22+-})\right]|E_{2,\mp}| \label{eq:dabsEdt2}
\end{align}

\begin{align}
\frac{\partial\phi_{21\pm\pm}}{\partial t} &= \frac{c\Gamma_{S}}{2n_{g}}\alpha\left(G_{21\pm} - G_{12\pm}\right) +\mu\cos(\phi_{21\pm\pm})\left(\frac{|E_{1,\pm}|}{|E_{2,\pm}|} -\frac{|E_{2,\pm}|}{|E_{1,\pm}|}\right) \nonumber \\
&+  \frac{c\Gamma_{S}}{2n_{g}}\Delta G_{\pm}\left[\alpha\cos(\phi_{21\pm\pm})\left(\frac{|E_{1,\pm}|}{|E_{2,\pm}|} - \frac{|E_{2,\pm}|}{|E_{1,\pm}|}\right) - \sin(\phi_{21\pm\pm})\left(\frac{|E_{1,\pm}|}{|E_{2,\pm}|} + \frac{|E_{2,\pm}|}{|E_{1,\pm}|}\right)\right] \nonumber \\
&+ \gamma_{p}\left[\cos(\phi_{11+-})\frac{|E_{1,\mp}|}{|E_{1,\pm}|} - \cos(\phi_{22+-})\frac{|E_{2,\mp}|}{|E_{2,\pm}|}\right] \mp \gamma_{a}\left[\sin(\phi_{11+-})\frac{|E_{1,\mp}|}{|E_{1,\pm}|} - \sin(\phi_{22+-})\frac{|E_{2,\mp}|}{|E_{2,\pm}|}\right]. \label{eq:dphi21pm}
\end{align}

\noindent and

\begin{align}
\frac{\partial\phi_{11+-}}{\partial t}  &= \frac{c\Gamma_{S}}{2n_{g}}\alpha\left(G_{12+} - G_{12-}\right) + \mu\left(\cos(\phi_{21++})\frac{|E_{2,+}|}{|E_{1,+}|} - \cos(\phi_{21--})\frac{|E_{2,-}|}{|E_{1,-}|}\right) \nonumber \\
&+ \frac{c\Gamma_{S}}{2n_{g}}\left[\Delta G_{+}\left(\alpha\cos(\phi_{21++}) + \sin(\phi_{21++})\right)\frac{|E_{2,+}|}{|E_{1,+}|} - \Delta G_{-}\left(\alpha\cos(\phi_{21--}) +\sin(\phi_{21--})\right)\frac{|E_{2,-}|}{|E_{1,-}|}\right] \nonumber \\
&+ \gamma_{a}\sin(\phi_{11+-})\left(\frac{|E_{1,+}|}{|E_{1,-}|} + \frac{|E_{1,-}|}{|E_{1,+}|}\right) + \gamma_{p}\cos(\phi_{11+-})\left(\frac{|E_{1,+}|}{|E_{1,-}|} - \frac{|E_{1,-}|}{|E_{1,+}|}\right) \label{eq:dphi11pm}
\end{align} 

\end{widetext}

\noindent Note that since $\phi_{22+-} = \phi_{21++} - \phi_{21--} + \phi_{11+-}$ (as may be derived from \eqref{eq:phi_ijpq}), we do not need an equation for this last phase variable.

In these equations, we have used

\begin{equation}
 \mu = \frac{\nu_{s} - \nu_{a}}{2}, \label{eq:mu}
\end{equation}

\noindent whilst the gain terms are defined by

\begin{equation}
 G_{12\pm} = \frac{\Gamma_{+}^{(1)}\overline{g}_{\pm}^{(1)} + \Gamma_{+}^{(2)}\overline{g}_{\pm}^{(2)}}{\Gamma_{S}}, \label{eq:G12}
\end{equation}

\begin{equation}
 G_{21\pm} = \frac{\Gamma_{-}^{(1)}\overline{g}_{\pm}^{(1)} + \Gamma_{-}^{(2)}\overline{g}_{\pm}^{(2)}}{\Gamma_{S}} \label{eq:G21}
\end{equation}

\noindent and

\begin{equation}
 \Delta G_{\pm} = \frac{\Delta\Gamma^{(1)}\overline{g}_{\pm}^{(1)} + \Delta\Gamma^{(2)}\overline{g}_{\pm}^{(2)}}{\Gamma_{S}}. \label{eq:DG}
\end{equation}

\noindent The $\Gamma$ terms introduced above are further defined in terms of the optical overlap factors via

\begin{equation}
 \Gamma_{\pm}^{(i)} = \frac{\Gamma_{ss}^{(i)} + \Gamma_{aa}^{(i)} \pm 2\Gamma_{sa}^{(i)}}{2} \label{eq:Gamma_pm}
\end{equation}

\noindent and

\begin{equation}
 \Delta\Gamma^{(i)} = \frac{\Gamma_{ss}^{(i)} - \Gamma_{aa}^{(i)}}{2}. \label{eq:DGamma}
\end{equation}

\noindent It is worth noting a general limiting behaviour as the separation between guides tends to infinity that $\Gamma_{ss}^{(i)} \to \Gamma_{aa}^{(i)} \equiv \Gamma^{(i)}$ and $\Gamma_{sa}^{(i)} \to 0$, where $\Gamma^{(i)}$ is half the optical confinement factor in an isolated guide. Hence, in this limit (recalling that the separation between the guides is $2d$), 

\begin{equation}
 \lim_{d\to\infty}\Gamma_{\pm}^{(i)} \equiv \Gamma_{\infty}^{(i)}~\mathrm{and}~\lim_{d\to\infty}\Delta\Gamma^{(i)} = 0. \label{eq:GamLim} 
\end{equation}

\noindent For equal guides, $2\Gamma_{\infty}^{(i)} = \Gamma_{S}$, the optical confinement factor of an isolated guide. For unequal guides, we may take $\Gamma_{S}$ to be an average of the optical confinement factors. This does not undermine the generality of \eqref{eq:dabsEdt1} to \eqref{eq:dphi11pm}, since in all cases the factor of $\Gamma_{S}$ in the denominator of the gain terms cancels with the factor multiplying it. The inclusion of $\Gamma_{S}$ here is one of convenience to elucidate the limiting behaviour of the rate equations, as discussed later in Section~\ref{sec:limit}.

\subsubsection{Carrier rate equations}
The carrier rate equations are straight-forward to render in the notation for the double guided structure. Earlier, these were found to be

\begin{equation}
\frac{\partial N_{\pm}^{(i)}}{\partial t} = -\frac{N_{\pm}^{(i)}}{\tau_{N}} + \Lambda_{\pm}^{(i)} - \frac{c}{n_{g}}\overline{g}_{\pm}^{(i)}I_{\pm}^{(i)} - \gamma_{J}\left(N_{\pm}^{(i)} - N_{\mp}^{(i)}\right),   \label{eq:dNdt}
\end{equation}

\noindent where

\begin{equation}
I_{\pm}^{(i)} =\sum_{k,k'}\Gamma_{kk'}^{(i)}\tilde{E}_{k,\pm}^{*}(t)\tilde{E}_{k',\pm}(t).  \label{eq:Ii2}
\end{equation}

\noindent Using $\Gamma_{sa}^{(i)} = \Gamma_{as}^{(i)}$, we can expand \eqref{eq:Ii2} in terms of $E_{1,\pm}$ and $E_{2,\pm}$ as

\begin{align}
I_{\pm}^{(i)} & = \frac{\Gamma_{ss}^{(i)} + 2\Gamma_{sa}^{(i)} +\Gamma_{aa}^{(i)}}{2}|E_{1,\pm}|^{2} \nonumber \\
&+ \left(\Gamma_{ss}^{(i)} - \Gamma_{aa}^{(i)}\right)|E_{1,\pm}||E_{2,\pm}|\cos(\phi_{21\pm\pm}) \nonumber \\
& + \frac{\Gamma_{ss}^{(i)} - 2\Gamma_{sa}^{(i)} +\Gamma_{aa}^{(i)}}{2}|E_{2,\pm}|^{2}. \nonumber
\end{align}

\noindent Expressing this in terms of  \eqref{eq:Gamma_pm} and \eqref{eq:DGamma}, we have

\begin{align}
I_{\pm}^{(i)} &= \Gamma_{+}^{(i)}|E_{1,\pm}|^{2} \nonumber \\
&+ 2\Delta\Gamma^{(i)}|E_{1,\pm}||E_{2,\pm}|\cos(\phi_{21\pm\pm}) + \Gamma_{-}^{(i)}|E_{2,\pm}|^{2}. \nonumber
\end{align}

\subsubsection{Normalised rate equations}
Upon normalisation, using the scheme described in Appendix~\ref{app:normal}, \eqref{eq:dNdt} may be re-written as

\begin{equation}
\frac{\partial M_{\pm}^{(i)}}{\partial t} = \gamma\left[\eta_{\pm}^{(i)} - \left(1 + \mathcal{I}_{\pm}^{(i)}\right)M_{\pm}^{(i)}\right] - \gamma_{J}\left(M_{\pm}^{(i)} - M_{\mp}^{(i)}\right), \label{eq:dNdt_norm}
\end{equation}

\noindent where the $M_{\pm}^{(i)}$ are the normalised carrier densities and 

\begin{align}
\mathcal{I}_{\pm}^{(i)} &= \frac{\Gamma_{+}^{(i)}}{\Gamma_{S}}|A_{1,\pm}|^{2} + 2\frac{\Delta\Gamma^{(i)}}{\Gamma_{S}}|A_{1,\pm}||A_{2,\pm}|\cos(\phi_{21\pm\pm}) \nonumber \\
&+ \frac{\Gamma_{-}^{(i)}}{\Gamma_{S}}|A_{2,\pm}|^{2}. \label{eq:Inorm}
\end{align}

\noindent Further defining

\begin{equation}
 M_{12\pm} = \frac{\Gamma_{+}^{(1)}M_{\pm}^{(1)} + \Gamma_{+}^{(2)}M_{\pm}^{(2)}}{\Gamma_{S}}, \label{eq:M12}
\end{equation}

\begin{equation}
 M_{21\pm} = \frac{\Gamma_{-}^{(1)}M_{\pm}^{(1)} + \Gamma_{-}^{(2)}M_{\pm}^{(2)}}{\Gamma_{S}} \label{eq:M21}
\end{equation}

\noindent and 

\begin{equation}
 \Delta M_{\pm}  = \frac{\Delta\Gamma^{(1)}M_{\pm}^{(1)} + \Delta\Gamma^{(2)}M_{\pm}^{(2)}}{\Gamma_{S}}, \label{eq:DM}
\end{equation}

\noindent and the cavity decay rate $\kappa = 1/(2\tau_{p})$, the normalised optical rate equations are

\begin{widetext}
 
\begin{align}
\frac{\partial |A_{1,\pm}|}{\partial t} &= \kappa\left(M_{12\pm} - 1\right)|A_{1,\pm}| + \left[\kappa\Delta M_{\pm}\left(\cos(\phi_{21\pm\pm}) - \alpha\sin(\phi_{21\pm\pm})\right) - \mu\sin(\phi_{21\pm\pm})\right]|A_{2,\pm}| \nonumber \\
&- \left[\gamma_{a}\cos(\phi_{11+-}) \pm \gamma_{p}\sin(\phi_{11+-})\right]|A_{1,\mp}|, \label{eq:dA1pmdt}
\end{align}

\begin{align}
\frac{\partial |A_{2,\pm}|}{\partial t} &= \kappa\left(M_{21\pm} - 1\right)|A_{2,\pm}| + \left[\kappa\Delta M_{\pm}\left(\cos(\phi_{21\pm\pm}) + \alpha\sin(\phi_{21\pm\pm})\right) + \mu\sin(\phi_{21\pm\pm})\right]|A_{1,\pm}| \nonumber \\
&- \left[\gamma_{a}\cos(\phi_{22+-}) \pm \gamma_{p}\sin(\phi_{22+-})\right]|A_{2,\mp}|, \label{eq:dA2pmdt}
\end{align}

\begin{align}
\frac{\partial\phi_{21\pm\pm}}{\partial t} &= \kappa\alpha\left(M_{21\pm} - M_{12\pm}\right) +\mu\cos(\phi_{21\pm\pm})\left(\frac{|A_{1,\pm}|}{|A_{2,\pm}|} -\frac{|A_{2,\pm}|}{|A_{1,\pm}|}\right) \nonumber \\
&+ \kappa\Delta M_{\pm}\left[\alpha\cos(\phi_{21\pm\pm})\left(\frac{|A_{1,\pm}|}{|A_{2,\pm}|} - \frac{|A_{2,\pm}|}{|A_{1,\pm}|}\right) - \sin(\phi_{21\pm\pm})\left(\frac{|A_{1,\pm}|}{|A_{2,\pm}|} + \frac{|A_{2,\pm}|}{|A_{1,\pm}|}\right)\right] \nonumber \\
&+ \gamma_{p}\left[\cos(\phi_{11+-})\frac{|A_{1,\mp}|}{|A_{1,\pm}|} - \cos(\phi_{22+-})\frac{|A_{2,\mp}|}{|A_{2,\pm}|}\right] \mp \gamma_{a}\left[\sin(\phi_{11+-})\frac{|A_{1,\mp}|}{|A_{1,\pm}|} - \sin(\phi_{22+-})\frac{|A_{2,\mp}|}{|A_{2,\pm}|}\right], \label{eq:dphi_21pp_norm}
\end{align}

\noindent and

\begin{align}
\frac{\partial\phi_{11+-}}{\partial t}  &= \kappa\alpha\left(M_{12+} - M_{12-}\right) + \mu\left(\cos(\phi_{21++})\frac{|A_{2,+}|}{|A_{1,+}|} - \cos(\phi_{21--})\frac{|A_{2,-}|}{|A_{1,-}|}\right) \nonumber \\
&+ \kappa\Delta M_{+}\left(\alpha\cos(\phi_{21++}) + \sin(\phi_{21++})\right)\frac{|A_{2,+}|}{|A_{1,+}|} - \kappa\Delta M_{-}\left(\alpha\cos(\phi_{21--}) +\sin(\phi_{21--})\right)\frac{|A_{2,-}|}{|A_{1,-}|} \nonumber \\
&+ \gamma_{a}\sin(\phi_{11+-})\left(\frac{|A_{1,+}|}{|A_{1,-}|} + \frac{|A_{1,-}|}{|A_{1,+}|}\right) + \gamma_{p}\cos(\phi_{11+-})\left(\frac{|A_{1,+}|}{|A_{1,-}|} - \frac{|A_{1,-}|}{|A_{1,+}|}\right). \label{eq:dphi_11pm_norm}
\end{align}

\end{widetext}

Equations \eqref{eq:dNdt_norm}, \eqref{eq:Inorm}, and \eqref{eq:dA1pmdt} to \eqref{eq:dphi_11pm_norm} constitute the dynamical model of double-guided structure. This may be applied to any waveguide geometry restricted to two guides and the two  lowest confined modes.   

\subsection{Limiting behaviour for widely separated guides}\label{sec:limit}
\subsubsection{Reduction to the spin-flip model (SFM)}
In this section, we assume equal guides and so employ the results for the identities and limiting behaviour of the overlap factors found in Section~\ref{sec:equal}. As $d \to \infty$, the factors defined earlier in \eqref{eq:Gamma_pm} and \eqref{eq:DGamma} become

\begin{equation*}
 \Gamma_{+}^{(1)} = \Gamma_{S},~\Gamma_{-}^{(1)} = 0,~\Gamma_{+}^{(2)} = 0,~\Gamma_{-}^{(2)} = \Gamma_{S}
\end{equation*}

\noindent and

\begin{equation*}
 \Delta\Gamma^{(i)} = 0. 
\end{equation*}

\noindent The carrier terms defined in \eqref{eq:M12} to \eqref{eq:DM} then become

\begin{equation}
 M_{12\pm} = M_{\pm}^{(1)},~M_{21\pm} = M_{\pm}^{(2)}~\mathrm{and}~\Delta M_{\pm} = 0
\end{equation}

\noindent and the expression for the optical intensity of \eqref{eq:Inorm} appearing in the carrier rate equations becomes

\begin{equation}
\mathcal{I}_{\pm}^{(i)} = |A_{i,\pm}|^{2}, 
\end{equation}

\noindent for $i = 1,2$. Also note that the term given earlier in \eqref{eq:mu} as $\mu = (\nu_{s} - \nu_{a})/2$ approaches zero as the guide separation approaches infinity and the frequencies of the normal modes become equal.

Hence, \eqref{eq:dA1pmdt} to \eqref{eq:dphi_11pm_norm} for the normalised optical rate equations reduce to

\begin{align}
\frac{\partial |A_{i,\pm}|}{\partial t} &= \kappa\left(M_{\pm}^{(i)} - 1\right)|A_{i,\pm}| \nonumber \\
&- \left(\gamma_{a}\cos(\phi_{ii+-}) \pm \gamma_{p}\sin(\phi_{ii+-})\right)|A_{i,\mp}|, \label{eq:dAidt_inf}
\end{align}

\noindent for $i = 1,2$,

\begin{widetext}

\begin{align}
\frac{\partial\phi_{21\pm\pm}}{\partial t} &= \left[\kappa\alpha M_{\pm}^{(2)}  - \left(\gamma_{p}\cos(\phi_{22+-}) \mp \gamma_{a}\sin(\phi_{22+-})\right)\frac{|A_{2,\mp}|}{|A_{2,\pm}|}\right] \nonumber \\
&- \left[\kappa\alpha M_{\pm}^{(1)} - \left(\gamma_{p}\cos(\phi_{11+-}) \mp \gamma_{a}\sin(\phi_{11+-})\right)\frac{|A_{1,\mp}|}{|A_{1,\pm}|}\right] \nonumber 
\end{align}

\noindent and

\begin{align}
\frac{\partial\phi_{11+-}}{\partial t}  &= \kappa\alpha\left(M_{+}^{(1)} - M_{-}^{(1)}\right) + \gamma_{a}\sin(\phi_{11+-})\left(\frac{|A_{1,+}|}{|A_{1,-}|} + \frac{|A_{1,-}|}{|A_{1,+}|}\right) + \gamma_{p}\cos(\phi_{11+-})\left(\frac{|A_{1,+}|}{|A_{1,-}|} - \frac{|A_{1,-}|}{|A_{1,+}|}\right) \label{eq:dphi11pm_inf}
\end{align}

\noindent Recalling that $\phi_{22+-} = \phi_{21++} - \phi_{21--} + \phi_{11+-}$, we find that

\begin{align}
\frac{\partial\phi_{22+-}}{\partial t}  &= \frac{\partial\phi_{21++}}{\partial t} - \frac{\partial\phi_{21++}}{\partial t} + \frac{\partial\phi_{11+-}}{\partial t}, \nonumber \\
&= \kappa\alpha\left(M_{+}^{(2)} - M_{-}^{(2)}\right) + \gamma_{a}\sin(\phi_{22+-})\left(\frac{|A_{2,+}|}{|A_{2,-}|} + \frac{|A_{2,-}|}{|A_{2,+}|}\right) + \gamma_{p}\cos(\phi_{22+-})\left(\frac{|A_{2,+}|}{|A_{2,-}|} - \frac{|A_{2,-}|}{|A_{2,+}|}\right), \label{eq:dphi22pm_inf}
\end{align}

\end{widetext}

\noindent Meanwhile, \eqref{eq:dNdt_norm} for the carrier rate equations becomes 

\begin{align}
\frac{\partial M_{\pm}^{(i)}}{\partial t} &= \gamma\left[\eta_{\pm}^{(i)} - \left(1 + |A_{i,\pm}|^{2}\right)M_{\pm}^{(i)}\right] \nonumber \\
&- \gamma_{J}\left(M_{\pm}^{(i)} - M_{\mp}^{(i)}\right), \label{eq:dNdt_norm_inf}
\end{align}

The guides are now completely uncoupled and we have two sets of equivalent equations for each. Defining new variables $N = (M_{+}^{(i)} + M_{-}^{(i)})/2$, $m = (M_{+}^{(i)} - M_{-}^{(i)})/2$, $|A_{\pm}| = |A_{i,\pm}|/\sqrt{2}$ and $\phi = \phi_{ii+-}$ for each guide, we may re-write \eqref{eq:dAidt_inf}, \eqref{eq:dphi11pm_inf}, \eqref{eq:dphi22pm_inf} and \eqref{eq:dNdt_norm_inf} as

\begin{align}
\frac{\partial |A_{\pm}|}{\partial t} &= \kappa\left(N \pm m - 1\right)|A_{\pm}| \nonumber \\
&- \left(\gamma_{a}\cos\phi \pm \gamma_{p}\sin\phi\right)|A_{\mp}|, \label{eq:dApm_SFM}
\end{align}

\begin{align}
\frac{\partial\phi}{\partial t}  &= 2\kappa\alpha m + \gamma_{a}\sin\phi\left(\frac{|A_{+}|}{|A_{-}|} + \frac{|A_{-}|}{|A_{+}|}\right) \nonumber \\
&+ \gamma_{p}\cos\phi\left(\frac{|A_{+}|}{|A_{-}|} - \frac{|A_{-}|}{|A_{+}|}\right), \label{eq:dphi_SFM}
\end{align}

\begin{align}
\frac{\partial N}{\partial t} &= \gamma\left[\eta - \left(1 + |A_{+}|^{2} + |A_{-}|^{2}\right)N\right. \nonumber \\
&- \left.\left(|A_{+}|^{2} - |A_{-}|^{2}\right)m\right] \label{eq:dN_SFM} 
\end{align}

\noindent and

\begin{align}
\frac{\partial m}{\partial t} &= \gamma\left[P\eta - \left(|A_{+}|^{2} - |A_{-}|^{2}\right)N\right. \nonumber \\
&- \left.\left(|A_{+}|^{2} + |A_{-}|^{2}\right)m\right] - \gamma_{s}m. \label{eq:dm_SFM} 
\end{align}

\noindent Here $\eta = (\eta_{+} + \eta_{-})/2$ (dropping the $(i)$ superscripts), we have defined an effective spin relaxation rate $\gamma_{s} = \gamma + 2\gamma_{J}$ and $P$ is the pump ellipticity defined by

\begin{equation}
P = \frac{\eta_{+} - \eta_{-}}{\eta_{+} + \eta_{-}}. \label{eq:pump_ellip} 
\end{equation}

\noindent Equations \eqref{eq:dApm_SFM} to \eqref{eq:dm_SFM} are the real form of the spin-flip model (SFM)~\cite{san1995light,martin1997polarization, gahl1999polarization}. Note that Gahl \emph{et al}~\cite{gahl1999polarization} have the factor of $(1 + i\alpha)$ multiplying the cavity loss term in the complex rate equations, which we take to be unphysical, so there will be a discrepancy between their expressions and those above.

\subsubsection{Reduction to the coupled mode model (CMM)}\label{sec:redCMM}
We may consider an alternative scenario in which we retain the coupling term $\mu$ between the guides but let the overlap factors take their limiting values as the guide separation tends to infinity. Additionally, we may remove the coupling between spin polarised components by setting $\gamma_{J} = \gamma_{a} = \gamma_{p} = 0$. In this case, \eqref{eq:dA1pmdt} to \eqref{eq:dphi_11pm_norm} reduce to

\begin{align}
\frac{\partial |A_{1,\pm}|}{\partial t} &= \kappa\left(M_{\pm}^{(1)} - 1\right)|A_{1,\pm}|  - \mu\sin(\phi_{21\pm\pm}) |A_{2,\pm}|, \label{eq:dA1dt_CMM}
\end{align}

\begin{align}
\frac{\partial |A_{2,\pm}|}{\partial t} &= \kappa\left(M_{\pm}^{(2)} - 1\right)|A_{2,\pm}| + \mu\sin(\phi_{21\pm\pm})|A_{1,\pm}|, \label{eq:dA2dt_CMM}
\end{align}

\begin{align}
\frac{\partial\phi_{21++}}{\partial t} &= \kappa\alpha\left(M_{+}^{(2)} - M_{+}^{(1)}\right) \nonumber \\
&+ \mu\cos(\phi_{21++})\left(\frac{|A_{1,+}|}{|A_{2,+}|} -\frac{|A_{2,+}|}{|A_{1,+}|}\right).  \label{eq:dphi21p_CMM}
\end{align}

\noindent and

\begin{align}
\frac{\partial\phi_{21--}}{\partial t} &= \frac{\partial\phi_{21++}}{\partial t} + \frac{\partial\phi_{11+-}}{\partial t} - \frac{\partial\phi_{22+-}}{\partial t}, \nonumber \\
&= \kappa\alpha\left(M_{-}^{(2)} - M_{-}^{(1)}\right) \nonumber \\
&+ \mu\cos(\phi_{21--})\left(\frac{|A_{1,-}|}{|A_{2,-}|} -\frac{|A_{2,-}|}{|A_{1,-}|}\right),  \label{eq:dphi21m_CMM}
\end{align}

\noindent whilst the carrier rate equations become

\begin{equation}
\frac{\partial M_{\pm}^{(i)}}{\partial t} = \gamma\left[\eta_{\pm}^{(i)} - \left(1 + |A_{i,\pm}|^{2}\right)M_{\pm}^{(i)}\right]. \label{eq:dNdt_infNP}
\end{equation}

\noindent These give us two independent sets of equations for the polarisation components, each of which reproduces the coupled mode model of Ref~\cite{adams2017effects} with real coupling coefficient $\mu$ and no frequency detuning (although with a difference in sign on the $\alpha$ factor due to opposite sign definitions). Of particular note is that the coupling coefficient $\mu$, given in terms of the difference between the mode frequencies in \eqref{eq:mu}, is consistent with the analysis of Marom \emph{et al}~\cite{Marom1984Relation}, who found the same relation between the coupled mode and normal mode models. 

\section{Steady state solutions}\label{sec:steady}
In general, analytical steady state solutions of the double-guided structure are not obtainable. However, both exact expressions and very good approximations are available in certain limiting cases. In this section, we continue to consider only the case of symmetric guides. 

\subsection{Effect of spin relaxation}\label{sec:spinrelax}

In the steady state, the carrier rate equations of \eqref{eq:dNdt_norm} may be written in matrix form as

\begin{align}
\left[\begin{array}{cc}
\gamma\left(I_{+}^{(i)} + 1\right) + \gamma_{J} & -\gamma_{J} \\
-\gamma_{J} & \gamma\left(I_{-}^{(i)} + 1\right) + \gamma_{J} \end{array}\right]\left[\begin{array}{c}
M_{+}^{(i)} \\
M_{-}^{(i)}\end{array}\right] \nonumber \\
= \gamma\left[\begin{array}{c}
\eta_{+}^{(i)} \\
\eta_{-}^{(i)}\end{array}\right], \nonumber
\end{align}

\noindent which has solutions

\begin{align}
M_{\pm}^{(i)} = \frac{\left(I_{\mp}^{(i)} + 1 + \gamma_{J}/\gamma\right)\eta_{\pm}^{(i)} + (\gamma_{J}/\gamma)\eta_{\mp}^{(i)}}{\left(I_{+}^{(i)} + 1\right)\left(I_{-}^{(i)} + 1\right) + (\gamma_{J}/\gamma)\left(I_{+}^{(i)} + I_{-}^{(i)} + 2\right)}. \nonumber \\
\label{eq:Mpm}
\end{align}

\noindent Alternatively, we may make the intensities the subject, giving

\begin{equation}
I_{\pm}^{(i)} = \frac{\eta_{\pm}^{(i)}}{M_{\pm}^{(i)}} - 1 - \frac{\gamma_{J}}{\gamma}\left(1 -\frac{M_{\mp}^{(i)}}{M_{\pm}^{(i)}}\right). \label{eq:Ipm}
\end{equation}

Close to threshold, we may take the optical intensities in \eqref{eq:Mpm} to be negligible and put these to zero, giving

\begin{equation}
M_{th\pm}^{(i)} \approx \frac{\left(1 + \gamma_{J}/\gamma\right)\eta_{th\pm}^{(i)} + (\gamma_{J}/\gamma)\eta_{th\mp}^{(i)}}{1 + 2(\gamma_{J}/\gamma)},~~~(\gamma_{J} \gg \gamma). \label{eq:Mthpm}
\end{equation}

\noindent Note that, in general, the different polarisation components of the intensity will not go to zero at the same overall pumping rate $\eta^{(i)} = \eta_{+}^{(i)} + \eta_{-}^{(i)}$, so \eqref{eq:Mthpm} is not exact. However, with a large spin relaxation rate $\gamma_{J} \gg \gamma$, \eqref{eq:Mthpm} is found to be a good approximation (in practice, $\gamma_{J} > 10\gamma$). From this we find

\begin{equation*}
M_{th+}^{(i)} + M_{th-}^{(i)} \approx \eta_{th+}^{(i)} + \eta_{th-}^{(i)},~~~(\gamma_{J} \gg \gamma) 
\end{equation*}

\noindent and, since the carrier concentrations are clamped at threshold, 

\begin{equation}
\frac{M_{+}^{(i)} - M_{-}^{(i)}}{M_{+}^{(i)} + M_{-}^{(i)}} \approx \frac{P^{(i)}}{1 + 2(\gamma_{J}/\gamma)},~~~(\gamma_{J} \gg \gamma), \label{eq:Mellip}
\end{equation}

\noindent where $P^{(i)}$ is the pump ellipticity in the $(i)$th guide, defined analogously to \eqref{eq:pump_ellip}. For small $\gamma_{J}$, the left-hand-side of \eqref{eq:Mellip} is no longer linear in $P^{(i)}$ and exhibits a bowing behaviour between $P^{(i)} = 0$ and $P^{(i)} = \pm1$. However, as we shall see in Section~\ref{sec:EPA}, the spin polarisation defined by the left-hand-side of \eqref{eq:Mellip} is only non-zero when we have direct coupling between the components of the optical polarisation. 

For $\gamma_{J} = 0$, \eqref{eq:Mpm} reduces to   

\begin{equation}
M_{\pm}^{(i)} = \frac{\eta_{\pm}^{(i)}}{I_{\pm}^{(i)} + 1},~~~(\gamma_{J} = 0). \label{eq:Mpm0}
\end{equation}

\subsection{Equal pumping}\label{sec:EPA}
A useful simplification to make is to assume equal pumping in each guide. That is, both the total pump power and pump ellipticity are the same $\eta_{\pm}^{(1)} = \eta_{\pm}^{(2)}$. Then, by symmetry, in the steady state we must have 

\begin{align}
 M_{\pm}^{(1)} = M_{\pm}^{(2)},~M_{21\pm} = M_{12\pm}, |A_{1,\pm}| = |A_{2,\pm}|, \nonumber \\
 \phi_{11+-} = \phi_{22+-}~\mathrm{and}~\phi_{21++} = \phi_{21--}. \label{eq:EPA}
\end{align}

\noindent The last of these relations follows from $\phi_{11+-} = \phi_{22+-}$ due to the fact that $\phi_{22+-} = \phi_{21++} - \phi_{21--} + \phi_{11+-}$. Under the equal pumping assumption (EPA), \eqref{eq:dphi_21pp_norm} reduces to

\begin{equation}
\frac{\partial\phi_{21\pm\pm}}{\partial t} = -2\kappa\Delta M_{\pm}\sin(\phi_{21\pm\pm}). \label{eq:dphi_21pp_red}
\end{equation}

\noindent In the steady state, this is satisfied for $\phi_{21\pm\pm} = 0, \pi$. On reduction to the coupled mode model~\cite{adams2017effects}, these are referred to as the `in-phase' and `out-of-phase' solutions respectively. 

Since $|A_{1,\pm}| = |A_{2,\pm}|$, \eqref{eq:dA1pmdt} and \eqref{eq:dA2pmdt} may be written as

\begin{align}
\frac{\partial |A_{i,\pm}|}{\partial t} &= \kappa\left(M_{ij\pm} - 1\right)|A_{i,\pm}| + (-1)^{n}\kappa\Delta M_{\pm}|A_{i,\pm}| \nonumber \\
&- \left[\gamma_{a}\cos(\phi_{ii+-}) \pm \gamma_{p}\sin(\phi_{ii+-})\right]|A_{i,\mp}|, \label{eq:dAipmdtEPA}
\end{align}

\noindent where $n = 0,1$ for the in-phase and out-of-phase solutions respectively and $i = 1,2$ for each guide.

We may simplify \eqref{eq:dAipmdtEPA} even further by assuming there is no direct optical coupling between the polarisation components. That is, the dichroism and birefringence rates are both zero, $\gamma_{a} = \gamma_{p} = 0$ (note that these components may still be indirectly coupled via the spin-polarised carrier concentrations). 

With this simplification, the steady state condition gives

\begin{equation}
 M_{ij\pm} + (-1)^{n}\Delta M_{\pm} = 1, \label{eq:MDM}
\end{equation}

\noindent Since $M_{\pm}^{(1)} = M_{\pm}^{(2)}$ for equal pumping, from \eqref{eq:M12} and \eqref{eq:M21}, $M_{12\pm} = \left(\Gamma_{+}^{(1)} + \Gamma_{+}^{(2)}\right)M_{\pm}^{(i)}/\Gamma_{S}$ and $M_{21\pm} = \left(\Gamma_{-}^{(1)} + \Gamma_{-}^{(2)}\right)M_{\pm}^{(i)}/\Gamma_{S}$. However, since $\Gamma_{kk}^{(1)} = \Gamma_{kk}^{(2)}$ and $\Gamma_{sa}^{(1)} = -\Gamma_{sa}^{(2)}$, from \eqref{eq:Gamma_pm} and \eqref{eq:DM}, we have

\begin{align}
 &M_{ij\pm} + (-1)^{n}\Delta M_{\pm} \nonumber \\
 &= \frac{1}{\Gamma_{S}}\left[\left(\Gamma_{ss}^{(i)} + \Gamma_{aa}^{(i)}\right) + (-1)^{n}\left(\Gamma_{ss}^{(i)} - \Gamma_{aa}^{(i)}\right)\right]M_{\pm}^{(i)}. \label{eq:MEPA} 
\end{align}

\noindent Hence, \eqref{eq:MDM} gives us 

\begin{equation}
 M_{\pm}^{(i)} = \frac{\Gamma_{S}}{2\Gamma_{ss}^{(i)}},~~~(\phi_{21\pm\pm} = 0; \gamma_{a}=\gamma_{p} = 0) \label{eq:Minphase}
\end{equation}

\noindent for the in-phase solution and 

\begin{equation}
 M_{\pm}^{(i)} = \frac{\Gamma_{S}}{2\Gamma_{aa}^{(i)}},~~~(\phi_{21\pm\pm} = \pi; \gamma_{a}=\gamma_{p} = 0) \label{eq:Moutphase}
\end{equation}

\noindent for the out-of-phase solution (at infinite separation, $\Gamma_{ss}^{(i)},\Gamma_{aa}^{(i)} \to \Gamma_{S}/2$ and we would have $M_{\pm}^{(i)} = 1$ in both cases).

Note that these solutions do not depend on carrier spin in any way, hence we also have $M_{\pm}^{(i)} = M_{\mp}^{(i)}$. The steady state spin polarisation of the carriers discussed in Section~\ref{sec:spinrelax} therefore only arises through direct optical coupling. The exception to this would be in the case of zero-spin relaxation if the carriers started off with a spin-polarised population. Otherwise, any spin-relaxation would cause the spin-up and spin-down concentrations to equalise in the steady state.  

Using Eq.~\eqref{eq:Inorm}, we find using the same assumptions that, 

\begin{equation}
\mathcal{I}_{\pm}^{(i)} = \frac{2\Gamma_{ss}^{(i)}}{\Gamma_{S}}|A_{i,\pm}|^{2},~~~\frac{2\Gamma_{aa}^{(i)}}{\Gamma_{S}}|A_{i,\pm}|^{2}, \nonumber
\end{equation}

\noindent for $\phi_{21\pm\pm} = 0,\pi$ respectively. Then, using Eq.~\eqref{eq:Ipm}, we have

\begin{equation}
|A_{i,\pm}| = \sqrt{\eta_{\pm}^{(i)} - \frac{\Gamma_{S}}{2\Gamma_{ss}^{(i)}}},~~~(\phi_{21\pm\pm} = 0; \gamma_{a}=\gamma_{p} = 0) \label{eq:AiinZOC}
\end{equation}

\noindent and

\begin{equation}
|A_{i,\pm}| = \sqrt{\eta_{\pm}^{(i)} - \frac{\Gamma_{S}}{2\Gamma_{aa}^{(i)}}},~~~(\phi_{21\pm\pm} = \pi;\gamma_{a}=\gamma_{p} = 0). \label{eq:AioutZOC}
\end{equation}

\noindent With these results for $|A_{i,\pm}|$, $M_{\pm}^{(i)}$ and $\phi_{21\pm\pm}$, we see that nothing depends on $\phi_{11+-}$ (moreover, it may also be shown that $\partial\phi_{ii+-}/\partial t = 0$ follows without assuming it to be so). Hence, in this simplified case, we have found exact, analytic steady state solutions for all variables (whilst $\phi_{11+-}$ may be set arbitrarily).

If we now allow $\gamma_{a}$ and $\gamma_{p}$ to be non-zero, incorporating the EPA steady state results above, \eqref{eq:dAipmdtEPA} gives

\begin{align}
M_{\pm}^{(i)} &= \frac{\Gamma_{S}}{2\Gamma_{kk}^{(i)}} \nonumber \\
&\times\left[1 + \frac{1}{\kappa}\left(\gamma_{a}\cos(\phi_{ii+-}) \pm 
\gamma_{p}\sin(\phi_{ii+-})\right)\frac{|A_{i,\mp}|}{|A_{i,\pm}|}\right], \label{eq:MpmEPA}
\end{align}

\noindent where $k = s,a$ for the in-phase and out-of-phase solutions respectively. 

Meanwhile, in the EPA, \eqref{eq:dphi_11pm_norm} reduces to

\begin{align}
\frac{\partial\phi_{ii+-}}{\partial t}  &= \kappa\alpha\left[\left(M_{ij+} + (-1)^{n}\Delta M_{+}\right)\right. \nonumber \\
&- \left.\left(M_{ij-} + (-1)^{n}\Delta M_{-}\right)\right] \nonumber \\
&+ \gamma_{a}\sin(\phi_{ii+-})\left(\frac{|A_{i,+}|}{|A_{i,-}|} + \frac{|A_{i,-}|}{|A_{i,+}|}\right) \nonumber \\
&+ \gamma_{p}\cos(\phi_{ii+-})\left(\frac{|A_{i,+}|}{|A_{i,-}|} - \frac{|A_{i,-}|}{|A_{i,+}|}\right). \label{eq:dphi_ii_EPA} 
\end{align}

\noindent Using \eqref{eq:MEPA}, this gives in the steady state

\begin{align}
&-\frac{2\Gamma_{kk}^{(1)}}{\Gamma_{S}}\kappa\alpha\left(M_{+}^{(i)} - M_{-}^{(i)}\right) \nonumber \\
&= \gamma_{a}\sin(\phi_{ii+-})\left(\frac{|A_{i,+}|}{|A_{i,-}|} + \frac{|A_{i,-}|}{|A_{i,+}|}\right) \nonumber \\
&+ \gamma_{p}\cos(\phi_{ii+-})\left(\frac{|A_{i,+}|}{|A_{i,-}|} - \frac{|A_{i,-}|}{|A_{i,+}|}\right). \label{eq:dphi_ii_EPA_SS} 
\end{align}

\noindent We may then use \eqref{eq:MpmEPA} to eliminate $M_{+}^{(i)}$ and $M_{-}^{(i)}$, obtaining

\begin{equation}
\tan\phi_{ii+-} = \left(\frac{\alpha\gamma_{a} - \gamma_{p}}{\alpha\gamma_{p} + \gamma_{a}}\right)\varepsilon^{(i)}, \label{eq:tanphi} 
\end{equation}

\noindent where we have defined the modal optical ellipticity in the $(i)$th guide via

\begin{equation}
\varepsilon^{(i)} = \frac{|A_{i,+}|^{2} - |A_{i,-}|^{2}}{|A_{i,+}|^{2} + |A_{i,-}|^{2}}. \label{eq:optellip_i} 
\end{equation}

\noindent We describe this as the `modal' ellipticity since it is terms of the composite mode amplitudes. Although this is defined for each guide, there is spatial dependence beyond this. Later, in Section~\ref{sec:spatial}, we shall define a spatially varying ellipticity, hence the reason for the specific nomenclature here. Note that since $\tan(\phi_{ii+-}) = \tan(m\pi + \phi_{ii+-})$, where $m$ is an integer, we have two possible solutions for $\phi_{ii+-}$ of $\phi_{0}$ and $\phi_{0} + \pi$, where $\phi_{0}$ is the solution of the arctangent of Eq.~\eqref{eq:tanphi} for $\phi_{0} \in [-\pi/2, \pi/2]$. The solutions $\phi_{ii+-} = \phi_{0} + \pi$ then flip the sign of $\cos$ and $\sin$ terms in \eqref{eq:MpmEPA}, so that we may re-write this as

\begin{align}
M_{\pm}^{(i)} &= \frac{\Gamma_{S}}{2\Gamma_{kk}^{(i)}} \left[1 + \frac{(-1)^{m}}{\kappa}\left(\gamma_{a}\cos(\phi_{ii+-})\right.\right. \nonumber \\
&\pm \left.\left.
\gamma_{p}\sin(\phi_{ii+-})\right)\frac{|A_{i,\mp}|}{|A_{i,\pm}|}\right], \label{eq:MpmEPAphase}
\end{align}

\noindent where $m = 0, 1$. In the reduction to the spin-flip model, these solutions are also referred to as being `in-phase' ($m = 0$) and `out-of-phase' ($m = 1$). However, we will not use this terminology here, to avoid confusion with the previously defined meaning of these terms in the context of the coupled mode model.

In this case, no closed form expression for the optical amplitudes can be found. However, the results of numerical simulation show that using \eqref{eq:AiinZOC} and \eqref{eq:AioutZOC} in conjunction with \eqref{eq:tanphi} and \eqref{eq:MpmEPA} provides a very good approximation for pump ellipticities of $|P^{(i)}| \lesssim 0.8$.

\section{Results and discussion}\label{sec:results}

\subsection{Numerical solutions of the rate equations}
For general solutions of the model we employ a combination of numerical methods. For time series simulations of \eqref{eq:dNdt_norm} and \eqref{eq:dA1pmdt} to \eqref{eq:dphi_11pm_norm}, we use an adaptive Runge-Kutta method of orders 4 and 5~\cite{dormand1980family, shampine1997matlab}. This is very useful for both finding both stable steady state solutions and simulating the temporal dynamics in regions of instability. For finding the unstable steady state solutions and computing the Jacobian, we use a nonlinear solver implementing a trust-region dogleg algorithm~\cite{powell1968fortran} based on the interior-reflective Newton method\cite{coleman1996interior,coleman1994convergence}. Where steady state solutions exist, this latter method is much faster than time series simulation, facilitating efficient routines for tracing stability boundaries and analysing the Jacobian eigenvalues to establish the nature of bifurcations.

In this section, we give results for weakly guided structures with $v = \pi/2$. This corresponds to the refractive index step of $\Delta n = n_{1} - n_{2} = 0.000971$. The waveguide parameters are listed in Table~\ref{tab:waveguide} whilst the specific laser parameters used are given in Table~\ref{tab:laser}. 

\begin{table}
\caption{\label{tab:laser}Laser parameters used in numerical simulations. Note, we tabulate the effective spin relaxation rate $\gamma_{s} = \gamma + 2\gamma_{J}$ to aid direct comparison with the SFM model.}
\begin{ruledtabular}
\begin{tabular}{llll}
Parameter & Value & Unit & Description\\
\hline
$\alpha$  & -2  &  & Linewidth enhancement\\
$\kappa$  & 70  & $\mathrm{ns}^{-1}$ & Cavity loss rate\\
$\gamma$  & 1  & $\mathrm{ns}^{-1}$ & Carrier loss rate\\
$\gamma_{a}$  & 0.1  & $\mathrm{ns}^{-1}$ & Dichroism rate\\
$\gamma_{p}$  & 2  & $\mathrm{ns}^{-1}$ & Birefringence rate\\
$\gamma_{s}$  & 100  & $\mathrm{ns}^{-1}$ & Effective spin relaxation rate\\
$N_{0}$  & $1.1\times10^{18}$  & $\mathrm{cm}^{-3}$ & Transparency density\\
$a_{diff}$  & $1.1\times10^{-15}$  & $\mathrm{cm}^{2}$ & Differential gain\\
$n_{g}$  & 3.4  &  & Group refractive index\\
\end{tabular}
\end{ruledtabular}
\end{table}

\subsection{Spatial profiles}\label{sec:spatial}
The spatial profiles of the intensity and output ellipticity presented in this section were calculated directly from the numerical solutions of the Helmholtz equation for circular guides (i.e. no empirical interpolation was used). The spatially-dependent output ellipticity is defined as

\begin{equation}
 \varepsilon(x,y) = \frac{\mathcal{I}_{+}(x,y) - \mathcal{I}_{-}(x,y)}{\mathcal{I}_{+}(x,y) + \mathcal{I}_{-}(x,y)}, \label{eq:spatial_ellip}
\end{equation}

\noindent where the circularly-polarised intensities are given in terms of the normal mode amplitudes $A_{k,\pm}$ by

\begin{equation}
 \mathcal{I}_{\pm}(x,y) = \left|A_{s,\pm}\Phi_{s}(x,y) + A_{a,\pm}\Phi_{a}(x,y)\right|^{2}. \label{eq:spatial_I}
\end{equation}

\noindent The normal mode amplitudes $A_{k,\pm}$ are reconstructed from the composite mode solutions as described in Appendix~\ref{app:recon}. In Fig.~\ref{fig:spatial_I} we show the intensity results with pump ellipticities of $P^{(1)} = 0$ and $P^{(2)} = 1$ and a total normalised pump power in each guide of $\eta^{(i)} = \eta_{+}^{(i)} + \eta_{-}^{(i)} = 100$ for circular guides of radius $a = 4~\mu\mathrm{m}$ and an edge-to-edge separation given by $d/a = 1$.

Here, the mid-line joining both guides is in the $y$-direction and guide $(2)$ is in the positive $y$ half of the plane (to the left in the diagrams). We note a residual component of left-circularly-polarised light in guide $(2)$. At this pumping power, this is largely due to spatial coupling between the guides. 

\begin{figure}[!ht]
\centering
\subfigure[\ Right-circularly polarised intensity $\mathcal{I}_{+}(x,y)$.]{
\includegraphics[width=0.45\textwidth]{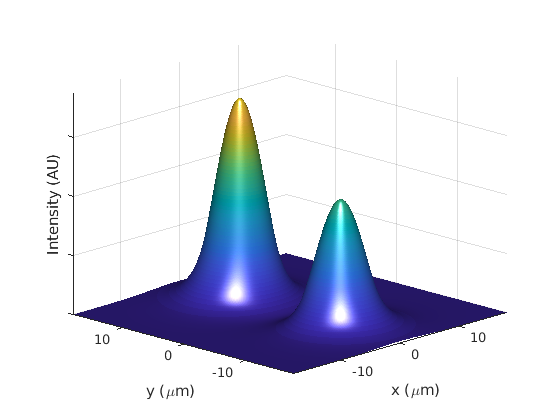}}
\subfigure[\ Left-circularly polarised intensity $\mathcal{I}_{-}(x,y)$.]{
\includegraphics[width=0.45\textwidth]{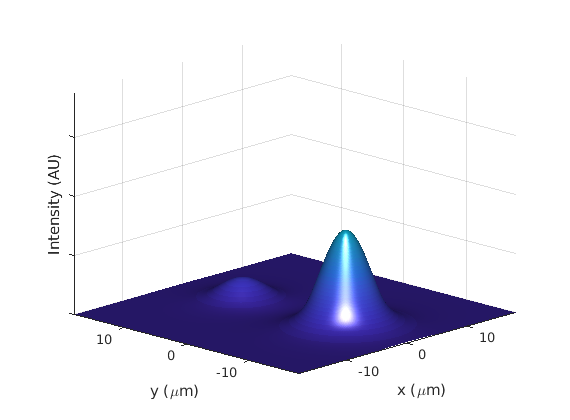}}
\caption{\label{fig:spatial_I} The spatial variation of the circularly-polarised components of the intensity, defined by \eqref{eq:spatial_I} for pump ellipticities of $P^{(1)} = 0$ and $P^{(2)} = 1$ with a total normalised pump power in each guide of $\eta^{(i)} = \eta_{+}^{(i)} + \eta_{-}^{(i)} = 100$. Guide $(2)$ is to the left in the diagrams (in the positive $y$ direction).}
\end{figure}

\begin{figure}[!ht]
\centering
\includegraphics[width=0.45\textwidth]{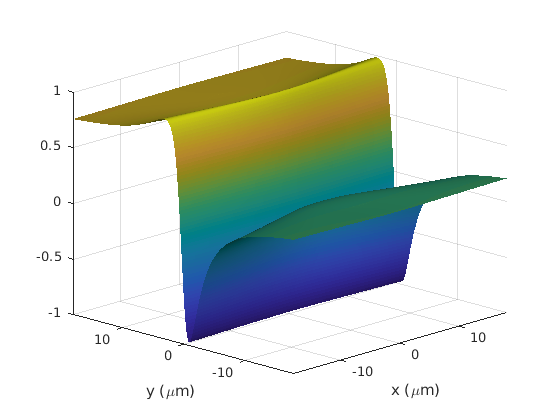}
\caption{\label{fig:spatial_ellip} The spatial variation of the optical ellipticity, defined by \eqref{eq:spatial_ellip} for pump ellipticities of $P^{(1)} = 0$ and $P^{(2)} = 1$ with a total normalised pump power in each guide of $\eta^{(i)} = \eta_{+}^{(i)} + \eta_{-}^{(i)} = 100$. Guide $(2)$ is to the left in the diagrams (in the positive $y$ direction).}
\end{figure}

\begin{figure}[!ht]
\centering
\includegraphics[width=0.45\textwidth]{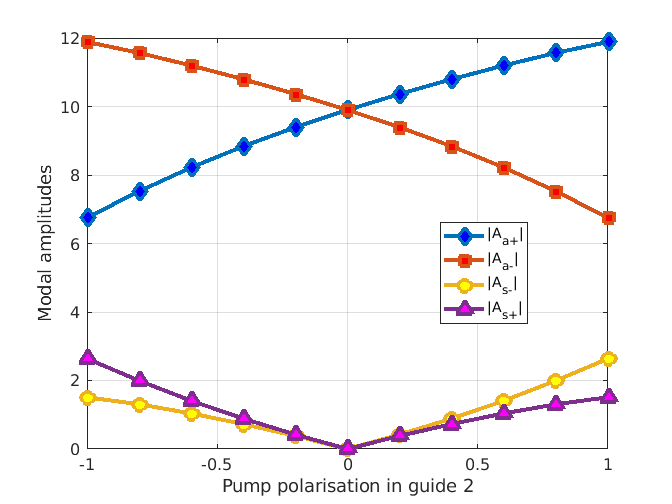}
\caption{\label{fig:modal_P2} Variation of the normal mode amplitudes $|A_{k,\pm}|$ against $P^{(2)}$, the pump ellipticity in guide $(2)$, for $P^{(1)} = 0$ and $\eta^{(i)} = \eta_{+}^{(i)} + \eta_{-}^{(i)} = 100$. Note that the polarisation is dominated by the anti-symmetric mode.}
\end{figure}

The corresponing output ellipticity for these intensities is shown in Fig.~\ref{fig:spatial_ellip}. We note a dip in between the guides where the ellipticity goes to -1, even though the pump polarisation is $P^{(2)} = 1$ in guide $(2)$ and $P^{(1)} = 0$ in guide $(1)$. The reason for this can be seen from Fig.~\ref{fig:modal_P2}, which plots the modal amplitudes against the pump ellipticity in guide $(2)$. Across the whole range, the optical polarisation is dominated by the anti-symmetric mode, which follows the ellipticity of $P^{(2)}$. However, at $y = 0$ the anti-symmetric mode goes through zero, so the only contribution at this point comes from the much smaller symmetric component, for which the left-circularly polarised amplitude is slightly larger. Hence we see this dramatic dip. Note, however, that the optical intensity for both components is very small in this region.

\subsection{Stability boundaries}\label{sec:stability}

\begin{figure}[!ht]
\centering
\includegraphics[width=0.45\textwidth]{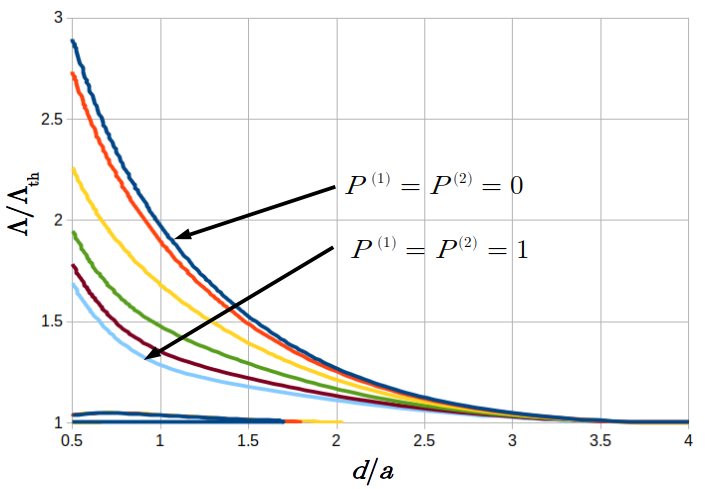}
\caption{\label{fig:P1equalsP2} Stability boundaries for equal pump ellipticities. The calculated curves are for $P^{(1)} = P^{(2)} \in \{0, 0.2, 0.4, 0.6, 0.8, 1\}$, with $P^{(1)} = 0$ at the top and $P^{(1)} = 1$ at the bottom.}
\end{figure}

\begin{figure}[!ht]
\centering
\includegraphics[width=0.45\textwidth]{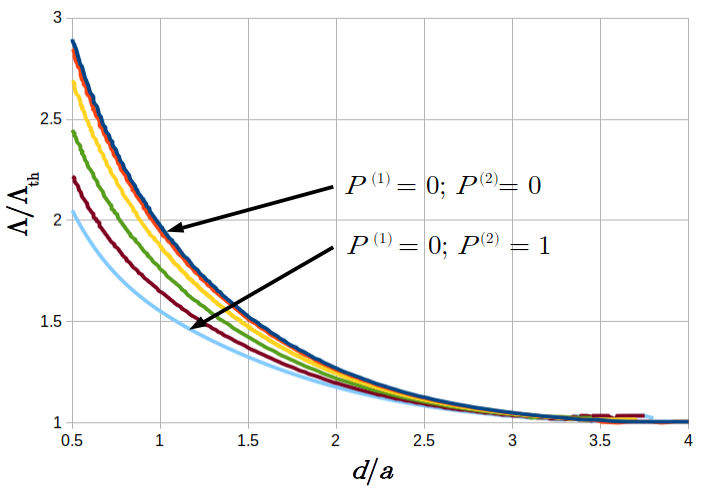}
\caption{\label{fig:P1equals0} Stability boundary for pump ellipticity $P^{(1)} = 0$ in guide $(1)$ and varying $P^{(2)}$ in guide $(2)$ from 0 to 1 in steps on 0.2.}
\end{figure}

\begin{figure}[!ht]
\centering
\includegraphics[width=0.45\textwidth]{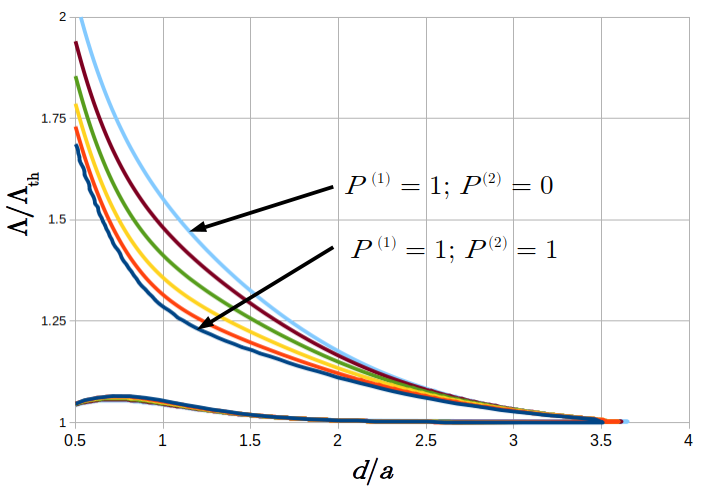}
\caption{\label{fig:P1equals1} Stability boundary for pump ellipticity $P^{(1)} = 1$ in guide $(1)$ and varying $P^{(2)}$ in guide $(2)$ from 0 to 1 in steps on 0.2.}
\end{figure}

\begin{figure}[!ht]
\centering
\includegraphics[width=0.45\textwidth]{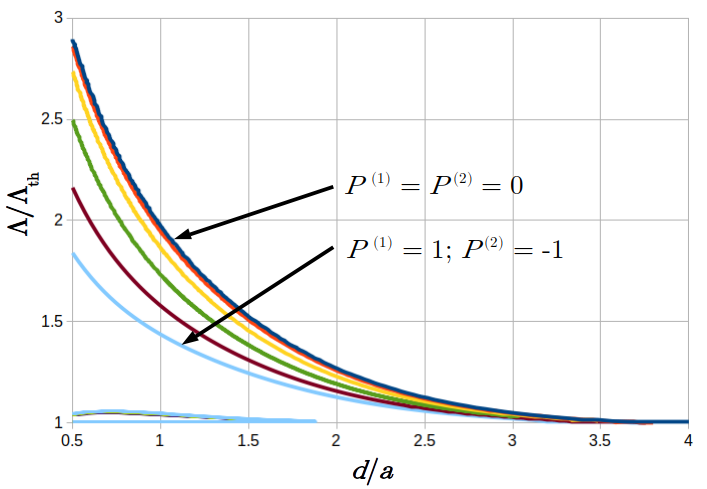}
\caption{\label{fig:P1equalsMinusP2} Stability boundaries for $P^{(1)} = -P^{(2)}$, varying $P^{(1)}$ from from 0 to 1 in steps on 0.2.}
\end{figure}

An initial comparison of the normal mode model to the coupled mode model~\cite{adams2017effects} (neglecting polarisation) has highlighted the importance of the overlap factors on the laser dynamics~\cite{vaughan2019stability}. This is more significant for the weakly-guided structures designed to support only the lowest optical modes. Here we extend this initial investigation to explore the effects of including both optical and carrier-spin polarisation. In particular, we consider the stability of the laser dynamics as a function of the ratio of the optical pump power to the threshold pump against the normalised guide separation. In all cases, we take the total pump power in both guides to be equal, that is $\eta^{(1)} = \eta_{+}^{(1)} + \eta_{-}^{(1)} = \eta^{(2)} = \eta_{+}^{(2)} + \eta_{-}^{(2)}$ and so we may drop the guide index.

Following Ref~\cite{adams2017effects} in the case of zero pump ellipticity, we may define the pump to pump threshold ratio in terms of $Q = (\eta_{+} + \eta_{-})/2$ (the factor of two accounts for the normalisation method used) and consider the threshold pump power $\Lambda_{th}$ at infinite guide separation. This gives us the relation

\begin{equation}
 \frac{\Lambda}{\Lambda_{th}} = \frac{Q + C_{Q}}{1 + C_{Q}}, \label{eq:Pumpratio}
\end{equation}

\noindent where $ C_{Q} = a_{diff}N_{0}/g_{th}$ and the threshold gain is $g_{th} = 2\kappa n_{g}/(c\Gamma_{S})$. The values of the  differential gain $a_{diff}$, transparency density $N_{0}$ and other laser parameters are given in Table~\ref{tab:laser}. With the optical confinement factor $\Gamma_{S}$ for a circular guide with refractive index step $\Delta n = 0.000971$ listed in Table~\ref{tab:circular}, this gives a value of $C_{Q} = 43.9$.  

For non-zero pump polarisation, we may expect the threshold pump to be affected. We retain the definition $Q = (\eta_{+} + \eta_{-})/2$ and again consider the behaviours at infinite separation. With no pump polarisation, the threshold pump would be $Q = 1$. In the same limiting conditions, our model is equivalent to the spin-flip model.

Following the same analysis that lead to the steady state threshold density expression \eqref{eq:MpmEPAphase} and noting that when the laser just turn on $Q_{th} = N_{th} = (M_{+}^{(i)} + M_{-}^{(i)})/2$, we find that

\begin{align}
Q_{th} &= 1 \pm \frac{1}{2\kappa}\left(\sqrt{\frac{1+\varepsilon}{1-\varepsilon}} + \sqrt{\frac{1-\varepsilon}{1+\varepsilon}}\right)\nonumber \\
&\times\left(\gamma_{a}\cos\phi_{0} - \varepsilon\gamma_{p}\sin\phi_{0}\right), \label{eq:Qth}
\end{align}

\noindent where $\varepsilon$ is the modal ellipticity in either guide, given by \eqref{eq:optellip_i} and $\phi_{0}$ is the solution of \eqref{eq:tanphi} for $\phi_{0} = \phi_{ii+-} \in [-\pi/2, \pi/2]$. Note that we have taken the limiting condition $\Gamma_{S}/(2\Gamma_{kk}^{(i)}) \to 1$.

Using the parameters in Table~\ref{tab:laser}, we find via numerical simulation that even for $P^{(i)} = 1$ the modal ellipticity does not exceed 0.6, and $Q_{th} - 1$ is of order 0.001. Hence, we may safely neglect the effect of the pump ellipticity on \eqref{eq:Pumpratio}. 

In the $\Lambda/\Lambda_{th} - d/a$ plane, we find a Hopf bifurcation separating stable steady state solutions in the upper right half of the plane from unstable solutions in the lower left. Regions of stability also appear at low pump powers and small separation, again separated by a Hopf bifurcation. In both regions, these stable solutions are the out-of-phase solutions, which have predominantly anti-symmetric modal components.

Solutions for slab waveguides neglecting the polarisation have been initially reported in Ref~\cite{vaughan2019stability} and compared to the coupled mode model results. It was found that generally the overlap factors tended to push the boundaries up in the direction of both increasing power and increasing separation, thus enlarging the regions of instability.

In this work, we focus on the effect of optical pump ellipticity and the simulation results described here are for weakly-guiding ($\Delta n = 0.000971$, $v = \pi/2$) circular guides. Starting with equal pump ellipticities $P^{(1)} = P^{(2)}$ in Fig.~\ref{fig:P1equalsP2}. Here we have plotted curves from $P^{(1)} = 0$ to $P^{(1)} = 0$ in steps of $0.2$. The instability region is reduced steadily as $P^{(1)} = P^{(2)}$ is increased. On the other hand, the boundary of the lower stability region remains insensitive to these changes. 

In Fig.~\ref{fig:P1equals0}, the ellipticity in guide $(1)$ is kept fixed at $P^{(1)} = 0$, whilst $P^{(2)}$ is varied from 0 to 1 in steps of 0.2. Here we see the same qualitative behaviour, with the stability moving towards the origin with increasing $P^{(2)}$, although not to the same extent as in Fig~\ref{fig:P1equalsP2}. Note that the lower stability region has not been plotted in this case.

Figure~\ref{fig:P1equals1} effectively shows the continuation of this set of results (but starting with $P^{(1)} = 1$ and $P^{(2)} = 0$) and increasing $P^{(2)}$ in steps of 0.2 to 1, ending with $P^{(1)} = P^{(2)} = 1$ as in Fig~\ref{fig:P1equalsP2}. Finally, \ref{fig:P1equalsMinusP2} shows the effect of putting $P^{(1)} = -P^{(2)}$ and increasing $P^{(1)}$ from 0 to 1. In this case, the final curve with $P^{(1)} = 1$ and $P^{(2)} = -1$ does not diminish the instability region to quite the same extent as $P^{(1)} = 1$ and $P^{(2)} = 1$, although we still see the same qualitative reduction of the region with increased pump ellipticity.  

\section{Conclusions}
We have derived a set of general rate equations for the laser dynamics in a waveguide of arbitrary geometry supporting any number of guiding regions and normal modes. The details of the geometry are encoded into the \emph{overlap factors} (generalisations of the optical confinement factor), which may then have a significant effect on the laser dynamics.

We have focused on the particular case of a double-guided structure in the case of just two supported normal modes and derived a set of real rate equations in terms of `composite modes'. This treatment is particularly useful for the consideration of symmetric guides (with reflection symmetry) and can be shown to reduce to both the spin flip model and the coupled mode model in the appropriate limiting case.

Assuming symmetric guides, we have found both exact and approximate analytical expressions for steady state solutions in certain simplified cases. In particular, we have looked at the case of equal power pumping in both guides and investigated the spatial solutions for the circularly polarised intensities (and optical ellipticity) and stability maps for different pump ellipticities. We have found that, in general, increasing the pump ellipticity reduces the region of instability in the pump power v normalised separation plane.   

Although the equations we have derived are general enough to deal with non-symmetric guides, where there is a large difference between guides in double guided structure, it may be more appropriate to use the general solutions, since the behaviour of the overlap factors tends to diverge from the symmetric case quite rapidly. We have not investigated the resulting dynamics associated with this in this paper and leave this matter to be addressed in more detail in future work.   

\begin{acknowledgments}
This research was funded by the Engineering and Physical Sciences Research Council (EPSRC) under grant No. EP/M024237/1.
\end{acknowledgments}

\appendix

\section{Derivation of the optical rate equations}\label{app:optical}
We begin with the wave equation derived from Maxwell's equations

\begin{equation}
\nabla^{2}\mathcal{E} -\mu_{0}\boldsymbol\sigma\frac{\partial\mathcal{E}}{\partial t} - \mu_{0}\epsilon_{0}\boldsymbol\epsilon\frac{\partial^{2}\mathcal{E}}{\partial t^{2}} = \mu_{0}\frac{\partial^{2}\Delta\mathcal{P}}{\partial t^{2}}, \label{eq:waveeq}
\end{equation}

\noindent for the electric field $\mathcal{E}(\mathbf{r},t)$ and the change in of material polarisation $\Delta\mathcal{P}(\mathbf{r},t)$ in the active area. The polarisation in the passive regions $\mathcal{P}(\mathbf{r},t)$ is taken up into the relative permittivity $\boldsymbol\epsilon = 1 + \boldsymbol\chi\left(\mathbf{r}\right)$, where $\boldsymbol\chi\left(\mathbf{r}\right)$ is the electric susceptibility tensor. $\boldsymbol\sigma(\mathbf{r})$ is the conductivity of the material medium, which we take to have the same spatial dependence as $\boldsymbol\chi\left(\mathbf{r}\right)$, whilst $\epsilon_{0}$ and $\mu_{0}$ are the permittivity and permeability of free space respectively. 

Here, we adopt the general approach of Sargent \emph{et al}~\cite{Sargent1974Laser} in deriving the rate equations from \eqref{eq:waveeq} in the slowly varying envelope approximation (SVEA). Accordingly, we shall assume that both $\mathcal{E}$ and $\Delta\mathcal{P}$ can be written in the form of a the product of a modulation function and a phase factor with propagation constant $\beta$ and a characteristic frequency $\Omega$

\begin{equation}
\mathcal{E}(\mathbf{r},t) = \mathbf{E}(x,y,t)e^{i(\beta z - \Omega t)} \label{eq:A1}
\end{equation}

\noindent and similarly for $\Delta\mathcal{P}(\mathbf{r},t)$ in terms of a modulation function $\Delta\mathbf{P}$. Furthermore, we assume that the total field may be decomposed into a superposition of the modes of the laser cavity via

\begin{equation}
\mathcal{E}(\mathbf{r}, t) = \sum_{k}\mathbf{E}_{k}(t)\Phi_{k}(x,y)e^{i(\beta z - \nu_{k})t}, \label{eq:E2}
\end{equation}

\noindent where $k$ labels the modes, $\nu_{k}$ are the modal frequencies and the $\mathbf{E}_{k}(t)$ depend only on time. For TE modes, the complex vector amplitudes may be given by two element Jones vectors of the form

\begin{equation}
 \mathbf{E}_{k}(t) = \left[\begin{array}{c}
                            E_{k,x} \\
                            E_{k,y}
                           \end{array}\right].
 \label{eq:Jones}
\end{equation}

\noindent The amplitudes $E_{k,i}$ incorporate phase information, which we shall deal with explicitly later. Note that, here, we have assumed that each mode $k$ has separate polarisations with the same spatial profile $\Phi_{k}(x,y)$ in the lateral directions. 

From \eqref{eq:A1}, the modulation function is then given by

\begin{equation}
\mathbf{E}(\mathbf{r}, t) = \sum_{k}\mathbf{E}_{k}(t)\Phi_{k}(x,y)e^{i(\Omega - \nu_{k})t} \label{eq:A2}
\end{equation}

\noindent and, again, similarly for the modulation function on the material polarisation.

Since $\mathbf{E}(\mathbf{r},t)$ is assumed to be slowly varying, we can take $\Omega$ to be the solution of the approximate Helmoholtz equation

\begin{equation}
\nabla^{2}\mathcal{E} = -\Omega^{2}\mu_{0}\epsilon_{0}\boldsymbol\epsilon\mathcal{E}, \label{eq:A_Helmholtz}
\end{equation}

\noindent where $\Omega$ will be determined in part by the boundary conditions on the ends of the cavity in the propagation direction, which fix the propagation constant $\beta$. Strictly, $\Omega$ will be different for different modal compositions of the modulation function. For instance, if we had $\mathbf{E}(\mathbf{r}, t) = \mathbf{E}_{k}(t)\Phi_{k}(x,y)e^{i(\Omega - \nu_{k})t}$, then we would have exactly $\Omega = \nu_{k}$. In the case of superpositions of modal solutions, for \eqref{eq:A_Helmholtz} to be satisfied exactly, $\Omega$ would have to be time varying, so treating it as a constant must be an approximation. It will prove to be convenient to take $\Omega$ to be the average of these modal frequencies.

Substituting, these results into \eqref{eq:waveeq}, we have 

\begin{align}
\sum_{k}\left(\nabla^{2} -\mu_{0}\boldsymbol\sigma\frac{\partial}{\partial t} - \mu_{0}\epsilon_{0}\boldsymbol\epsilon\frac{\partial^{2}}{\partial t^{2}}\right)\mathbf{E}_{k}\Phi_{k}e^{-i\nu_{k}t} \nonumber \\
= \mu_{0}\sum_{k}\frac{\partial^{2}\Delta\mathbf{P}_{k}}{\partial t^{2}}\Phi_{k}e^{-i\nu_{k}t}. \label{eq:waveeqB}
\end{align}

We may deal with the Laplacian term in \eqref{eq:waveeqB} by noting that the spatial modes $\Phi_{k}$ satisfy the Helmholtz eigenmode equation, given earlier in \eqref{eq:Helmholtz}. Since we are making the approximation that for each $k$, the spatial modes are the same for each polarisation, using $\mu_{0}\epsilon_{0} = 1/c^{2}$, we may put

\begin{equation}
 \nabla^{2}\mathbf{E}_{k,p}\Phi_{k} = -\Omega^{2}\mu_{0}\epsilon_{0}\boldsymbol\epsilon\mathbf{E}_{k,p}\Phi_{k}, 
\end{equation}

\noindent so now we have

\begin{align}
-\sum_{k}\left[\mu_{0}\boldsymbol\sigma\frac{\partial}{\partial t} + \mu_{0}\epsilon_{0}\boldsymbol\epsilon\left(\frac{\partial^{2}}{\partial t^{2}} + \Omega^{2}\right)\right]\mathbf{E}_{k}\Phi_{k}e^{-i\nu_{k}t} \nonumber \\
= \mu_{0}\sum_{k}\frac{\partial^{2}\Delta\mathbf{P}_{k}}{\partial t^{2}}\Phi_{k}e^{-i\nu_{k}t}. \label{eq:waveeqC}
\end{align}

In the active region, the interaction with the material is diagonal in a basis of circularly polarized optical components. It is therefore convenient to transform the optical polarisation vectors to this basis. Since the $z$ components of the electric field remain unchanged, we may describe the transformation in terms of a 2 dimensional matrix 

\begin{equation*}
 \mathbf{T} = \frac{1}{\sqrt{2}}\left[\begin{array}{cc}
               1 & i \\
               1 & -i
              \end{array}\right],
\end{equation*}

\noindent so that a vector in the $x$, $y$ basis is transformed according to

\begin{equation*}
 \left[\begin{array}{c}
               A_{+} \\
               A_{-}
              \end{array}\right] = \frac{1}{\sqrt{2}}\left[\begin{array}{cc}
               1 & i \\
               1 & -i
              \end{array}\right]\left[\begin{array}{c}
               A_{x} \\
               A_{y}
              \end{array}\right]. 
\end{equation*}

\noindent This gives $A_{\pm} = (A_{x} \pm iA_{y})/\sqrt{2}$ as required. The transformation of an equation in the form of $\mathbf{P} = \epsilon_{0}\boldsymbol\chi\mathbf{E}$, where $\boldsymbol\chi$ is a 2-dimensional matrix, then becomes $\mathbf{T}\mathbf{P} = \mathbf{P}_{\pm} = \epsilon_{0}\mathbf{T}\boldsymbol\chi\mathbf{T}^{-1}\mathbf{T}\mathbf{E} = \epsilon_{0}\boldsymbol\chi_{\pm}\mathbf{E}_{\pm}$. If $\boldsymbol\chi$ is diagonal, it then takes the form

\begin{equation*}
 \boldsymbol\chi_{\pm} = \frac{1}{2}\left[\begin{array}{cc}
               1 & i \\
               1 & -i
              \end{array}\right]\left[\begin{array}{cc}
               \chi_{xx} & 0 \\
               0 & \chi_{yy}
              \end{array}\right]\left[\begin{array}{cc}
               1 & 1 \\
               -i & i
              \end{array}\right]  = \left[\begin{array}{cc}
            \chi_{0} & \delta\chi \\
            \delta\chi & \chi_{0}
        \end{array}\right]. 
\end{equation*}

\noindent where we have defined $\chi_{0} = \left(\chi_{xx} + \chi_{yy}\right)/2$ and $\delta\chi = \left(\chi_{xx} - \chi_{yy}\right)/2$. Hence, putting $\epsilon = 1 + \chi_{0}$, the circularly polarised components of $\boldsymbol\epsilon\mathbf{E}_{k}$ are $\left(\boldsymbol\epsilon\mathbf{E}_{k}\right)_{\pm} = \epsilon E_{k,\pm} + \delta\chi E_{k,\mp}$.

Similarly, defining $\sigma_{0} = \left(\sigma_{xx} + \sigma_{yy}\right)/2$ and $\delta\sigma = \frac{1}{2}\left(\sigma_{xx} - \sigma_{yy}\right)$ for the conductivity, the components arising out of terms of the form $\boldsymbol\sigma\mathbf{A}$ will be given by $\left(\boldsymbol\sigma\mathbf{A}\right)_{\pm} = \sigma_{0}A_{\pm} + \delta\sigma A_{\mp}$. Applying the transformation $\mathbf{T}$ to \eqref{eq:waveeqC}, we may then write the wave equation in circularly polarised component form as

\begin{align}
&-\sum_{k}\left[\mu_{0}\frac{\partial}{\partial t}\left(\sigma_{0}E_{k,\pm} + \delta\sigma E_{k,\mp}\right)\right. \nonumber \\
&+ \left.\mu_{0}\epsilon_{0}\left(\frac{\partial^{2}}{\partial t^{2}} + \Omega^{2}\right)\left(\epsilon E_{k,\pm} + \delta\chi E_{k,\mp}\right)\right]\Phi_{k}e^{-i\nu_{k}t} \nonumber \\
&= \mu_{0}\sum_{k}\frac{\partial^{2}\Delta P_{k,\pm}}{\partial t^{2}}\Phi_{k}e^{-i\nu_{k}t}. \label{eq:waveeqD}
\end{align}

Putting $E_{k,\pm} = |E_{k,\pm}|e^{-i\varphi_{k,\pm}}$ and $\Delta P_{k,\pm} = \Delta\mathcal{P}_{k,\pm}e^{-i\varphi_{k,\pm}}$, the time derivatives of the optical field components are

\begin{align}
 &\frac{\partial}{\partial t}\left(E_{k,\pm}e^{-i\nu_{k}t}\right) \nonumber \\
 &= \left[|\dot{E}_{k,\pm}| - i\left(\nu_{k} + \dot{\varphi}_{k,\pm}\right)|E_{k,\pm}|\right]e^{-i(\nu_{k}t + \varphi_{k,\pm})}  \label{eq:dEdt}
\end{align}

\noindent and

\begin{align}
 &\frac{\partial^{2}}{\partial t^{2}}\left(E_{k,\pm}e^{-i\nu_{k}t}\right) \nonumber \\
 & = \left\{|\ddot{E}_{k,\pm}| - i2\left[\nu_{k} + \dot{\varphi}_{k,\pm}\right]|\dot{E}_{k,\pm}| \right. \nonumber \\
 & - \left.\left[\left(\nu_{k} + \dot{\varphi}_{k,\pm}\right)^{2} + i\ddot{\varphi}_{k,\pm}\right]|E_{k,\pm}|\right\}e^{-i(\nu_{k}t + \varphi_{k,\pm})} \nonumber \\
 & \approx -\left[i2\nu_{k}|\dot{E}_{k,\pm}| + \left(\nu_{k} + \dot{\varphi}_{k,\pm}\right)^{2}|E_{k,\pm}|\right]e^{-i(\nu_{k}t + \varphi_{k,\pm})}\label{eq:d2Edt2}
\end{align}

\noindent where, in accordance with the slowly varying envelope approximation, we may neglect all second derivatives and products of first derivatives. A similar expression to \eqref{eq:d2Edt2} may be obtained for the second derivative of the material polarisation terms.

Inserting \eqref{eq:dEdt} and \eqref{eq:d2Edt2} into \eqref{eq:waveeqD} introduces several terms that will be negligible under the SVEA. Specifically, products involving the components of the conductivity and time derivatives $\sigma_{0} |\dot{E}_{k,\pm}|$, $\delta\sigma|\dot{E}_{k,\pm}|$ will be small, as will $\delta\chi|\dot{E}_{k,\pm}|$ and time derivatives of $\Delta P_{k,\pm}$. Neglecting these, we get

\begin{align}
& i2\epsilon_{0}\epsilon\sum_{k}\nu_{k}\bf\left(|\dot{E}_{k,\pm}|e^{-i\varphi_{k,\pm}}    \right. \nonumber \\
&+ \left\{\frac{\sigma_{0}}{2\epsilon_{0}\epsilon} - \frac{i}{2\nu_{k}}\left[\left(\nu_{k} + \dot{\varphi}_{k,\pm}\right)^{2} - \Omega^{2}\right]\right\}|E_{k,\pm}|e^{-i\varphi_{k,\pm}} \nonumber \\
&+ \left\{\frac{\delta\sigma}{2\epsilon_{0}\epsilon} - \left.\frac{i\delta\chi}{2\epsilon\nu_{k}}\left[\left(\nu_{k} + \dot{\varphi}_{k,\mp}\right)^{2} - \Omega^{2}\right]\right\}|E_{k,\mp}|e^{-i\varphi_{k,\mp}}\right)\nonumber \\
&\times\Phi_{k}e^{-i\nu_{k}t} \nonumber \\
&= -\sum_{k}\left(\nu_{k} + \dot{\varphi}_{k,\pm}\right)^{2}\Delta \mathcal{P}_{k,\pm}\Phi_{k}e^{-i(\nu_{k}t + \varphi_{k,\pm})}. \label{eq:waveeqE}
\end{align}

Now, we may assume $\Omega = \nu_{k} + \Delta$, where $\Delta \ll \nu_{k}$. Then, since $\dot{\varphi}_{k,\pm}^{2}$ is also small, we find that

\begin{align}
\left(\nu_{k} + \dot{\varphi}_{k,\pm}\right)^{2} - \Omega^{2} &\approx 2\nu_{k}\left(\dot{\varphi}_{k,\pm} - \Delta\right) \nonumber \\
&= 2\nu_{k}\left(\nu_{k} + \dot{\varphi}_{k,\pm} - \Omega\right). 
\end{align}

\noindent Making the further approximation $\left(\nu_{k} + \dot{\varphi}_{k,p}\right)^{2}\Delta\mathcal{P}_{k,p} \approx \nu_{k}^{2}\Delta\mathcal{P}_{k,p}$ and neglecting the small term involving $\delta\chi\dot{\varphi}_{k,\mp}$, the wave equation then reads

\begin{align}
& \epsilon\sum_{k}\nu_{k}\left\{|\dot{E}_{k,\pm}|e^{-i\varphi_{k,\pm}}  \right. \nonumber \\
&+ \left[\frac{\sigma_{0}}{2\epsilon_{0}\epsilon}- i\left(\nu_{k} + \dot{\varphi}_{k,\pm} - \Omega\right)\right]|E_{k,\pm}|e^{-i\varphi_{k,\pm}}   \nonumber \\
&+ \left.\left[\frac{\delta\sigma}{2\epsilon_{0}\epsilon} + \frac{i\delta\chi}{\epsilon}\left(\Omega - \nu_{k}\right)\right]|E_{k,\mp}|e^{-i\varphi_{k,\mp}}\right\}\Phi_{k}e^{-i\nu_{k}t} \nonumber \\
&= i\sum_{k}\frac{\nu_{k}^{2}}{2\epsilon_{0}}\Delta \mathcal{P}_{k,\pm}\Phi_{k}e^{-i(\nu_{k}t + \varphi_{k,\pm})}. \label{eq:waveeqF}
\end{align}

We may simplify \eqref{eq:waveeqF} further by identifying characteristic rates. The term involving $\sigma_{0}$ may be interpreted as the cavity loss rate $\kappa = 1/(2\tau_{p}) = \sigma_{0}/(2\epsilon_{0}\epsilon)$, where $\tau_{p}$ is the photon lifetime. The term involving $\delta\chi$ may be identified as the (induced) birefringence rate $\gamma_{p} = \delta\chi\left(\Omega - \nu_{k}\right)/\epsilon$ and that involving $\delta\sigma$ as the dichroism rate $\gamma_{a} = \delta\sigma/(2\epsilon_{0}\epsilon)$. We then have

\begin{align}
& \epsilon\sum_{k}\nu_{k}\left\{|\dot{E}_{k,\pm}|e^{-i\varphi_{k,\pm}} \right. \nonumber \\
&+ \left[\frac{1}{2\tau_{p}} - i\left(\nu_{k} + \dot{\varphi}_{k,\pm} - \Omega\right)\right]|E_{k,\pm}|e^{-i\varphi_{k,\pm}}    \nonumber \\
&+ \left.\left[\gamma_{a} + i\gamma_{p}\right]|E_{k,\mp}|e^{-i\varphi_{k,\mp}}\right\}\Phi_{k}e^{-i\nu_{k}t} \nonumber \\
&= i\sum_{k}\frac{\nu_{k}^{2}}{2\epsilon_{0}}\Delta \mathcal{P}_{k,\pm}\Phi_{k}e^{-i(\nu_{k}t + \varphi_{k,\pm})}. \label{eq:waveeqG}
\end{align}

As eigenfunctions of the Helmholtz equation, the spatial modes $\Phi_{k}$ are orthogonal to one another. Moreover, for symmetric guides, the orthogonality relations

\begin{equation}
 \int \epsilon(\mathbf{r})\Phi_{k}(\mathbf{r})\Phi_{k'}(\mathbf{r})~d^{2}\mathbf{r} = \mathcal{N}\delta_{kk'}, \label{eq:orthog}
\end{equation}

\noindent also apply, where $\mathcal{N}$ is a constant characterising the permittivity and $\delta_{kk'}$ is the Kronecker delta. These relations will also hold approximately for weakly guided structures in the case of non-symmetric guides. We may therefore multiply Eq.~\eqref{eq:waveeqG} by a particular spatial mode $\Phi_{k}$ and integrate over space to obtain

\begin{align}
& |\dot{E}_{k,\pm}| + \left[\frac{1}{2\tau_{p}} - i\left(\nu_{k} + \dot{\varphi}_{k,\pm} - \Omega\right)\right]|E_{k,\pm}| \nonumber \\
&+ \left(\gamma_{a} + i\gamma_{p}\right)|E_{k,\mp}|e^{i\Delta\phi_{kk\pm}} \nonumber \\
&= i\sum_{k'}\frac{\nu_{k'}e^{i\Delta\Theta_{kk'\pm}}}{2\epsilon_{0}\mathcal{N}}\int\Delta \mathcal{P}_{k',\pm}\Phi_{k}\Phi_{k'}~d^{2}\mathbf{r}, \label{eq:waveeqH}
\end{align}

\noindent where we have used $\nu_{k'} \approx \nu_{k}$, $\Delta\phi_{kk\pm} = \varphi_{k,\pm} - \varphi_{k,\mp}$ and $\Delta\Theta_{kk'\pm} = \nu_{k}t + \varphi_{k,\pm} - \nu_{k'}t - \varphi_{k',\pm}$. Note that the summation on the right-hand-side does not disappear due to the spatial dependence of the $\Delta\mathcal{P}_{k',\pm}$.

Now, we may express the change in material polarisation due to the change in carrier density in the active region as $\Delta \mathcal{P}_{k,\pm} = \epsilon_{0}\Delta\chi_{\pm}|E_{k,\pm}|$, where $\Delta\chi_{\pm}$ is the change in electric susceptibility for the component of circularly polarised light (recall that the susceptibility is diagonal in this basis in the active region). This may be written in terms of real and imaginary parts $\Delta\chi_{\pm}'$ and $\Delta\chi_{\pm}''$ respectively as

\begin{equation}
 i\Delta\chi_{\pm} = -\Delta\chi_{\pm}''\left(1 - i\frac{\Delta\chi_{\pm}'}{\Delta\chi_{\pm}''}\right) = -\Delta\chi_{\pm}''\left(1 + i\alpha\right), \label{eq:chi}
\end{equation}

\noindent where $\alpha = -\Delta\chi_{\pm}'/\Delta\chi_{\pm}''$ is the linewidth enhancement factor. The material gain $g_{\pm}(v_{k})$ is then related to the imaginary part of the susceptibility via

\begin{equation}
 g_{\pm}(\nu_{k}) = -\frac{\nu_{k}\Delta\chi_{\pm}''}{nc}, \label{eq:gain}
\end{equation}

\noindent where $n$ is the refractive index in the guide. Incorporating \eqref{eq:chi} and \eqref{eq:gain} into \eqref{eq:waveeqH} and putting $\mathcal{N} \approx nn_{g}$, we have

\begin{align}
& |\dot{E}_{k,\pm}| + \left[\frac{1}{2\tau_{p}} - i\left(\nu_{k} + \dot{\varphi}_{k,\pm} - \Omega\right)\right]|E_{k,\pm}| \nonumber \\
&+ \left(\gamma_{a} + i\gamma_{p}\right)|E_{k,\mp}|e^{i\Delta\phi_{kk\pm}} \nonumber \\
&= \sum_{k'}\frac{c}{2n_{g}}\left(1 + i\alpha\right)|E_{k',\pm}|e^{i\Delta\Theta_{kk'\pm}}\int g_{\pm}(\mathbf{r})\Phi_{k}\Phi_{k'}~d^{2}\mathbf{r}. \label{eq:waveeqI}
\end{align}

Here, we have indicated that the gain has a spatial dependence. Specifically, the increase in carriers giving rise to the gain will be localised to the guides. We may therefore define optical confinement factors $\Gamma_{kk'}^{(i)}$ for the $i$th guide via 

\begin{equation*}
 \overline{g}_{\pm}^{(i)}\Gamma_{kk'}^{(i)} \equiv \int_{(i)} g_{\pm}(\mathbf{r})\Phi_{k}\Phi_{k'}~d^{2}\mathbf{r}, 
\end{equation*}

\noindent where $\overline{g}_{\pm}^{(i)}$ is the average gain in guide $(i)$ and the integral is over the $i$th guide. Hence, the integral appearing in \eqref{eq:waveeqI} will be

\begin{equation}
 \int g_{\pm}(\mathbf{r})\Phi_{k}\Phi_{k'}~d^{2}\mathbf{r} =  \sum_{i}\overline{g}_{\pm}^{(i)}\Gamma_{kk'}^{(i)}. \label{eq:sumGamma_kk}
\end{equation}

\noindent We may then re-write \eqref{eq:waveeqI} as

\begin{align}
 & |\dot{E}_{k,\pm}| - i\dot{\varphi}_{k,\pm}|E_{k,\pm}| \nonumber \\
 &= \left[i\left(\nu_{k}  - \Omega\right) - \frac{1}{2\tau_{p}}\right]|E_{k,\pm}| \nonumber \\
 &- \left(\gamma_{a} + i\gamma_{p}\right)|E_{k,\mp}|e^{i\Delta\phi_{kk\pm}} \nonumber \\
&+ \sum_{k'}\frac{c}{2n_{g}}\left(1 + i\alpha\right)|E_{k',\pm}|e^{i\Delta\Theta_{kk'\pm}}\sum_{i}\overline{g}_{\pm}^{(i)}\Gamma_{kk'}^{(i)}. \label{eq:waveeqJ}
\end{align}

\noindent Noting that 

\begin{equation*}
 \frac{\partial E_{k,\pm}}{\partial t} = \left(|\dot{E}_{k,\pm}| - i\dot{\varphi}_{k,\pm}|E_{k,\pm}|\right)e^{-i\varphi_{k,\pm}}, 
\end{equation*}

\noindent we may write \eqref{eq:waveeqJ} in complex form as

\begin{align*}
 \frac{\partial E_{k,\pm}}{\partial t} &= \left[i\left(\nu_{k}  - \Omega\right) - \frac{1}{2\tau_{p}}\right]E_{k,\pm} - \left(\gamma_{a} + i\gamma_{p}\right)E_{k,\mp}  \\
&+ \sum_{k'}\frac{c}{2n_{g}}\left(1 + i\alpha\right)E_{k',\pm}e^{i\Delta\nu_{kk'}t}\sum_{i}\overline{g}_{\pm}^{(i)}\Gamma_{kk'}^{(i)}, 
\end{align*}

\noindent where $\Delta\nu_{kk'} = \nu_{k} - \nu_{k'}$. This is the result for the optical rate equations as given in the main text. 

\section{Derivation of the carrier rate equations}\label{app:carrier}
After scaling the electric field components to match the dimensions of \eqref{eq:dNpm}, we may obtain the circularly polarised components of the normal mode superposition in \eqref{eq:E} via

\begin{equation}
E_{\pm}(\mathbf{r},t) = \mathbf{e}_{\pm}^{\dagger}\mathbf{E}(\mathbf{r},t) = \sum_{k}E_{k,\pm}(t)\Phi_{k}(\mathbf{r})e^{-i\nu_{k}t}, \label{eq:super}
\end{equation}

\noindent where the $\mathbf{e}_{\pm}$ are the unit Jones vectors for circularly polarised light 

\begin{equation*}
\mathbf{e}_{\pm} = \frac{1}{\sqrt{2}}\left[\begin{array}{c}
                                            1 \\
                                            \mp i
                                           \end{array}\right] 
\end{equation*}

\noindent and the dagger superscript `$^\dagger$' indicates the operation of complex transposition. Taking the squared modulus of \eqref{eq:super} gives

\begin{equation}
 \left|E_{\pm}(\mathbf{r},t)\right|^{2} = \sum_{k,k'}E_{k,\pm}^{*}(t)E_{k',\pm}(t)\Phi_{k}(\mathbf{r})\Phi_{k'}(\mathbf{r})e^{i\Delta\nu_{kk'}t}. \label{eq:Esqmod}
\end{equation}

We shall assume that the shape of the spatial profile for the carriers in a given guide is a constant and that only the overall amplitude for each guide changes. We further assume that the spatial profiles in each guide are the same for each spin-population, as given by \eqref{eq:Nsum}. As described in the main text, $N_{\pm}^{(i)}(t)$ is the time dependent carrier concentration in the $(i)$th guide and $\xi^{(i)}(x,y) = 0$ everywhere outside of the $i$th guide. Since the gain depends on the carrier concentration, the gain may also be decomposed into separate regions: 

\begin{align}
 g_{\pm}(N_{\pm}) & =  \sum_{i}g_{\pm}\left(N^{(i)}\xi^{(i)}(x,y)\right) \equiv  \sum_{i}g_{\pm}^{(i)}(x,y). \label{eq:gsum}
\end{align}

\noindent Similarly, we may also define different pumping terms in each guide

\begin{equation}
 \Lambda(x,y) = \sum_{i}\Lambda^{(i)}(x,y). \label{eq:Lsum}
\end{equation}

\noindent Incorporating \eqref{eq:Esqmod} to \eqref{eq:Lsum} into \eqref{eq:dNpm}, we then have

\begin{align}
 \sum_{i}\frac{\partial N_{\pm}^{(i)}}{\partial t}\xi^{(i)} &=  \sum_{i}\left[-\frac{N_{\pm}^{(i)}}{\tau_{N}}\xi^{(i)} + \Lambda_{\pm}^{(i)} \right. \nonumber \\
 &- \gamma_{J}\left(N_{\pm}^{(i)} - N_{\mp}^{(i)}\right)\xi^{(i)}  \nonumber \\
 &- \left. \frac{c}{n_{g}}g_{\pm}^{(i)}\sum_{k,k'}E_{k,\pm}^{*}E_{k',\pm}\Phi_{k}\Phi_{k'}e^{i\Delta\nu_{kk'}t}\right]. \label{eq:dNpmsum}
\end{align}

The simplest model for the spatial profiles of the carriers, gain and pumping is that of a step profile. We set $\xi^{(i)} = 1$ inside guide $(i)$ and zero elsewhere. Similarly, $g_{\pm}^{(i)}$ and $\Lambda_{\pm}^{(i)}$ may be set to their average values over the $i$th guide inside the guide and zero outside. We may then multiply \eqref{eq:dNpmsum} by a particular step profile $\xi^{(i)}$ and integrate over $x$ and $y$ to obtain

\begin{align}
 \frac{\partial N_{\pm}^{(i)}}{\partial t} &=  -\frac{N_{\pm}^{(i)}}{\tau_{N}} + \Lambda_{\pm}^{(i)} - \gamma_{J}\left(N_{\pm}^{(i)} - N_{\mp}^{(i)}\right)  \nonumber \\
 &-  \frac{c}{n_{g}}\sum_{k,k'}E_{k,\pm}^{*}E_{k',\pm}\overline{g}_{\pm}^{(i)}\int_{(i)}\Phi_{k}\Phi_{k'}~dxdy~e^{i\Delta\nu_{kk'}t}. \nonumber
\end{align}

\noindent However, the integral appearing in this expression has already been defined in terms of the confinement factors in \eqref{eq:Gamma_kk}, so we may re-write this as

\begin{align*}
 \frac{\partial N_{\pm}^{(i)}}{\partial t} &=  -\frac{N_{\pm}^{(i)}}{\tau_{N}} + \Lambda_{\pm}^{(i)} - \gamma_{J}\left(N_{\pm}^{(i)} - N_{\mp}^{(i)}\right) \\
 &-  \frac{c}{n_{g}}\sum_{k,k'}E_{k,\pm}^{*}E_{k',\pm}\overline{g}_{\pm}^{(i)}\Gamma_{kk'}^{(i)}e^{i\Delta\nu_{kk'}t},
\end{align*}

\noindent yielding the carrier rate equations as given in the text.

\section{Derivation of the real form of the optical rate equations}\label{app:real}

In the main text, we defined the amplitudes of composite modes via \eqref{eq:Eplus} and \eqref{eq:Eminus}. We may also define new summations analogously 

\begin{equation}
\Sigma_{1,\pm} = \frac{1}{\sqrt{2}}\left(\tilde{\Sigma}_{s,\pm} + \tilde{\Sigma}_{a,\pm}\right) \label{eq:Splus}
\end{equation}

\noindent and

\begin{equation}
\Sigma_{2,\pm} = \frac{1}{\sqrt{2}}\left(\tilde{\Sigma}_{s,\pm} - \tilde{\Sigma}_{a,\pm}\right). \label{eq:Sminus}
\end{equation}

Rearranging \eqref{eq:Eplus}, \eqref{eq:Eminus}, \eqref{eq:Splus} and \eqref{eq:Sminus}, we have

\begin{equation*}
 \tilde{E}_{s,\pm} = \frac{1}{\sqrt{2}}\left(E_{1,\pm} + E_{2,\pm}\right),~\tilde{E}_{a,\pm} = \frac{1}{\sqrt{2}}\left(E_{1,\pm} - E_{2,\pm}\right) 
\end{equation*}

\noindent and 

\begin{equation*}
 \tilde{\Sigma}_{s,\pm} = \frac{1}{\sqrt{2}}\left(\Sigma_{1,\pm} + \Sigma_{2,\pm}\right),~\tilde{\Sigma}_{a,\pm} = \frac{1}{\sqrt{2}}\left(\Sigma_{1,\pm} - \Sigma_{2,\pm}\right). 
\end{equation*}

\noindent Substituting these expressions into \eqref{eq:dEdtnew} and adding and subtracting $\partial\tilde{E}_{s,\pm}/\partial t$ and $\partial\tilde{E}_{a,\pm}/\partial t$, we obtain

\begin{align}
\frac{\partial E_{1,\pm}}{\partial t} &= i\mu E_{2,\pm} - \frac{1}{2\tau_{p}}E_{1,\pm} + \frac{c}{2n_{g}}\left(1 + i\alpha\right)\Sigma_{1,\pm} \nonumber \\
&- \left(\gamma_{a} + i\gamma_{p}\right)E_{1,\mp} \label{eq:dEdt1}
\end{align}

\noindent and

\begin{align}
\frac{\partial E_{2,\pm}}{\partial t} &= i\mu E_{1,\pm} - \frac{1}{2\tau_{p}}E_{2,\pm} + \frac{c}{2n_{g}}\left(1 + i\alpha\right)\Sigma_{2,\pm} \nonumber \\
&- \left(\gamma_{a} + i\gamma_{p}\right)E_{2,\mp}, \label{eq:dEdt2}
\end{align}

\noindent where $\mu$ is given in terms of the modal frequency difference by $(\nu_{s} - \nu_{a})/2$ as given earlier in \eqref{eq:mu}.

Putting $E_{i,\pm} = \left|E_{i,\pm}\right|e^{i\phi_{i,\pm}}$, we may define phase differences 

\begin{equation}
 \phi_{ijpq} = \phi_{i,p} - \phi_{j,q}, \label{eq:phi_ijpq}
\end{equation}

\noindent where $i,j \in \{1,2\}$ and $p,q \in \{+, -\}$. Equations~\eqref{eq:dEdt1} and \eqref{eq:dEdt2} may then be written as

\begin{align}
\frac{\partial |E_{i,\pm}|}{\partial t}  &= -\mu \sin(\phi_{ji\pm\pm})|E_{j,\pm}| - \frac{1}{2\tau_{p}}|E_{i,\pm}| \nonumber \\
&+ \frac{c}{2n_{g}}\Re\left[\left(1 + i\alpha\right)e^{-i\phi_{i,\pm}}\Sigma_{i,\pm}\right] \nonumber \\
&- \left(\gamma_{a}\cos(\phi_{ii\pm\mp}) + \gamma_{p}\sin(\phi_{ii\pm\mp})\right)|E_{i,\mp}| \label{eq:dE_ipm_1}
\end{align}

\noindent and

\begin{align}
\frac{\partial\phi_{i,\pm}}{\partial t}  &= \mu\cos(\phi_{ji\pm\pm})\frac{|E_{j,\pm}|}{|E_{i,\pm}|} \nonumber \\
&+ \frac{c}{2n_{g}}\Im\left[\left(1 + i\alpha\right)e^{-i\phi_{i,\pm}}\frac{\Sigma_{i,\pm}}{|E_{i,\pm}|}\right] \nonumber \\
&- \left(\gamma_{a}\sin(\phi_{ii\mp\pm}) + \gamma_{p}\cos(\phi_{ii\mp\pm})\right)\frac{|E_{i,\mp}|}{|E_{i,\pm}|}, \label{eq:dphi_ipm_1}
\end{align}

\noindent where $\Re[z]$ and $\Im[z]$ denote the real and imaginary parts of the argument $z$ respectively.

We may combine \eqref{eq:dphi_ipm_1} to find 4 independent equations for the phase differences of \eqref{eq:phi_ijpq}, namely $\phi_{11+-}$, $\phi_{22+-}$, $\phi_{21++}$ and $\phi_{21--}$. Noting that $\phi_{ii+-} = -\phi_{ii-+}$, we have

\begin{align}
&\frac{\partial\phi_{ii+-}}{\partial t}  = \frac{\partial\phi_{i,+}}{\partial t} - \frac{\partial\phi_{i,-}}{\partial t}, \nonumber \\
&= \mu\left(\cos(\phi_{21++})\frac{|E_{j,+}|}{|E_{i,+}|} - \cos(\phi_{21--})\frac{|E_{j,-}|}{|E_{i,-}|}\right) \nonumber \\
&+ \frac{c}{2n_{g}}\Im\left[\left(1 + i\alpha\right)\left(e^{-i\phi_{i,+}}\frac{\Sigma_{i,+}}{|E_{i,+}|} - e^{-i\phi_{i,-}}\frac{\Sigma_{i,-}}{|E_{i,-}|}\right)\right] \nonumber \\
&+ \gamma_{a}\sin(\phi_{ii+-})\left(\frac{|E_{i,+}|}{|E_{i,-}|} + \frac{|E_{i,-}|}{|E_{i,+}|}\right) \nonumber \\
&+ \gamma_{p}\cos(\phi_{ii+-})\left(\frac{|E_{i,+}|}{|E_{i,-}|} - \frac{|E_{i,-}|}{|E_{i,+}|}\right) \nonumber \\ \label{eq:dphiiipm_1}
\end{align}

\noindent and

\begin{align}
&\frac{\partial\phi_{21\pm\pm}}{\partial t}  = \frac{\partial\phi_{2,\pm}}{\partial t} - \frac{\partial\phi_{1,\pm}}{\partial t}, \nonumber \\
&= \mu\cos(\phi_{21\pm\pm})\left(\frac{|E_{1,\pm}|}{|E_{2,\pm}|} -\frac{|E_{2,\pm}|}{|E_{1,\pm}|}\right) \nonumber \\
&+ \frac{c}{2n_{g}}\Im\left[\left(1 + i\alpha\right)\left(e^{-i\phi_{2\pm}}\frac{\Sigma_{2,\pm}}{\left|E_{2,\pm}\right|} - e^{-i\phi_{1\pm}}\frac{\Sigma_{1,\pm}}{\left|E_{1,\pm}\right|}\right)\right] \nonumber \\
&+ \gamma_{p}\left(\cos(\phi_{11+-})\frac{|E_{1,\mp}|}{|E_{1,\pm}|} - \cos(\phi_{22+-})\frac{|E_{2,\mp}|}{|E_{2,\pm}|}\right) \nonumber \\
&\mp \gamma_{a}\left(\sin(\phi_{11+-})\frac{|E_{1,\mp}|}{|E_{1,\pm}|} - \sin(\phi_{22+-})\frac{|E_{2,\mp}|}{|E_{2,\pm}|}\right). \label{eq:dphi21pp_1}
\end{align}

We can now find $\Sigma_{1,\pm}$ and $\Sigma_{2,\pm}$ by expanding the sums $\tilde{\Sigma}_{s,\pm}$ and $\tilde{\Sigma}_{a,\pm}$. These are

\begin{align}
 \tilde{\Sigma}_{s,\pm} &= \frac{E_{1,\pm}}{\sqrt{2}}\left[\left(\Gamma_{ss}^{(1)} + \Gamma_{sa}^{(1)}\right)\overline{g}_{\pm}^{(1)} + \left(\Gamma_{ss}^{(2)} + \Gamma_{sa}^{(2)}\right)\overline{g}_{\pm}^{(2)}\right] \nonumber  \\
 & + \frac{E_{2,\pm}}{\sqrt{2}}\left[\left(\Gamma_{ss}^{(1)} - \Gamma_{sa}^{(1)}\right)\overline{g}_{\pm}^{(1)} + \left(\Gamma_{ss}^{(2)} - \Gamma_{sa}^{(2)}\right)\overline{g}_{\pm}^{(2)}\right] \label{eq:Ss}
\end{align}

\noindent and

\begin{align}
 \tilde{\Sigma}_{a,\pm} & = \frac{E_{1,\pm}}{\sqrt{2}}\left[\left(\Gamma_{aa}^{(1)} + \Gamma_{sa}^{(1)}\right)\overline{g}_{\pm}^{(1)} + \left(\Gamma_{aa}^{(2)} + \Gamma_{sa}^{(2)}\right)\overline{g}_{\pm}^{(2)}\right] \nonumber  \\
 & - \frac{E_{2,\pm}}{\sqrt{2}}\left[\left(\Gamma_{aa}^{(1)} - \Gamma_{sa}^{(1)}\right)\overline{g}_{\pm}^{(1)} + \left(\Gamma_{aa}^{(2)} - \Gamma_{sa}^{(2)}\right)\overline{g}_{\pm}^{(2)}\right]. \label{eq:Sa}
\end{align}

\noindent Adding and subtracting \eqref{eq:Ss} and \eqref{eq:Sa} then gives

\begin{align}
 \Sigma_{1,\pm} & = \frac{E_{1,\pm}}{2}\left[\left(\Gamma_{ss}^{(1)} + \Gamma_{aa}^{(1)} + 2\Gamma_{sa}^{(1)}\right)\overline{g}_{\pm}^{(1)} \right. \nonumber \\
 &+ \left.\left(\Gamma_{ss}^{(2)} + \Gamma_{aa}^{(2)} + 2\Gamma_{sa}^{(2)}\right)\overline{g}_{\pm}^{(2)}\right] \nonumber  \\
 & + \frac{E_{2,\pm}}{2}\left[\left(\Gamma_{ss}^{(1)} - \Gamma_{aa}^{(1)}\right)\overline{g}_{\pm}^{(1)} + \left(\Gamma_{ss}^{(2)} - \Gamma_{aa}^{(2)}\right)\overline{g}_{\pm}^{(2)}\right]\label{eq:S1}
\end{align}

\noindent and

\begin{align}
 \Sigma_{2,\pm} & = \frac{E_{1,\pm}}{2}\left[\left(\Gamma_{ss}^{(1)} - \Gamma_{aa}^{(1)}\right)\overline{g}_{\pm}^{(1)} + \left(\Gamma_{ss}^{(2)} - \Gamma_{aa}^{(2)}\right)\overline{g}_{\pm}^{(2)}\right] \nonumber  \\
 & + \frac{E_{2,\pm}}{2}\left[\left(\Gamma_{ss}^{(1)} + \Gamma_{aa}^{(1)} - 2\Gamma_{sa}^{(1)}\right)\overline{g}_{\pm}^{(1)} \right. \nonumber \\
 &+ \left.\left(\Gamma_{ss}^{(2)} + \Gamma_{aa}^{(2)} - 2\Gamma_{sa}^{(2)}\right)\overline{g}_{\pm}^{(2)}\right]. \label{eq:S2}
\end{align}

For convenience of expression, we define terms $\Gamma_{\pm}^{(i)}$ and $\Delta\Gamma^{(i)}$ via \eqref{eq:Gamma_pm} and \eqref{eq:DGamma} respectively. Equations~\eqref{eq:S1} and \eqref{eq:S2} may then be written more simply as

\begin{align}
 \Sigma_{1,\pm} &= E_{1,\pm}\left(\Gamma_{+}^{(1)}\overline{g}_{\pm}^{(1)} + \Gamma_{+}^{(2)}\overline{g}_{\pm}^{(2)}\right) \nonumber \\
 &+ E_{2,\pm}\left(\Delta\Gamma^{(1)}\overline{g}_{\pm}^{(1)} + \Delta\Gamma^{(2)}\overline{g}_{\pm}^{(2)}\right) \label{eq:S1pm}
\end{align}

\noindent and

\begin{align}
 \Sigma_{2,\pm} &= E_{1,\pm}\left(\Delta\Gamma^{(1)}\overline{g}_{\pm}^{(1)} + \Delta\Gamma^{(2)}\overline{g}_{\pm}^{(2)}\right) \nonumber \\
 &+ E_{2,\pm}\left(\Gamma_{-}^{(1)}\overline{g}_{\pm}^{(1)} + \Gamma_{-}^{(2)}\overline{g}_{\pm}^{(2)}\right). \label{eq:S2pm}
\end{align}

Introducing the gain terms defined in \eqref{eq:G12}, \eqref{eq:G21} and \eqref{eq:DG} and using $E_{k} = |E_{k}|e^{i\phi_{k\pm}}$ as earlier, we may write \eqref{eq:S1pm} and \eqref{eq:S2pm} as

\begin{equation*}
 \Sigma_{1,\pm}e^{-i\phi_{1\pm}} = \Gamma_{S}\left(|E_{1,\pm}|G_{12\pm} + |E_{2,\pm}|\Delta G_{\pm}e^{i\phi_{21\pm\pm}}\right) 
\end{equation*}

\noindent and

\begin{equation*}
 \Sigma_{2,\pm}e^{-i\phi_{2\pm}} = \Gamma_{S}\left(|E_{1,\pm}|\Delta G_{\pm}e^{-i\phi_{21\pm\pm}} + |E_{2,\pm}|G_{21\pm}\right). 
\end{equation*}

\noindent Using these expressions, we find

\begin{align}
&\Re\left[\left(1 + i\alpha\right)e^{-i\phi_{1\pm}}\Sigma_{1,\pm}\right] = \Gamma_{S}G_{12\pm}|E_{1,\pm}| \nonumber \\
&+ \Gamma_{S}\Delta G_{\pm}\left(\cos(\phi_{21\pm\pm}) - \alpha\sin(\phi_{21\pm\pm})\right)|E_{2,\pm}|, \nonumber
\end{align}

\begin{align}
&\Re\left[\left(1 + i\alpha\right)e^{-i\phi_{2\pm}}\Sigma_{2,\pm}\right] = \Gamma_{S}G_{21\pm}|E_{2,\pm}| \nonumber \\
&+ \Gamma_{S}\Delta G_{\pm}\left(\cos(\phi_{21\pm\pm}) + \alpha\sin(\phi_{21\pm\pm})\right)|E_{1,\pm}|,  \nonumber
\end{align}

\begin{widetext}

\begin{align}
& \Im\left[\left(1 + i\alpha\right)\left(e^{-i\phi_{i+}}\frac{\Sigma_{i,+}}{\left|E_{i,+}\right|} - e^{-i\phi_{i-}}\frac{\Sigma_{i,-}}{\left|E_{i,-}\right|}\right)\right] \nonumber \\
&=  \alpha\Gamma_{S}\left[G_{ij+} - G_{ij-} + \Delta G_{+}\cos(\phi_{ji++})\frac{|E_{j,+}|}{|E_{i,+}|}- \Delta G_{-}\cos(\phi_{ji--})\frac{|E_{j,-}|}{|E_{i,-}|}\right] \nonumber \\
&+ \Gamma_{S}\left(\Delta G_{+}\sin(\phi_{ji++})\frac{|E_{j,+}|}{|E_{i,+}|} - \Delta G_{-}\sin(\phi_{ji--})\frac{|E_{j,-}|}{|E_{i,-}|}\right). \nonumber 
\end{align}

\noindent and

\begin{align}
& \Im\left\{\left(1 + i\alpha\right)\left(e^{-i\phi_{2\pm}}\frac{\Sigma_{2,\pm}}{\left|E_{2,\pm}\right|} - e^{-i\phi_{1\pm}}\frac{\Sigma_{1,\pm}}{\left|E_{1,\pm}\right|}\right)\right\} \nonumber \\
&=  \alpha\Gamma_{S}\left[\left(G_{21\pm} - G_{12\pm}\right) + \Delta G_{\pm}\cos(\phi_{21\pm\pm})\left(\frac{|E_{1,\pm}|}{|E_{2,\pm}|} - \frac{|E_{2,\pm}|}{|E_{1,\pm}|}\right)\right] \nonumber \\
&- \Gamma_{S}\Delta G_{\pm}\sin(\phi_{21\pm\pm})\left(\frac{|E_{1,\pm}|}{|E_{2,\pm}|} + \frac{|E_{2,\pm}|}{|E_{1,\pm}|}\right). \nonumber 
\end{align}

\end{widetext}

\noindent Inserting these results into \eqref{eq:dE_ipm_1}, \eqref{eq:dphiiipm_1} and \eqref{eq:dphi21pp_1} and noting that $\phi_{21\pm\pm} = -\phi_{12\pm\pm}$, we arrive at the real form of the optical rate equations for the double guided structure given in \eqref{eq:dabsEdt1}, \eqref{eq:dabsEdt2}   \eqref{eq:dphi21pm} and \eqref{eq:dphi11pm}.

\subsection{Reconstructing the normal modes}\label{app:recon}
The normal modes of the structure are related to the composite modes via

\begin{equation}
 |A_{s,\pm}|e^{-i\varphi_{s,\pm}} = \frac{e^{i\phi_{1,\pm}}}{\sqrt{2}}\left(|A_{1,\pm}| + |A_{2,\pm}|e^{i\phi_{21\pm\pm}}\right) \label{eq:Asrecon}
\end{equation}

\noindent and

\begin{equation}
 |A_{a,\pm}|e^{-i\varphi_{a,\pm}} = \frac{e^{i\phi_{1,\pm}}}{\sqrt{2}}\left(|A_{1,\pm}| - |A_{2,\pm}|e^{i\phi_{21\pm\pm}}\right), \label{eq:Aarecon}
\end{equation}

\noindent from which the amplitudes $|A_{s,\pm}|$ and $|A_{a,\pm}|$ may be found easily by taking the absolute value of each side. Dividing \eqref{eq:Asrecon} by \eqref{eq:Aarecon}, we have

\begin{equation*}
 \frac{|A_{s,\pm}|}{|A_{a,\pm}|}e^{-i(\varphi_{s,\pm}-\varphi_{a,\pm})} = \frac{|A_{1,\pm}| + |A_{2,\pm}|e^{i\phi_{21\pm\pm}}}{|A_{1,\pm}| - |A_{2,\pm}|e^{i\phi_{21\pm\pm}}} 
\end{equation*}

\noindent so, since only the relative phase is of importance, we may put

\begin{equation*}
 A_{s,\pm} = \frac{|A_{1,\pm}| + |A_{2,\pm}|e^{i\phi_{21\pm\pm}}}{|A_{1,\pm}| - |A_{2,\pm}|e^{i\phi_{21\pm\pm}}}|A_{a,\pm}| 
\end{equation*}

\noindent and

\begin{equation*}
 A_{a,\pm} = |A_{a,\pm}|. 
\end{equation*}

\section{Normalisation of the rate equations}\label{app:normal}
In this section, we shall describe the normalisation of the general model given by \eqref{eq:optical_rate} and \eqref{eq:carrier_rate}. The same scheme may be simply applied to the double-guided structure using the same procedure.

Before normalising \eqref{eq:optical_rate} and \eqref{eq:carrier_rate}, it is useful to reduce them to the unpolarised case of a 1D waveguide structure. This is simply done by dropping the $\pm$ subscripts and setting the birefringence, dichroism and spin relaxation rates to zero. For a single guide, \eqref{eq:optical_rate} becomes

\begin{equation*}
\frac{\partial \tilde{E}_{k}}{\partial t} =   \left[i\left(\nu_{k} - \Omega\right) - \frac{1}{2\tau_{p}}\right]\tilde{E}_{k} +   \sum_{k'}\frac{c\Gamma_{kk'}}{2n_{g}}\left(1 + i\alpha\right)\tilde{E}_{k'}\overline{g}. 
\end{equation*}

\noindent For a single-moded guide, we may put $\Omega = \nu_{k}$ and the summation reduces to a single term with $\Gamma_{kk} = \Gamma_{S}$, the optical confinement factor for the guide. Hence, for $\alpha = 0$, we have the well-known rate equation for a single guide 

\begin{equation}
\frac{\partial \tilde{E}}{\partial t} = \left[- \frac{1}{2\tau_{p}} + \frac{c\Gamma_{S}}{2n_{g}}\overline{g}\right]\tilde{E} \label{eq:single_opt}
\end{equation}

\noindent (for non-zero values of $\alpha$, there is no steady-state solution for the phase of $\tilde{E}$).

In the same limiting case, the carrier rate equation reduces to 

\begin{equation}
 \frac{\partial N}{\partial t} = -\frac{N}{\tau_{N}} + \Lambda - \frac{c\Gamma_{S}}{n_{g}}\overline{g}\left|\tilde{E}\right|^{2}. \label{eq:single_car}
\end{equation}

\noindent Here, the appearance of the confinement factor $\Gamma_{S}$ is not usual. This occurs because we have performed an explicit integration over space, which is not ordinarily carried out. More typically, the carrier rate equation would be given in the form of \eqref{eq:dNpm}, where all variables are still spatially dependent. The implicit assumption would then be that these variables are constant across the guide. 

In the linear gain model, we may put

\begin{equation}
 \overline{g} = a_{diff}\left(N - N_{0}\right), \label{eq:lingain}
\end{equation}

\noindent where $a_{diff}$ is the differential gain and $N_{0}$ is the transparency concentration. Hence, in the steady state, we find the threshold carrier concentration $N_{S}$ for the single guide to be given by

\begin{equation}
 N_{S} - N_{0} = \frac{n_{g}}{\Gamma_{S}a_{diff}c\tau_{p}}. \label{eq:Nth}
\end{equation}

We may then define normalised variables via

\begin{equation*}
M = \frac{N - N_{0}}{N_{S} - N_{0}},
\end{equation*}

\begin{equation*}
\eta = \frac{\tau_{N}\Lambda -  N_{0}}{N_{S} - N_{0}} 
\end{equation*}

\noindent and 

\begin{equation}
\left|A\right|^{2} = \frac{\Gamma_{S}a_{diff}c\tau_{N}}{n_{g}} \left|\tilde{E}\right|^{2}.
\end{equation}

\noindent Putting $\kappa = 1/(2\tau_{p})$ for the cavity loss rate and $\gamma = 1/\tau_{N}$ for the recombination rate, \eqref{eq:single_opt} and \eqref{eq:single_car} may then be written in normalised form as

\begin{equation*}
\frac{\partial A}{\partial t} = \left[M - \kappa\right]A 
\end{equation*}

\noindent and

\begin{equation*}
 \frac{\partial M}{\partial t} = \left\{\eta - \left[1 + \left|A\right|^{2}\right]M\right\}. 
\end{equation*}

We may apply a similar normalisation scheme to the general rate equations. The linear gain may now be written

\begin{equation}
 \overline{g}_{\pm}^{(i)} = a_{diff}\left(N_{\pm}^{(i)} - N_{0}\right), \label{eq:lingain_i}
\end{equation}

\noindent (where $a_{diff}$ and $N_{0}$ are assumed to be the same for both spin polarisations) and the normalised variables are defined by

\begin{equation*}
M_{\pm}^{(i)} = \frac{a_{diff}c\tau_{p}}{n_{g}}\left(N_{\pm}^{(i)} - N_{0}\right), 
\end{equation*}

\begin{equation*}
\eta_{\pm}^{(i)} = \frac{a_{diff}c\tau_{p}}{n_{g}}\left(\tau_{N}\Lambda_{\pm}^{(i)} -  N_{0}\right)
\end{equation*}

\noindent and 

\begin{equation*}
\left|A_{k,\pm}\right|^{2} = \frac{\Gamma_{S}a_{diff}c\tau_{N}}{n_{g}} \left|\tilde{E}_{k,\pm}\right|^{2}. 
\end{equation*}

\noindent Note that, in general, $N_{S}$ will no longer be the threshold carrier density in any particular guide. Moreover, for structures involving guides of different widths, $\Gamma_{S}$ will not be uniquely defined and it may be useful to define this as the average confinement factor for each isolated guide.  

The normalised form of the general rate equations is then found to be

\begin{align}
\frac{\partial M_{\pm}^{(i)}}{\partial t} & = \gamma\left[\eta_{\pm}^{(i)} - \left(1 + \sum_{k,k'}\frac{\Gamma_{kk'}^{(i)}}{\Gamma_{S}}A_{k,\pm}^{*}A_{k',\pm}\right)M_\pm^{(i)}\right] \nonumber \\
&- \gamma_{J}\left(M_{\pm}^{(i)} - M_{\mp}^{(i)}\right) \label{eq:carrier_norm}
\end{align}

\noindent and

\begin{align}
\frac{\partial A_{k,\pm}}{\partial t} & = \left[i\left(\nu_{k} - \Omega\right)- \kappa\right]A_{k,\pm} - \left[\gamma_{a} + i\gamma_{p}\right]A_{k,\mp}\nonumber \\
& + \kappa\left(1 + i\alpha\right)\sum_{k'}A_{k',\pm}\sum_{i}M_{\pm}^{(i)}\frac{\Gamma_{kk'}^{(i)}}{\Gamma_{S}}. \label{eq:optical_norm}
\end{align}

\section{Spatial solutions of the Helmholtz equation}\label{app:Helmholtz}
\subsection{Slab waveguides}
For the 1D slab waveguides for transverse electric (TE) polarisation, we have used a semi-analytical method to find normal mode frequencies and spatial profiles - the latter being required to calculate the overlap factors defined earlier in \eqref{eq:Gamma_kk}. 

The electromagnetic wave is take to be propagating in the $z$-direction with propagation constant $\beta
=n\pi/L$ (for integral $n$), fixed by the round trip boundary conditions along the cavity of length $L$. The guided modes in the $x$-direction are found by solving the 1D form of \eqref{eq:Helmholtz}

\begin{equation}
\frac{\partial^{2}\Phi_{k}(x)}{\partial x^{2}} + \left(\frac{n^{2}(x)\nu_{k}^{2}}{c^{2}} - \beta^{2}\right)\Phi_{k}(x)  = 0, \label{eq:Helmholtz_x}
\end{equation}

\noindent where, for a stepped refractive index profile, $n(x) = n_{1}$ in the core regions and $n(x) = n_{2}$ elsewhere. For guided mode solutions, we define

\begin{align} 
\kappa_{0}^{2} & \equiv  \frac{n_{1}^{2}\nu_{k}^{2}}{c^{2}} - \beta^{2} > 0, \nonumber \\
\gamma_{0}^{2} & \equiv  \beta^{2} - \frac{n_{2}^{2}\nu_{k}^{2}}{c^{2}} > 0. \label{eq:gamkap0}
\end{align}

\noindent This constrains the modal frequency to lie in the interval $c\beta/n_{2} > \nu_{k} > c\beta/n_{1}$. 

We consider the general case of a double guided structure with two guide regions of width $w_{1}$ and $w_{2}$, separated by a distance of $2d$. The equation we must solve for $\nu_{k}$ is then

\begin{align}
&\tanh(2\gamma_{0} d) - \frac{\kappa_{0}}{\gamma_{0}}\left(\tan\Psi_{1} + \tan\Psi_{2}\right) \nonumber \\
&+ \left(\frac{\kappa_{0}}{\gamma_{0}}\right)^{2}\tan\Psi_{1}\tan\Psi_{2}\tanh(2\gamma_{0} d) = 0, \label{eq:fbeta}
\end{align}

\noindent where $\Psi_{i} = \kappa_{0} w_{i} - \phi$ and $\phi = \tan^{-1}(\gamma_{0}/\kappa_{0})$. In practice, we use a bisection method to achieve this. The solutions of Eqs.~\eqref{eq:Helmholtz_x} then have the form

\begin{align}
&\Phi_{I} =  Ae^{\gamma_{0} x}, \nonumber \\
&\Phi_{II} =  AB\cos\left(\kappa_{0}\left[x + w_{1} + d\right] - \phi\right), \nonumber \\
&\Phi_{III} =  AB\left(\cos\Psi_{1}\cosh\left[\gamma_{0}(x+d)\right]\right. \nonumber \\
&- \left. (\kappa_{0}/\gamma_{0})\sin\Psi_{1}\sinh\left[\gamma_{0}(x+d)\right]\right), \nonumber \\
&\Phi_{IV} =  ABC\cos\left(\kappa_{0}\left[x - w_{2} - d\right] + \phi\right), \nonumber \\
&\Phi_{V} =  ACe^{\gamma_{0}(w_{2} - w_{1} - x)}, \label{eq:psisol2}
\end{align}

\noindent where regions II and IV are the waveguide cores, region III is the region between the guides, and regions I and V are those outside the waveguides. Here $A$ is a normalisation constant,

\begin{equation*}
B = e^{-\gamma_{0}(w_{1} + d)}\sqrt{1 + \frac{\gamma_{0}^{2}}{\kappa_{0}^{2}}} 
\end{equation*}

\noindent and

\begin{equation*}
 C = e^{\gamma_{0} 2d}\frac{\cos\Psi_{1} - (\kappa_{0}/\gamma_{0})\sin\Psi_{1}}{\cos\Psi_{2} + (\kappa_{0}/\gamma_{0})\sin\Psi_{2}}. 
\end{equation*}

In this current work, we restrict ourselves to the case of equal width guides, putting $w_{1} = w_{2} = 2a$. In this case, we obtain symmetric and anti-symmetric solutions. It may be shown that, for the symmetric solutions, $C \to 1$ and \eqref{eq:fbeta} reduces to

\begin{equation}
 \tan\left(2\kappa_{0}a - \phi\right) = \frac{\gamma_{0}}{\kappa_{0}}\mathrm{tanh}\left(\gamma_{0} d\right), \label{eq:sym_sol}
\end{equation}

\noindent whilst for the anti-symmetric solutions, $C \to -1$ and \eqref{eq:fbeta} reduces to

\begin{equation}
 \tan\left(2\kappa_{0}a - \phi\right) = \frac{\gamma_{0}}{\kappa_{0}}\mathrm{coth}\left(\gamma_{0} d\right). \label{eq:asym_sol}
\end{equation}

\bibliography{spinVCSEL}

\begin{thebibliography}{32}%
\makeatletter
\providecommand \@ifxundefined [1]{%
 \@ifx{#1\undefined}
}%
\providecommand \@ifnum [1]{%
 \ifnum #1\expandafter \@firstoftwo
 \else \expandafter \@secondoftwo
 \fi
}%
\providecommand \@ifx [1]{%
 \ifx #1\expandafter \@firstoftwo
 \else \expandafter \@secondoftwo
 \fi
}%
\providecommand \natexlab [1]{#1}%
\providecommand \enquote  [1]{``#1''}%
\providecommand \bibnamefont  [1]{#1}%
\providecommand \bibfnamefont [1]{#1}%
\providecommand \citenamefont [1]{#1}%
\providecommand \href@noop [0]{\@secondoftwo}%
\providecommand \href [0]{\begingroup \@sanitize@url \@href}%
\providecommand \@href[1]{\@@startlink{#1}\@@href}%
\providecommand \@@href[1]{\endgroup#1\@@endlink}%
\providecommand \@sanitize@url [0]{\catcode `\\12\catcode `\$12\catcode
  `\&12\catcode `\#12\catcode `\^12\catcode `\_12\catcode `\%12\relax}%
\providecommand \@@startlink[1]{}%
\providecommand \@@endlink[0]{}%
\providecommand \url  [0]{\begingroup\@sanitize@url \@url }%
\providecommand \@url [1]{\endgroup\@href {#1}{\urlprefix }}%
\providecommand \urlprefix  [0]{URL }%
\providecommand \Eprint [0]{\href }%
\providecommand \doibase [0]{http://dx.doi.org/}%
\providecommand \selectlanguage [0]{\@gobble}%
\providecommand \bibinfo  [0]{\@secondoftwo}%
\providecommand \bibfield  [0]{\@secondoftwo}%
\providecommand \translation [1]{[#1]}%
\providecommand \BibitemOpen [0]{}%
\providecommand \bibitemStop [0]{}%
\providecommand \bibitemNoStop [0]{.\EOS\space}%
\providecommand \EOS [0]{\spacefactor3000\relax}%
\providecommand \BibitemShut  [1]{\csname bibitem#1\endcsname}%
\let\auto@bib@innerbib\@empty
%</preamble>
\bibitem [{\citenamefont {San~Miguel}\ \emph {et~al.}(1995)\citenamefont
  {San~Miguel}, \citenamefont {Feng},\ and\ \citenamefont
  {Moloney}}]{san1995light}%
  \BibitemOpen
  \bibfield  {author} {\bibinfo {author} {\bibfnamefont {M.}~\bibnamefont
  {San~Miguel}}, \bibinfo {author} {\bibfnamefont {Q.}~\bibnamefont {Feng}}, \
  and\ \bibinfo {author} {\bibfnamefont {J.~V.}\ \bibnamefont {Moloney}},\
  }\href@noop {} {\bibfield  {journal} {\bibinfo  {journal} {Phys. Rev. A}\
  }\textbf {\bibinfo {volume} {52}},\ \bibinfo {pages} {1728} (\bibinfo {year}
  {1995})}\BibitemShut {NoStop}%
\bibitem [{\citenamefont {Martin-Regalado}\ \emph
  {et~al.}(1997{\natexlab{a}})\citenamefont {Martin-Regalado}, \citenamefont
  {Prati}, \citenamefont {San~Miguel},\ and\ \citenamefont
  {Abraham}}]{martin1997polarizationProperties}%
  \BibitemOpen
  \bibfield  {author} {\bibinfo {author} {\bibfnamefont {J.}~\bibnamefont
  {Martin-Regalado}}, \bibinfo {author} {\bibfnamefont {F.}~\bibnamefont
  {Prati}}, \bibinfo {author} {\bibfnamefont {M.}~\bibnamefont {San~Miguel}}, \
  and\ \bibinfo {author} {\bibfnamefont {N.}~\bibnamefont {Abraham}},\
  }\href@noop {} {\bibfield  {journal} {\bibinfo  {journal} {IEEE J. of Quantum
  Electron.}\ }\textbf {\bibinfo {volume} {33}},\ \bibinfo {pages} {765}
  (\bibinfo {year} {1997}{\natexlab{a}})}\BibitemShut {NoStop}%
\bibitem [{\citenamefont {Travagnin}\ \emph {et~al.}(1996)\citenamefont
  {Travagnin}, \citenamefont {Van~Exter}, \citenamefont {Van~Doorn},\ and\
  \citenamefont {Woerdman}}]{travagnin1996role}%
  \BibitemOpen
  \bibfield  {author} {\bibinfo {author} {\bibfnamefont {M.}~\bibnamefont
  {Travagnin}}, \bibinfo {author} {\bibfnamefont {M.}~\bibnamefont
  {Van~Exter}}, \bibinfo {author} {\bibfnamefont {A.~J.}\ \bibnamefont
  {Van~Doorn}}, \ and\ \bibinfo {author} {\bibfnamefont {J.}~\bibnamefont
  {Woerdman}},\ }\href@noop {} {\bibfield  {journal} {\bibinfo  {journal}
  {Phys. Rev. A}\ }\textbf {\bibinfo {volume} {54}},\ \bibinfo {pages} {1647}
  (\bibinfo {year} {1996})}\BibitemShut {NoStop}%
\bibitem [{\citenamefont {Travagnin}\ \emph {et~al.}(1997)\citenamefont
  {Travagnin}, \citenamefont {van Exter}, \citenamefont {Van~Doorn},\ and\
  \citenamefont {Woerdman}}]{travagnin1997erratum}%
  \BibitemOpen
  \bibfield  {author} {\bibinfo {author} {\bibfnamefont {M.}~\bibnamefont
  {Travagnin}}, \bibinfo {author} {\bibfnamefont {M.}~\bibnamefont {van
  Exter}}, \bibinfo {author} {\bibfnamefont {A.~J.}\ \bibnamefont {Van~Doorn}},
  \ and\ \bibinfo {author} {\bibfnamefont {J.}~\bibnamefont {Woerdman}},\
  }\href@noop {} {\bibfield  {journal} {\bibinfo  {journal} {Phys. Rev. A}\
  }\textbf {\bibinfo {volume} {55}},\ \bibinfo {pages} {4641} (\bibinfo {year}
  {1997})}\BibitemShut {NoStop}%
\bibitem [{\citenamefont {Balle}\ \emph {et~al.}(1999)\citenamefont {Balle},
  \citenamefont {Tolkachova}, \citenamefont {San~Miguel}, \citenamefont
  {Tredicce}, \citenamefont {Martin-Regalado},\ and\ \citenamefont
  {Gahl}}]{balle1999mechanisms}%
  \BibitemOpen
  \bibfield  {author} {\bibinfo {author} {\bibfnamefont {S.}~\bibnamefont
  {Balle}}, \bibinfo {author} {\bibfnamefont {E.}~\bibnamefont {Tolkachova}},
  \bibinfo {author} {\bibfnamefont {M.}~\bibnamefont {San~Miguel}}, \bibinfo
  {author} {\bibfnamefont {J.~R.}\ \bibnamefont {Tredicce}}, \bibinfo {author}
  {\bibfnamefont {J.}~\bibnamefont {Martin-Regalado}}, \ and\ \bibinfo {author}
  {\bibfnamefont {A.}~\bibnamefont {Gahl}},\ }\href@noop {} {\bibfield
  {journal} {\bibinfo  {journal} {Opt. Lett.}\ }\textbf {\bibinfo {volume}
  {24}},\ \bibinfo {pages} {1121} (\bibinfo {year} {1999})}\BibitemShut
  {NoStop}%
\bibitem [{\citenamefont {Sondermann}\ \emph {et~al.}(2004)\citenamefont
  {Sondermann}, \citenamefont {Ackemann}, \citenamefont {Balle}, \citenamefont
  {Mulet},\ and\ \citenamefont {Panajotov}}]{sondermann2004experimental}%
  \BibitemOpen
  \bibfield  {author} {\bibinfo {author} {\bibfnamefont {M.}~\bibnamefont
  {Sondermann}}, \bibinfo {author} {\bibfnamefont {T.}~\bibnamefont
  {Ackemann}}, \bibinfo {author} {\bibfnamefont {S.}~\bibnamefont {Balle}},
  \bibinfo {author} {\bibfnamefont {J.}~\bibnamefont {Mulet}}, \ and\ \bibinfo
  {author} {\bibfnamefont {K.}~\bibnamefont {Panajotov}},\ }\href@noop {}
  {\bibfield  {journal} {\bibinfo  {journal} {Opt. Commun.}\ }\textbf {\bibinfo
  {volume} {235}},\ \bibinfo {pages} {421} (\bibinfo {year}
  {2004})}\BibitemShut {NoStop}%
\bibitem [{\citenamefont {Panajotov}\ and\ \citenamefont
  {Pratl}(2012)}]{Panajotov2012Polarization}%
  \BibitemOpen
  \bibfield  {author} {\bibinfo {author} {\bibfnamefont {K.}~\bibnamefont
  {Panajotov}}\ and\ \bibinfo {author} {\bibfnamefont {F.}~\bibnamefont
  {Pratl}},\ }in\ \href@noop {} {\emph {\bibinfo {booktitle} {VCSELs:
  Fundamentals, Technology and Applications of Vertical-Cavity Surface-Emitting
  Lasers}}},\ \bibinfo {series} {Springer Series in Optical Sciences}, Vol.\
  \bibinfo {volume} {166},\ \bibinfo {editor} {edited by\ \bibinfo {editor}
  {\bibfnamefont {R.}~\bibnamefont {Michalzik}}}\ (\bibinfo  {publisher}
  {Springer},\ \bibinfo {year} {2012})\ Chap.~\bibinfo {chapter}
  {6}\BibitemShut {NoStop}%
\bibitem [{\citenamefont {Mulet}\ and\ \citenamefont
  {Balle}(2002)}]{mulet2002spatio}%
  \BibitemOpen
  \bibfield  {author} {\bibinfo {author} {\bibfnamefont {J.}~\bibnamefont
  {Mulet}}\ and\ \bibinfo {author} {\bibfnamefont {S.}~\bibnamefont {Balle}},\
  }\href@noop {} {\bibfield  {journal} {\bibinfo  {journal} {IEEE J. Quantum
  Electron.}\ }\textbf {\bibinfo {volume} {38}},\ \bibinfo {pages} {291}
  (\bibinfo {year} {2002})}\BibitemShut {NoStop}%
\bibitem [{\citenamefont {Masoller}\ and\ \citenamefont
  {Torre}(2008)}]{masollera2008modeling}%
  \BibitemOpen
  \bibfield  {author} {\bibinfo {author} {\bibfnamefont {C.}~\bibnamefont
  {Masoller}}\ and\ \bibinfo {author} {\bibfnamefont {M.}~\bibnamefont
  {Torre}},\ }\href@noop {} {\bibfield  {journal} {\bibinfo  {journal} {Opt.
  Express}\ }\textbf {\bibinfo {volume} {16}},\ \bibinfo {pages} {21282}
  (\bibinfo {year} {2008})}\BibitemShut {NoStop}%
\bibitem [{\citenamefont {Gerhardt}\ and\ \citenamefont
  {Hofmann}(2012)}]{gerhardt2012spin}%
  \BibitemOpen
  \bibfield  {author} {\bibinfo {author} {\bibfnamefont {N.~C.}\ \bibnamefont
  {Gerhardt}}\ and\ \bibinfo {author} {\bibfnamefont {M.~R.}\ \bibnamefont
  {Hofmann}},\ }\href@noop {} {\bibfield  {journal} {\bibinfo  {journal} {Adv.
  Opt. Technol.}\ }\textbf {\bibinfo {volume} {2012}} (\bibinfo {year}
  {2012})}\BibitemShut {NoStop}%
\bibitem [{\citenamefont {Gahl}\ \emph {et~al.}(1999)\citenamefont {Gahl},
  \citenamefont {Balle},\ and\ \citenamefont {Miguel}}]{gahl1999polarization}%
  \BibitemOpen
  \bibfield  {author} {\bibinfo {author} {\bibfnamefont {A.}~\bibnamefont
  {Gahl}}, \bibinfo {author} {\bibfnamefont {S.}~\bibnamefont {Balle}}, \ and\
  \bibinfo {author} {\bibfnamefont {M.~S.}\ \bibnamefont {Miguel}},\
  }\href@noop {} {\bibfield  {journal} {\bibinfo  {journal} {IEEE J. Quantum
  Electron.}\ }\textbf {\bibinfo {volume} {35}},\ \bibinfo {pages} {342}
  (\bibinfo {year} {1999})}\BibitemShut {NoStop}%
\bibitem [{\citenamefont {Gerhardt}\ \emph {et~al.}(2006)\citenamefont
  {Gerhardt}, \citenamefont {Hovel}, \citenamefont {Hofmann}, \citenamefont
  {Yang}, \citenamefont {Reuter},\ and\ \citenamefont
  {Wieck}}]{gerhardt2006enhancement}%
  \BibitemOpen
  \bibfield  {author} {\bibinfo {author} {\bibfnamefont {N.}~\bibnamefont
  {Gerhardt}}, \bibinfo {author} {\bibfnamefont {S.}~\bibnamefont {Hovel}},
  \bibinfo {author} {\bibfnamefont {M.}~\bibnamefont {Hofmann}}, \bibinfo
  {author} {\bibfnamefont {J.}~\bibnamefont {Yang}}, \bibinfo {author}
  {\bibfnamefont {D.}~\bibnamefont {Reuter}}, \ and\ \bibinfo {author}
  {\bibfnamefont {A.}~\bibnamefont {Wieck}},\ }\href@noop {} {\bibfield
  {journal} {\bibinfo  {journal} {Electron. Lett.}\ }\textbf {\bibinfo {volume}
  {42}},\ \bibinfo {pages} {88} (\bibinfo {year} {2006})}\BibitemShut {NoStop}%
\bibitem [{\citenamefont {Adams}\ and\ \citenamefont
  {Alexandropoulos}(2009)}]{adams2009parametric}%
  \BibitemOpen
  \bibfield  {author} {\bibinfo {author} {\bibfnamefont {M.~J.}\ \bibnamefont
  {Adams}}\ and\ \bibinfo {author} {\bibfnamefont {D.}~\bibnamefont
  {Alexandropoulos}},\ }\href@noop {} {\bibfield  {journal} {\bibinfo
  {journal} {IEEE J. Quantum Electron.}\ }\textbf {\bibinfo {volume} {45}},\
  \bibinfo {pages} {744} (\bibinfo {year} {2009})}\BibitemShut {NoStop}%
\bibitem [{\citenamefont {Adams}\ \emph {et~al.}(2018)\citenamefont {Adams},
  \citenamefont {Li}, \citenamefont {Cemlyn}, \citenamefont {Susanto},\ and\
  \citenamefont {Henning}}]{adams2018algebraic}%
  \BibitemOpen
  \bibfield  {author} {\bibinfo {author} {\bibfnamefont {M.}~\bibnamefont
  {Adams}}, \bibinfo {author} {\bibfnamefont {N.}~\bibnamefont {Li}}, \bibinfo
  {author} {\bibfnamefont {B.}~\bibnamefont {Cemlyn}}, \bibinfo {author}
  {\bibfnamefont {H.}~\bibnamefont {Susanto}}, \ and\ \bibinfo {author}
  {\bibfnamefont {I.}~\bibnamefont {Henning}},\ }\href@noop {} {\bibfield
  {journal} {\bibinfo  {journal} {Semicond. Sci. Technol.}\ }\textbf {\bibinfo
  {volume} {33}},\ \bibinfo {pages} {064002} (\bibinfo {year}
  {2018})}\BibitemShut {NoStop}%
\bibitem [{\citenamefont {Li}\ \emph {et~al.}(2010)\citenamefont {Li},
  \citenamefont {J{\"a}hme}, \citenamefont {Soldat}, \citenamefont {Gerhardt},
  \citenamefont {Hofmann},\ and\ \citenamefont
  {Ackemann}}]{li2010birefringence}%
  \BibitemOpen
  \bibfield  {author} {\bibinfo {author} {\bibfnamefont {M.}~\bibnamefont
  {Li}}, \bibinfo {author} {\bibfnamefont {H.}~\bibnamefont {J{\"a}hme}},
  \bibinfo {author} {\bibfnamefont {H.}~\bibnamefont {Soldat}}, \bibinfo
  {author} {\bibfnamefont {N.}~\bibnamefont {Gerhardt}}, \bibinfo {author}
  {\bibfnamefont {M.}~\bibnamefont {Hofmann}}, \ and\ \bibinfo {author}
  {\bibfnamefont {T.}~\bibnamefont {Ackemann}},\ }\href@noop {} {\bibfield
  {journal} {\bibinfo  {journal} {Appl. Phys. Lett.}\ }\textbf {\bibinfo
  {volume} {97}},\ \bibinfo {pages} {191114} (\bibinfo {year}
  {2010})}\BibitemShut {NoStop}%
\bibitem [{\citenamefont {Lindemann}\ \emph {et~al.}(2016)\citenamefont
  {Lindemann}, \citenamefont {Pusch}, \citenamefont {Michalzik}, \citenamefont
  {Gerhardt},\ and\ \citenamefont {Hofmann}}]{lindemann2016frequency}%
  \BibitemOpen
  \bibfield  {author} {\bibinfo {author} {\bibfnamefont {M.}~\bibnamefont
  {Lindemann}}, \bibinfo {author} {\bibfnamefont {T.}~\bibnamefont {Pusch}},
  \bibinfo {author} {\bibfnamefont {R.}~\bibnamefont {Michalzik}}, \bibinfo
  {author} {\bibfnamefont {N.~C.}\ \bibnamefont {Gerhardt}}, \ and\ \bibinfo
  {author} {\bibfnamefont {M.~R.}\ \bibnamefont {Hofmann}},\ }\href@noop {}
  {\bibfield  {journal} {\bibinfo  {journal} {Appl. Phys. Lett.}\ }\textbf
  {\bibinfo {volume} {108}},\ \bibinfo {pages} {042404} (\bibinfo {year}
  {2016})}\BibitemShut {NoStop}%
\bibitem [{\citenamefont {Torre}\ \emph {et~al.}(2017)\citenamefont {Torre},
  \citenamefont {Susanto}, \citenamefont {Li}, \citenamefont {Schires},
  \citenamefont {Salvide}, \citenamefont {Henning}, \citenamefont {Adams},\
  and\ \citenamefont {Hurtado}}]{torre2017high}%
  \BibitemOpen
  \bibfield  {author} {\bibinfo {author} {\bibfnamefont {M.~S.}\ \bibnamefont
  {Torre}}, \bibinfo {author} {\bibfnamefont {H.}~\bibnamefont {Susanto}},
  \bibinfo {author} {\bibfnamefont {N.}~\bibnamefont {Li}}, \bibinfo {author}
  {\bibfnamefont {K.}~\bibnamefont {Schires}}, \bibinfo {author} {\bibfnamefont
  {M.}~\bibnamefont {Salvide}}, \bibinfo {author} {\bibfnamefont
  {I.}~\bibnamefont {Henning}}, \bibinfo {author} {\bibfnamefont
  {M.}~\bibnamefont {Adams}}, \ and\ \bibinfo {author} {\bibfnamefont
  {A.}~\bibnamefont {Hurtado}},\ }\href@noop {} {\bibfield  {journal} {\bibinfo
   {journal} {Opt. Lett.}\ }\textbf {\bibinfo {volume} {42}},\ \bibinfo {pages}
  {1628} (\bibinfo {year} {2017})}\BibitemShut {NoStop}%
\bibitem [{\citenamefont {Hendriks}\ \emph {et~al.}(1996)\citenamefont
  {Hendriks}, \citenamefont {Van~Exter}, \citenamefont {Woerdman},\ and\
  \citenamefont {Van~der Poel}}]{hendriks1996phase}%
  \BibitemOpen
  \bibfield  {author} {\bibinfo {author} {\bibfnamefont {R.}~\bibnamefont
  {Hendriks}}, \bibinfo {author} {\bibfnamefont {M.}~\bibnamefont {Van~Exter}},
  \bibinfo {author} {\bibfnamefont {J.}~\bibnamefont {Woerdman}}, \ and\
  \bibinfo {author} {\bibfnamefont {C.}~\bibnamefont {Van~der Poel}},\
  }\href@noop {} {\bibfield  {journal} {\bibinfo  {journal} {Appl. Phys.
  Lett.}\ }\textbf {\bibinfo {volume} {69}},\ \bibinfo {pages} {869} (\bibinfo
  {year} {1996})}\BibitemShut {NoStop}%
\bibitem [{\citenamefont {Ebeling}\ \emph {et~al.}(2018)\citenamefont
  {Ebeling}, \citenamefont {Michalzik},\ and\ \citenamefont
  {Moench}}]{ebeling2018vertical}%
  \BibitemOpen
  \bibfield  {author} {\bibinfo {author} {\bibfnamefont {K.~J.}\ \bibnamefont
  {Ebeling}}, \bibinfo {author} {\bibfnamefont {R.}~\bibnamefont {Michalzik}},
  \ and\ \bibinfo {author} {\bibfnamefont {H.}~\bibnamefont {Moench}},\
  }\href@noop {} {\bibfield  {journal} {\bibinfo  {journal} {Japan. J. Appl.
  Phys.}\ }\textbf {\bibinfo {volume} {57}},\ \bibinfo {pages} {08PA02}
  (\bibinfo {year} {2018})}\BibitemShut {NoStop}%
\bibitem [{\citenamefont {Czyszanowski}\ \emph {et~al.}(2013)\citenamefont
  {Czyszanowski}, \citenamefont {Sarza{\l}a}, \citenamefont {Dems},
  \citenamefont {Walczak}, \citenamefont {Wasiak}, \citenamefont {Nakwaski},
  \citenamefont {Iakovlev}, \citenamefont {Volet},\ and\ \citenamefont
  {Kapon}}]{czyszanowski2013spatial}%
  \BibitemOpen
  \bibfield  {author} {\bibinfo {author} {\bibfnamefont {T.}~\bibnamefont
  {Czyszanowski}}, \bibinfo {author} {\bibfnamefont {R.~P.}\ \bibnamefont
  {Sarza{\l}a}}, \bibinfo {author} {\bibfnamefont {M.}~\bibnamefont {Dems}},
  \bibinfo {author} {\bibfnamefont {J.}~\bibnamefont {Walczak}}, \bibinfo
  {author} {\bibfnamefont {M.}~\bibnamefont {Wasiak}}, \bibinfo {author}
  {\bibfnamefont {W.}~\bibnamefont {Nakwaski}}, \bibinfo {author}
  {\bibfnamefont {V.}~\bibnamefont {Iakovlev}}, \bibinfo {author}
  {\bibfnamefont {N.}~\bibnamefont {Volet}}, \ and\ \bibinfo {author}
  {\bibfnamefont {E.}~\bibnamefont {Kapon}},\ }\href@noop {} {\bibfield
  {journal} {\bibinfo  {journal} {IEEE J. Select. Topics in Quantum Electron.}\
  }\textbf {\bibinfo {volume} {19}},\ \bibinfo {pages} {1} (\bibinfo {year}
  {2013})}\BibitemShut {NoStop}%
\bibitem [{\citenamefont {Blackbeard}\ \emph {et~al.}(2014)\citenamefont
  {Blackbeard}, \citenamefont {Wieczorek}, \citenamefont {Erzgr{\"a}ber},\ and\
  \citenamefont {Dutta}}]{blackbeard2014synchronisation}%
  \BibitemOpen
  \bibfield  {author} {\bibinfo {author} {\bibfnamefont {N.}~\bibnamefont
  {Blackbeard}}, \bibinfo {author} {\bibfnamefont {S.}~\bibnamefont
  {Wieczorek}}, \bibinfo {author} {\bibfnamefont {H.}~\bibnamefont
  {Erzgr{\"a}ber}}, \ and\ \bibinfo {author} {\bibfnamefont {P.~S.}\
  \bibnamefont {Dutta}},\ }\href@noop {} {\bibfield  {journal} {\bibinfo
  {journal} {Physica D}\ }\textbf {\bibinfo {volume} {286}},\ \bibinfo {pages}
  {43} (\bibinfo {year} {2014})}\BibitemShut {NoStop}%
\bibitem [{\citenamefont {Vaughan}\ \emph {et~al.}(2019)\citenamefont
  {Vaughan}, \citenamefont {Susanto}, \citenamefont {Li}, \citenamefont
  {Henning},\ and\ \citenamefont {Adams}}]{vaughan2019stability}%
  \BibitemOpen
  \bibfield  {author} {\bibinfo {author} {\bibfnamefont {M.}~\bibnamefont
  {Vaughan}}, \bibinfo {author} {\bibfnamefont {H.}~\bibnamefont {Susanto}},
  \bibinfo {author} {\bibfnamefont {N.}~\bibnamefont {Li}}, \bibinfo {author}
  {\bibfnamefont {I.}~\bibnamefont {Henning}}, \ and\ \bibinfo {author}
  {\bibfnamefont {M.}~\bibnamefont {Adams}},\ }\href@noop {} {\bibfield
  {journal} {\bibinfo  {journal} {Photonics}\ }\textbf {\bibinfo {volume}
  {6}},\ \bibinfo {pages} {74} (\bibinfo {year} {2019})}\BibitemShut {NoStop}%
\bibitem [{\citenamefont {Adams}\ \emph {et~al.}(2017)\citenamefont {Adams},
  \citenamefont {Li}, \citenamefont {Cemlyn}, \citenamefont {Susanto},\ and\
  \citenamefont {Henning}}]{adams2017effects}%
  \BibitemOpen
  \bibfield  {author} {\bibinfo {author} {\bibfnamefont {M.}~\bibnamefont
  {Adams}}, \bibinfo {author} {\bibfnamefont {N.}~\bibnamefont {Li}}, \bibinfo
  {author} {\bibfnamefont {B.}~\bibnamefont {Cemlyn}}, \bibinfo {author}
  {\bibfnamefont {H.}~\bibnamefont {Susanto}}, \ and\ \bibinfo {author}
  {\bibfnamefont {I.}~\bibnamefont {Henning}},\ }\href@noop {} {\bibfield
  {journal} {\bibinfo  {journal} {Phys. Rev. A}\ }\textbf {\bibinfo {volume}
  {95}},\ \bibinfo {pages} {053869} (\bibinfo {year} {2017})}\BibitemShut
  {NoStop}%
\bibitem [{\citenamefont {Martin-Regalado}\ \emph
  {et~al.}(1997{\natexlab{b}})\citenamefont {Martin-Regalado}, \citenamefont
  {Balle}, \citenamefont {San~Miguel}, \citenamefont {Valle},\ and\
  \citenamefont {Pesquera}}]{martin1997polarization}%
  \BibitemOpen
  \bibfield  {author} {\bibinfo {author} {\bibfnamefont {J.}~\bibnamefont
  {Martin-Regalado}}, \bibinfo {author} {\bibfnamefont {S.}~\bibnamefont
  {Balle}}, \bibinfo {author} {\bibfnamefont {M.}~\bibnamefont {San~Miguel}},
  \bibinfo {author} {\bibfnamefont {A.}~\bibnamefont {Valle}}, \ and\ \bibinfo
  {author} {\bibfnamefont {L.}~\bibnamefont {Pesquera}},\ }\href@noop {}
  {\bibfield  {journal} {\bibinfo  {journal} {Quantum Semiclass. Opt.}\
  }\textbf {\bibinfo {volume} {9}},\ \bibinfo {pages} {713} (\bibinfo {year}
  {1997}{\natexlab{b}})}\BibitemShut {NoStop}%
\bibitem [{\citenamefont {Ogawa}(1977)}]{ogawa1977simplified}%
  \BibitemOpen
  \bibfield  {author} {\bibinfo {author} {\bibfnamefont {K.}~\bibnamefont
  {Ogawa}},\ }\href@noop {} {\bibfield  {journal} {\bibinfo  {journal} {Bell
  Syst. Tech. J.}\ }\textbf {\bibinfo {volume} {56}},\ \bibinfo {pages} {729}
  (\bibinfo {year} {1977})}\BibitemShut {NoStop}%
\bibitem [{\citenamefont {Marom}\ \emph {et~al.}(1984)\citenamefont {Marom},
  \citenamefont {Ramer},\ and\ \citenamefont {Ruschin}}]{Marom1984Relation}%
  \BibitemOpen
  \bibfield  {author} {\bibinfo {author} {\bibfnamefont {E.}~\bibnamefont
  {Marom}}, \bibinfo {author} {\bibfnamefont {O.}~\bibnamefont {Ramer}}, \ and\
  \bibinfo {author} {\bibfnamefont {S.}~\bibnamefont {Ruschin}},\ }\href@noop
  {} {\bibfield  {journal} {\bibinfo  {journal} {IEEE J. Quantum Electron.}\
  }\textbf {\bibinfo {volume} {20}},\ \bibinfo {pages} {1311} (\bibinfo {year}
  {1984})}\BibitemShut {NoStop}%
\bibitem [{\citenamefont {Dormand}\ and\ \citenamefont
  {Prince}(1980)}]{dormand1980family}%
  \BibitemOpen
  \bibfield  {author} {\bibinfo {author} {\bibfnamefont {J.~R.}\ \bibnamefont
  {Dormand}}\ and\ \bibinfo {author} {\bibfnamefont {P.~J.}\ \bibnamefont
  {Prince}},\ }\href@noop {} {\bibfield  {journal} {\bibinfo  {journal} {J.
  Comput. Appl. Math}\ }\textbf {\bibinfo {volume} {6}},\ \bibinfo {pages} {19}
  (\bibinfo {year} {1980})}\BibitemShut {NoStop}%
\bibitem [{\citenamefont {Shampine}\ and\ \citenamefont
  {Reichelt}(1997)}]{shampine1997matlab}%
  \BibitemOpen
  \bibfield  {author} {\bibinfo {author} {\bibfnamefont {L.~F.}\ \bibnamefont
  {Shampine}}\ and\ \bibinfo {author} {\bibfnamefont {M.~W.}\ \bibnamefont
  {Reichelt}},\ }\href@noop {} {\bibfield  {journal} {\bibinfo  {journal} {SIAM
  J. Sci. Comput.}\ }\textbf {\bibinfo {volume} {18}},\ \bibinfo {pages} {1}
  (\bibinfo {year} {1997})}\BibitemShut {NoStop}%
\bibitem [{\citenamefont {Powell}(1970)}]{powell1968fortran}%
  \BibitemOpen
  \bibfield  {author} {\bibinfo {author} {\bibfnamefont {M.~J.}\ \bibnamefont
  {Powell}},\ }in\ \href@noop {} {\emph {\bibinfo {booktitle} {Numerical
  Methods for Nonlinear Algebraic Equations}}},\ \bibinfo {editor} {edited by\
  \bibinfo {editor} {\bibfnamefont {P.}~\bibnamefont {Rabinowitz}}}\ (\bibinfo
  {year} {1970})\BibitemShut {NoStop}%
\bibitem [{\citenamefont {Coleman}\ and\ \citenamefont
  {Li}(1996)}]{coleman1996interior}%
  \BibitemOpen
  \bibfield  {author} {\bibinfo {author} {\bibfnamefont {T.~F.}\ \bibnamefont
  {Coleman}}\ and\ \bibinfo {author} {\bibfnamefont {Y.}~\bibnamefont {Li}},\
  }\href@noop {} {\bibfield  {journal} {\bibinfo  {journal} {SIAM J. Optimiz.}\
  }\textbf {\bibinfo {volume} {6}},\ \bibinfo {pages} {418} (\bibinfo {year}
  {1996})}\BibitemShut {NoStop}%
\bibitem [{\citenamefont {Coleman}\ and\ \citenamefont
  {Li}(1994)}]{coleman1994convergence}%
  \BibitemOpen
  \bibfield  {author} {\bibinfo {author} {\bibfnamefont {T.~F.}\ \bibnamefont
  {Coleman}}\ and\ \bibinfo {author} {\bibfnamefont {Y.}~\bibnamefont {Li}},\
  }\href@noop {} {\bibfield  {journal} {\bibinfo  {journal} {Math. Program.}\
  }\textbf {\bibinfo {volume} {67}},\ \bibinfo {pages} {189} (\bibinfo {year}
  {1994})}\BibitemShut {NoStop}%
\bibitem [{\citenamefont {Sargent~III}\ \emph {et~al.}(1974)\citenamefont
  {Sargent~III}, \citenamefont {Scully},\ and\ \citenamefont
  {Lamb}}]{Sargent1974Laser}%
  \BibitemOpen
  \bibfield  {author} {\bibinfo {author} {\bibfnamefont {M.}~\bibnamefont
  {Sargent~III}}, \bibinfo {author} {\bibfnamefont {M.}~\bibnamefont {Scully}},
  \ and\ \bibinfo {author} {\bibfnamefont {J.}~\bibnamefont {Lamb}},\
  }\href@noop {} {\emph {\bibinfo {title} {Laser physics}}}\ (\bibinfo
  {publisher} {Addison-Wesley, Massachusetts},\ \bibinfo {year}
  {1974})\BibitemShut {NoStop}%
\end{thebibliography}%

\end{document}